\documentclass{aa}

\usepackage{amsmath}
\usepackage{subfig,graphicx}
\usepackage{flushend}
\usepackage[varg]{txfonts}
\usepackage{hyperref}
\newcommand{\lya}{Ly$\alpha$}

\begin{document}  


\title{Determining the fraction of reddened quasars in COSMOS with multiple selection techniques from X-ray to radio wavelengths\thanks{Partly based on observations made with the Nordic Optical Telescope, operated by the Nordic Optical Telescope Scientific Association at the Observatorio del Roque de los Muchachos, La Palma, Spain, of the Instituto de Astrofisica de Canarias.}}

\author{
K.~E.~Heintz\inst{1},
J.~P.~U.~Fynbo\inst{1},
P. M\o ller\inst{2},
B.~Milvang-Jensen\inst{1},
J.~Zabl\inst{1},
N.~Maddox\inst{3},
J.-K.~Krogager\inst{1,4},
S.~Geier\inst{5,6},
M.~Vestergaard\inst{1,7},
P.~Noterdaeme\inst{4},
C.~Ledoux\inst{8}
}
\institute{
Dark Cosmology Centre, Niels Bohr Institute, Copenhagen University, Juliane Maries Vej 30, 2100 Copenhagen \O, Denmark\\
\email{heintz@dark-cosmology.dk}
\and
European Southern Observatory, Karl-Schwarzschildstrasse 2, D-85748 Garching, Germany
\and
ASTRON, the Netherlands Institute for Radio Astronomy, Postbus 2, 7990 AA, Dwingeloo, The Netherlands
\and
Institut d'Astrophysique de Paris, CNRS-UPMC, UMR7095, 98bis bd Arago, 75014 Paris, France
\and
Gran Telescopio Canarias (GRANTECAN), Cuesta de San Jos\'e s/n, E-38712 , Bre\~na Baja, La Palma, Spain 
\and
Instituto de Astrof\'isica de Canarias, V\'ia L\'actea s/n, E38200, La Laguna, Tenerife, Spain
\and
Steward Observatory, University of Arizona, 933 N. Cherry Avenue, Tucson AZ 85721, USA
\and
European Southern Observatory, Alonso de C\'ordova 3107, Vitacura, Casilla 19001, Santiago 19, Chile
}

\date{accepted for publication in \textit{Astronomy \& Astrophysics}, August 19, 2016}

\abstract{The sub-population of quasars reddened by
intrinsic or intervening clouds of dust are known to be underrepresented in
optical quasar surveys. By defining a complete parent sample of the brightest
and spatially unresolved quasars in the COSMOS field, we quantify to which extent
this sub-population is fundamental to our understanding of the true
population of quasars. By using the available multiwavelength data of various
surveys in the COSMOS field, we built a parent sample of 33 quasars brighter than
$J=20$ mag, identified by
reliable X-ray to radio wavelength selection techniques. Spectroscopic
follow-up with the NOT/ALFOSC was carried out for four candidate quasars that had not been
targeted previously to obtain a 100\% redshift
completeness of the sample. The population of high $A_V$ quasars (HAQs), a
specific sub-population of quasars selected from optical/near-infrared photometry, some of which were shown to be missed in large optical surveys such as SDSS, is
found to contribute $21\%^{+9}_{-5}$ of the parent sample. The full population of
bright spatially unresolved quasars represented by our parent sample 
consists of $39\%^{+9}_{-8}$ reddened quasars defined by having $A_V>0.1$,
and $21\%^{+9}_{-5}$ of the sample having $E(B-V)>0.1$ assuming the extinction
curve of the Small Magellanic Cloud. 
We show that the HAQ selection works
well for selecting reddened quasars, but some are missed because their
optical spectra are too blue to pass the $g-r$ color cut in the HAQ selection. This is
either due to a low degree of dust reddening or anomalous spectra.
We find that the fraction of quasars with contributing 
light from the host galaxy, causing observed extended spatial morphology, is most 
dominant at $z \lesssim 1$. At higher redshifts the
population of spatially unresolved quasars selected by our parent sample is found to 
be representative of the full population of bright active galactic
nuclei at $J<20$ mag. This work 
quantifies the bias against reddened quasars in studies that
are based solely on optical surveys.
}

\keywords{quasars:~general -- galaxies:~active -- surveys -- dust, extinction}

\titlerunning{HAQs in COSMOS}
\authorrunning{Heintz et al. 2016}

\maketitle

\section{Introduction}     \label{sec:introduction}

The nature of quasars and their cosmological applications
are of central importance in contemporary astrophysics. To properly understand
the origin and physical nature of quasars in a general context, great caution has
to be applied when defining the techniques with which these objects are
selected and samples are built. For many applications the assembly of large
samples of quasars has to be representative of the general population since
selection bias will give a distorted picture of the 
full underlying population. Most large-area quasar surveys have relied on optical
photometric selection, for instance, the Sloan Digital Sky Survey
\citep[SDSS;][]{York00,Richards02} and the 2dF quasar redshift survey
\citep[2QZ;][]{Croom04}. However, they have been found to be systematically
biased against intrinsically dust-reddened quasars
\citep[e.g.,][]{Richards03,Krawczyk15} and intervening dust-rich
absorbers reddening the background quasar
\citep{Noterdaeme09b,Noterdaeme10,Noterdaeme12,Kaplan10,Fynbo11,Wang12,Krogager16}.
Moreover, optical surveys contain other biases, for example, against quasars above
redshifts of $z>2.5$, at which point the Lyman-$\alpha$ (\lya) forest (\lya
F) enters the optical $u$-band, effectively removing the observed UV excess
(UVX) of quasars compared to that of stars \citep{Fan99}.

It it therefore important to understand to what extent various quasar samples suffer
from biases against dust-reddened systems and quasars at high redshifts so that we
can build more complete and representative samples. A large number of studies
targeting reddened quasars have been executed in the past two decades \citep[see, e.g.,][]{Webster95,Warren00,Gregg02,Richards03,Hopkins04,Glikman07,Glikman12,Glikman13,Maddox08,Maddox12,Urrutia09,Banerji12,Fynbo13,Krogager15},
all trying to probe this underrepresented subset of quasars using a range of selection
techniques that are less sensitive to dust obscuration. 
With the advent of
large-area near- and mid-infrared surveys such as UKIDSS, \textit{WISE,} and \textit{Spitzer}
\citep[][]{Wright10,Cutri13,Lawrence07,Warren07,Werner04,Fazio04}, the selection of quasars based
on infrared (IR) photometry alone has been made possible as well
\citep{Donley08,Donley12,Peth11,Stern12,Mateos12,Secrest15}. Quasars identified by
their mid-infrared colors can be selected without requiring X-ray or radio
detections and blue optical colors. In addition, other types of
selection techniques have recently been proposed that are unbiased in terms of colors
\citep{Schmidt10,Graham14,Heintz15}, to independently probe the full underlying
population of quasars.

How large the missing population of dust-reddened quasars is compared to the
full population is still debated. Recent studies estimate fractions from below
10\% to above 40\%
\citep{Richards03,Richards06,Glikman04,Glikman12,Gibson09,Urrutia09,Allen11,Maddox08,Maddox12}.
The majority of objects in these samples consist of broad absorption line
(BAL) quasars. The search for quasars reddened by
intervening absorbers, dubbed the high $A_V$ quasar
\citep[HAQ;][]{Fynbo13,Krogager15} survey, reveals that a large portion of
intrinsically reddened quasars are missing in optical samples as well. 

In this paper we examine the fraction of the HAQs and in general the dust-reddened population of quasars (which we define as having $A_V>0.1$) compared to
the total. The aim is to determine whether this missing sub-population
constitutes a significant fraction of the true population of quasars and if the reddest of the most
bright and spatially unresolved quasars are indeed identified by the tailored
optical and near-infrared (NIR) color criteria used in the HAQ survey. 
We have decided to focus on spatially unresolved sources because the selection of point sources is a central part of the HAQ survey. In the same vein, we
focused on relatively bright quasars, $J<20$ mag because we are interested in the population 
of intrinsically bright quasars that are typically used for 
quasar absorption lines studies, for example. To
securely determine the fraction, an effective selection of the complete
population of quasars within the appropriate limit in brightness has to be
obtained. We assume that this can be done using multiple selection techniques
covering the different observed features of quasars.

The paper is structured as follows. In Sect.~\ref{sec:data} we briefly describe
the different data sets obtained from the COSMOS surveys and how these
were combined to apply all the multiwavelength information to each object
within our defined region of the COSMOS field. In Sect.~\ref{sec:selmet} we
define the quasar selection techniques used to build our parent sample,
representing the full population of the brightest spatially unresolved quasars. In
Sect.~\ref{sec:obs} we describe our follow-up observations, and in
Sect.~\ref{sec:res} we present the results. In Sect.~\ref{sec:disc} we discuss
our analysis, and in Sect.~\ref{sec:conc} we conclude on the implication of our
work in a general context.  We assume a standard flat $\Lambda CDM$ cosmology
with $H_0=70$ km s$^{-1}$ Mpc$^{-1}$, $\Omega_M=0.3$ and
$\Omega_{\Lambda}=0.7$. Unless otherwise stated, the AB magnitude system
\citep{Oke74} is used throughout this paper.

\section{Data description}    \label{sec:data}

By using the extensive multiwavelength data sets available from the combined
COSMOS \citep{Scoville07a} surveys, all covering the $Chandra$ COSMOS
\citep[C-COSMOS;][]{Elvis09,Civano12} field, the different methods for quasar
selection (see Sect.~\ref{subsec:haqsel}-\ref{subsec:sbsel}) can be applied. 
This sub-field of COSMOS covers the central 0.9 deg$^2$ of the
approximately two square degrees of the COSMOS field (see, e.g., Fig.~1 in \citet{Elvis09} 
and Fig.~1 in \citet{McCracken12} for the specific outline of the field). 
This region is one of the most observed fields
on the sky, making it excellent for detailed studies of general quasar properties
and populations, and it is therefore ideal for our study. By building a parent sample
consisting of the quasars selected with a range of different methods, the fraction of the HAQs compared
to the full underlying population, that is, the extent to which the HAQs
contribute to the defined parent sample of quasars, can be determined. Although
extensive spectroscopic analysis is available in this field, a minority
of the candidate quasars from the parent sample were without existing spectra, so that
follow-up spectroscopy was required to confirm their nature for the purpose of this study.

\subsection{COSMOS field: from X-ray to radio wavelengths} \label{subsec:xraytoradio}

The COSMOS field was originally defined by \cite{Scoville07a} as a two deg$^2$ region with multiwavelength coverage, including 
HST/ACS F814W imaging over 1.8 deg$^2$ \citep{Scoville07b}. It was observed using the Advanced Camera for Surveys (ACS) Wide Field Channel (WFC) and imaged in the F814W filter which is broad $I$-band filter centered on an effective wavelength of 7940\AA with a resolution of $\sim 0\farcs1$. For a description, see for example \cite{Koekemoer07} for how the HST-ACS mosaic was produced and \cite{Leauthaud07} for how the mosaic was used to assemble a source catalog. This catalog consists of approximately $1.2\times 10^6$ objects down to a limiting magnitude of $F814W=26.6$ mag.

In our study we used the X-ray information from the initial $Chandra$ COSMOS \citep[C-COSMOS;][]{Elvis09,Civano12} and the XMM-COSMOS
surveys \citep{Hasinger07,Cappelluti09,Brusa10}. C-COSMOS, an extensive 1.8 Ms \textit{Chandra} program, has imaged the central 0.9 deg$^2$ of the COSMOS field down to the limiting fluxes of $1.9\times 10^{-16}$ erg cm$^{-2}$ s$^{-1}$ in the soft (0.5 - 2 keV) band, $7.3\times 10^{-16}$ erg cm$^{-2}$ s$^{-1}$ in the hard (2 - 10 keV) band and $5.7\times 10^{-16}$ erg cm$^{-2}$ s$^{-1}$ in the full (0.5 - 10 keV) band \citep{Elvis09}. Complementary to {\it Chandra} are the X-ray data from the XMM-COSMOS mission, which is a contiguous 2~deg$^2$ \textit{XMM-Newton} survey of the COSMOS field. The limiting fluxes are $5\times 10^{-16}$, $3\times 10^{-15}$ and $7\times 10^{-17}$ erg cm$^{-2}$ s$^{-1}$ in the $0.5-2$, $2-10$ and $5-10$ keV bands, respectively \citep{Hasinger07}. The flux limits of the two surveys are not reached over the complete area but vary across the field because of irregular exposure
times. The sensitivity of C-COSMOS due to its sub-arcsecond imaging is three times below the corresponding flux limits for the XMM-COSMOS survey.

The initial C-COSMOS survey used in this paper consists of 1761 X-ray detected point sources, where the XMM-COSMOS survey have observed 1797 X-ray point sources. \cite{Elvis09} reported that 70 sources from the XMM-COSMOS catalog were absent
from the C-COSMOS sample, while 24 XMM-COSMOS sources have been resolved into two distinct sources in the C-COSMOS catalog. The optical/near-infrared COSMOS counterparts were found for each of the X-ray point source detections with \textit{Chandra} and \textit{XMM-Newton} \citep[see][respectively]{Civano12,Brusa10}. Based on these counterparts, the photometric redshifts of each of the sources have been determined by \cite{Salvato09,Salvato11}.

In addition to the two X-ray surveys, we used the photometric data from the multiband catalog described in \cite{Ilbert13}. This catalog consists of multiband photometry covering the near-ultraviolet (NUV) to the mid-infrared (MIR) effective wavelengths. The NUV-optical data sets are from previous releases, described in \citet{Capak07,Ilbert09}. The catalog by \cite{Capak07} also contains optical photometry in the $u,g,r,i,z$ bands from the SDSS data release 2 \citep[DR2;][]{Abazajian04}, produced by facilitating photometric measurements from a mosaic of the SDSS placed on the same grid as the other COSMOS data. We used the star/galaxy classifier from the \cite{Capak07} catalog, which allowed us to only target optical point sources (spatially unresolved quasars).
At wavelengths redward of the UltraVISTA ($\sim 1-2~\mu$m) coverage, we used the MIR data from the \textit{Wide-field Infrared Survey Explorer} \citep[\textit{WISE};][]{Wright10} most recent data release \citep[AllWISE;][]{Cutri13}. This all-sky mission has imaged the sky in four different MIR filters with effective wavelengths ranging from $3.4~\mu$m to $22~\mu$m. Finally, we included the radio ($\sim 21$ cm) data from the deep and joint VLA-COSMOS catalog \citep{Schinnerer07,Schinnerer10}. Detailed imaging from the various COSMOS surveys is publicly available at the NASA/IPAC Infrared Science Archive (IRSA) web page\footnote{{\tt http://irsa.ipac.caltech.edu/data/COSMOS/index\_cutouts.html}}.

\subsection{Matching procedure}\label{subsec:photcat}

We required that each of the sources from the various COSMOS data sets are within 2 arcsec from the objects listed in the photometric catalog by \citet[][hereafter {\tt I13}]{Ilbert13}. Since multiple objects can be within this region, we selected
only the nearest match. For the \textit{Chandra/XMM-Newton} detected X-ray point sources, we matched the optical/NIR counterparts found by \citet{Civano12,Brusa10}, respectively, to the photometric catalog by {\tt I13}. In the first COSMOS optical/NIR catalog by \cite{Capak07}, the morphology information from the CFHT+Subaru $i$-band images was included. Therefore only this catalog had to be matched to the photometric catalog by {\tt I13}, again within 2 arcsec. This was done primarily to attach the morphological data to the UltraVISTA DR1 sources, but also to obtain the SDSS five-band magnitudes for each of the objects. Following the same procedure, we paired the objects detected in the AllWISE and VLA-COSMOS data sets to the optical/NIR counterparts in the photometric catalog ({\tt I13}) as well.

\section{Selection techniques}    \label{sec:selmet}

\begin{figure} 
        \centering
            \includegraphics[width=1.0\columnwidth]{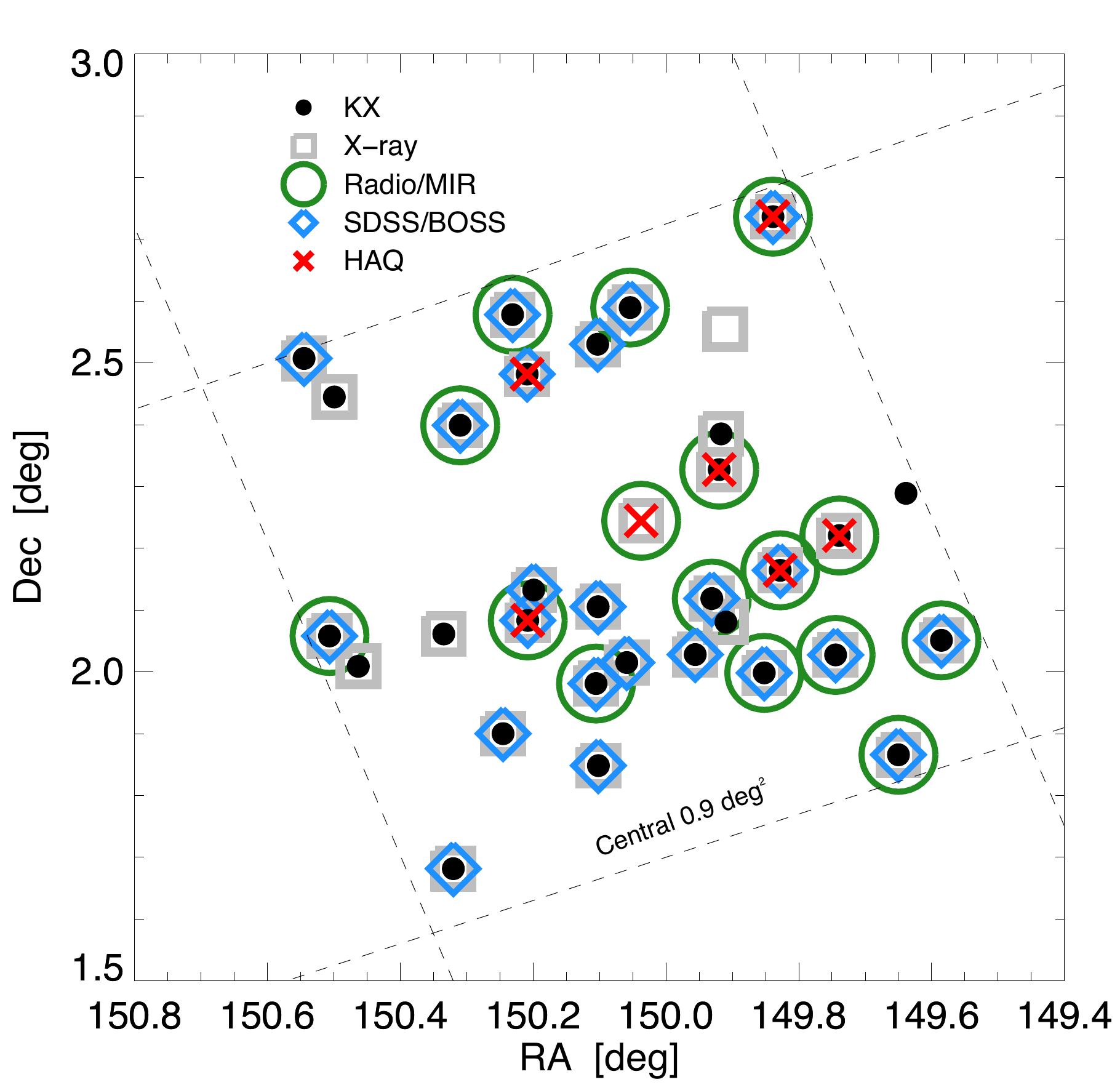}
            \caption[Network2]%
            {Location of the 34 confirmed and candidate quasars from the parent sample. Overplotted are the cuts in right ascension (RA) and declination (Dec) used to isolate the C-COSMOS field. The symbols represent individual selection methods used to identify each candidate quasar (see Sect.~\ref{subsec:haqsel}-\ref{subsec:sbsel}).}
            \label{fig:radeccuts}
    \end{figure}

Since the publicly available C-COSMOS survey is limited to the central $0.9$ deg$^2$ of the COSMOS field, see for instance Fig.~1 of \cite{Elvis09}, we isolated this region from the rest of the data sets. In Fig.~\ref{fig:radeccuts} the confirmed and the candidate quasars (specifics of the subsample symbols are defined below) within this specific region is shown, where the gray dashed lines represent the coverage of the C-COSMOS survey, indicating our targeted region of the COSMOS field. Applying these boundaries furthermore ensures that the whole sub-field examined is covered by all the multiwavelength surveys.

\subsection{Morphology and brightness}

In the original HAQ criteria (and most other optical/NIR color selection approaches) the targeted objects were defined as optical and NIR point sources. In this study we follow the same procedure and select spatially unresolved sources at observed optical wavelengths only (using the star/galaxy classifier of \cite{Capak07}, based on ground-based photometry from the CFHT+Subaru $i$-band images), while discarding the sources with dominating light from the host galaxy. This means that we only considered the optically unresolved quasar population and not the overall/general active galactic
nucleus (AGN) population. We discuss the implication of this specific morphology cut in detail in Sect.~\ref{sec:disc}. Moreover, we also defined an upper magnitude limit of $J_{\mathrm{AB}} < 20$ mag to avoid being biased against the depth of each of the individual surveys. This magnitude limit furthermore ensures a fair comparison of each of the respective quasar populations within a certain brightness. This provides a more secure detection and low errors on the photometric data. In the following sections we describe how each of the specific quasar selection techniques were applied. We only considered the optically unresolved sources that were brighter than our magnitude limit before the distinctive selection techniques.

\subsection{HAQ selection} \label{subsec:haqsel}

This tailored search for reddened quasars was originally designed to search for dust-rich damped \lya~absorbers (DLAs) along the line of sight toward the background source. Dust-rich DLAs are known to redden the background source to the extent where optical color selection fails to classify them as quasars \citep{Noterdaeme09b,Noterdaeme10,Noterdaeme12,Fynbo11,Kaplan10,Krogager16}.
To select HAQ candidates from the COSMOS photometric catalog by {\tt I13}, we adopted the same criteria as in \cite{Krogager15}, all on the AB magnitude system:
\begin{align}
\begin{split}
&J-K_s > 0, ~~~ H-K_s > 0, ~~~ J-H < 0.4,\\
&0.5 < g-r < 1.0, ~~~ 0.1 < r-i < 0.7~~.
\end{split}
\label{eq:haq}
\end{align}
These criteria were defined based on the pilot study executed before the HAQ survey, described in \cite{Fynbo13}. They were found to identify quasars (not necessarily all having high $A_V$,
as the name suggests) with a purity of $P \sim 97 \%$ (154/159) from the set of candidates selected from optical/NIR photometry, where the remaining objects were two dwarf stars and three so-far unidentified objects. While only a handful of DLAs were found, numerous intrinsically reddened quasars were detected with a pronounced amount of extinction from dust, quantified by the parameter $A_V$. Only $\sim$45\% of the HAQ candidates from the original study were spectroscopically classified as quasars in DR7 before observation. A small fraction of the spectroscopically verified HAQs were later observed by the SDSS-III/BOSS \citep[DR10Q;][]{Paris14} program, which relies on more advanced selection algorithms than the original optical color criteria of SDSS-I/II \citep[see, e.g.,][and Sect.~\ref{subsec:sbsel}]{Ross12,Dawson13}. To distinguish the HAQ selected quasars from the general population of reddened quasars, we use the term HAQs for the quasars that were selected by the same (and very limiting) optical/NIR criteria of \cite{Krogager15}. We found that seven objects satisfy the HAQ criteria. The HAQ selected candidate quasars are marked with an "H" in Col. 3 in Table~\ref{tab:comptab} and with red crosses in Fig.~\ref{fig:radeccuts}. Owing to different redshifts, intrinsic slopes, and other absorption features, not all quasars selected in this way do in fact have high $A_V$, but we adopted the original name for consistency. The quasars with $A_V>0.1$ are referred to as red quasars or the general reddened population of quasars. Defining this lower limit ensures that quasars with natural variability \citep[typically $\pm 0.05$ mag, as inferred from][]{Krogager15} are not considered as part of the reddened population of quasars.

\subsection{KX selection}

Another approach to identify the missing reddened quasars, regardless of redshift and amount of dust reddening, is to exploit the general $K$-band excess (KX) shown by all quasars \citep{Warren00,Croom01,Maddox08,Maddox12} hereafter referred to as KX. This approach is comparable to the HAQ selection criteria with the NIR $K$-band excess of quasars being the primary quasar/star classifier, although not as tailored for targeting objects with red optical colors. This selection was therefore not as specialized in selecting quasars with intervening DLAs as the HAQ survey, but is unbiased toward those objects in the same way, and will therefore be more complete in detecting quasars. The KX selected candidate quasars were selected from the COSMOS photometric catalog by {\tt I13} as well, where we converted the original selection criteria from Vega to AB magnitudes. We refined the magnitude range used in their study of $14.0 < K_{Vega} < 16.6$ to $K_{s,AB}>15.85,~J_{AB} \leq 20$ mag. Here, the lower limit was converted from Vega into AB magnitudes \citep[following the formulation of][]{Blanton07}, and the upper limit is $\sim 0.5-1$ magnitude fainter than the original to match our overall magnitude limit. The refined criteria, that is, the selection lines, are then as follows, again all on the AB magnitude system:
\begin{equation}
g-J<\begin{cases}
    4.4(J-K_s)-2.08, & \text{for $J-K_s \leq -0.04$, $g-J \leq 2.01$}\\
    20.83(J-K_s)-2.83, & \text{for $J-K_s \ge -0.04$, $g-J \ge 2.01$}
  \end{cases}
  \label{eq:kx}
.\end{equation}
These can be directly compared to the selection criteria and the selection lines in Fig.~2 of \cite{Maddox08,Maddox12}. Following the original KX selection procedure, we proceeded to run the candidate quasars without existing spectra, detected by their $K$-band excess, through the custom-made photo-$z$ algorithm described in \cite{Maddox08,Maddox12}. This algorithm is based on a standard model quasar template with seven reddened models created from the base model with increasing amounts of Small Magellanic Cloud (SMC)-type dust. To exclude compact galaxies, seven galaxy templates spanning E through Scd types were included. Star templates were also incorporated obtained from the Bruzual–Persson–Gunn–Stryker atlas\footnote{{\tt http://www.stsci.edu/hst/observatory/cdbs/bpgs.html}}. By first applying the $gJK$ color cut, about $95\%$ of the objects were removed in the original sample. Then, by applying the photo-$z$ code to the remaining objects, most of the contaminants were excluded, with colors inconsistent with standard quasars. After the photometric data of the candidates with no pre-existing spectra were processed by the photo-$z$ code, the sample of KX selected candidate quasars consisted of 31 objects. The KX selected candidate quasars are marked with a K in Col. 3 in Table~\ref{tab:comptab} and with black dots in Fig.~\ref{fig:radeccuts}.

\subsection{X-ray detection}

Selection at X-ray wavelengths has been found to effectively identify quasars, or AGNs in general \citep{Brandt05}, with an insignificant amount of contamination from star-forming galaxies, which are otherwise present in optical or infrared surveys \citep{Donley08,Donley12}. Stars with strong X-ray emission can also appear in X-ray samples, but because of their distinct stellar colors, they can easily be rejected as non-quasar candidates. Moreover, X-ray surveys are efficient in detecting low-luminosity and obscured quasars below the \textit{Compton-thick} limit. To select quasars detected by \textit{Chandra/XMM-Newton,} we required that the optical/NIR counterparts of the X-ray point sources were unresolved in the CFHT+Subaru $i$-band images. The flux limits of the two X-ray surveys and the matching to the optical/NIR counterpart to apply the morphological information are described in Sects.~\ref{subsec:xraytoradio} and \ref{subsec:photcat}, respectively.

A significant amount of the selected objects were stars with strong X-ray emission. We discarded the X-ray point sources determined to be Galactic stellar sources by \cite{Salvato09,Salvato11}, based on either spectroscopically confirmed spectra or photometrically computed stellar fits by their $\chi^2$ analysis. No requirements were imposed on the optical/IR photometry, except for the general defined magnitude limit. In Sect.~\ref{sec:disc} we discuss in
more detail the morphological context of the objects that are point sources at X-ray wavelengths but not in the optical, or
in other words, the general AGN X-ray population. We found 32 objects detected with \textit{Chandra/XMM-Newton} all having a spatially unresolved optical/NIR counterpart in the photometric catalog by {\tt I13}. The X-ray selected candidate quasars are marked with an X in Col. 3 in Table~\ref{tab:comptab} and with gray squares in Fig.~\ref{fig:radeccuts}.

\subsection{Radio detection}

Apparent optical stellar sources with extreme radio emission represent the first detected subpopulation of radio-loud quasars \citep{Schmidt63,Matthews63}. This selection is not affected by extinction at optical wavelengths and would therefore not be biased against dust-reddened quasars, see, for instance, \cite{Becker97}. 
By definition, only the radio-loud quasars will be targeted by this approach, so that the sample in itself is not complete \citep[see, e.g.,][for a description of the radio emission from quasars and the empirical observations of the two sub-populations, respectively]{Urry95,Sikora07}. Two recent radio selections of optical spatially unresolved quasars \citep{Ivezic02,Balokovic12} used data from the Faint Images of the Radio Sky at Twenty centimeters \citep[FIRST;][]{Becker95} mission together with the SDSS program. A similar approach was executed based on source detection with the FIRST matched to the NIR counterparts of the Two Micron All Sky Survey \citep[2MASS;][]{Skrutskie06} by \citet{Glikman07,Glikman12,Urrutia09}. Equivalent to these, we selected candidate quasars detected by VLA by only requiring that the optical/NIR counterpart of the radio-detected sources were unresolved in the CFHT+Subaru $i$-band images. No further constraints were set on the VLA data, and we found that nine sources were selected in this way. The radio-selected candidate quasars are marked with an R in Col. 3 in Table~\ref{tab:comptab} and with green circles in Fig.~\ref{fig:radeccuts}.

\subsection{Mid-infrared selection}

Recently, with the launch of the complementary MIR \textit{Spitzer} and \textit{WISE} missions, multiple new approaches of selecting quasars using MIR colors alone have been developed. See, for
example, the \textit{Spitzer} two-color criteria of \cite{Donley12}, the \textit{WISE} one-color criteria of \citet{Stern12,Assef13}, and the \textit{WISE} two-color criteria of \cite{Mateos12}. All these criteria were defined by the unique colors of quasars in MIR color space, primarily tracing the torus around the super-massive black hole (SMBH) of the AGN. A selection based on the MIR wavelength range is insensitive to dust extinction and will therefore not be biased toward reddened quasars \citep{Mateos13,LaMassa16}. We chose the \textit{WISE} two-color criteria of \cite{Mateos12} as used by \cite{Secrest15} to represent the MIR selected quasars. They defined the AGN locus in Vega magnitudes by
\begin{equation}
W1-W2 = 0.315\times (W2-W3),
  \label{eq:mir}
\end{equation}
where the top and bottom boundaries of this two-color wedge are obtained by adding y-axis intercepts of +0.796 and -0.222, respectively \citep[see Eqs. 3 and 4 of][]{Mateos12}.
This selection, compared to the other MIR selection approaches, was found to show less contamination from brown dwarfs and to be more complete. Following the same procedure as for the X-ray and radio-detected candidate quasars, we discarded objects that were selected by the MIR two-color criteria but were spatially resolved in the CFHT+Subaru $i$-band images. Again, the implication for this specific selection is discussed in Sect.~\ref{sec:disc}. By using these criteria, we found ten sources selected by their MIR colors and spatially unresolved counterpart in {\tt I13}. The MIR-selected candidate quasars are marked with an M in Col.
3 in Table~\ref{tab:comptab} and with green circles in Fig.~\ref{fig:radeccuts}.

\subsection{SDSS-III/BOSS DR12Q sample}  \label{subsec:sbsel}

The DR12Q sample is a combination of the pre-BOSS SDSS-I/II quasar selection program \citep[see, e.g.,][for the selection criteria and the full pre-BOSS sample, respectively]{Richards02,Schneider10}, which focused on quasars with high UVX and objects with outlying optical colors compared to that of stars. The SDSS-III/BOSS program depended on numerous other more advanced selection algorithms \citep[see][for a description of the selection algorithms and the spectroscopic survey of the SDSS-III/BOSS programme]{Ross12,Dawson13}, specifically designed to target quasars at $z > 2.2$.
We included all the confirmed quasars from the pre-BOSS (SDSS-I/II) and SDSS-III/BOSS DR12Q samples (P\^{a}ris et al. 2016, in prep.) within the central region of the COSMOS field with the same requirements of morphology and brightness from the COSMOS photometric catalog by {\tt I13}. In Sect.~\ref{sec:disc} we directly show which of the SDSS/BOSS-selected quasars we discarded due to these cuts. Some of the spectroscopically verified quasars have been observed by the SDSS/BOSS programs only because they were detected in the \textit{ROSAT, FIRST, Chandra,} or \textit{XMM} missions. These were not included as SDSS- or BOSS-selected quasars in our sample, but with the corresponding X-ray or radio detection instead. This was done to examine the photometric selection of these two selection functions alone, since X-ray and radio selection is a part of our study by design. Notes on how each of the SDSS/BOSS quasars were selected are given in the appendix, Fig.~\ref{fig:specimgquasar}. We found that 23 sources were selected based on their photometry by the SDSS/BOSS programs, where 15 quasars are from the SDSS and 8 quasars are from the BOSS survey. The SDSS/BOSS photometrically selected candidate quasars are marked with either an S or B for SDSS or BOSS selection, respectively, in Col. 3 in Table~\ref{tab:comptab} and with blue diamonds in Fig.~\ref{fig:radeccuts}.

\subsection{Parent sample} \label{subsec:ps}

Combining the confirmed and the candidate quasars obtained from all six selection
approaches, we found a total number of 34 objects brighter than $J=20$
magnitude, all located in the C-COSMOS field. The majority of the quasars (30/34)
have existing spectra in the literature (see, e.g., the SDSS DR12
database\footnote{{\tt http://skyserver.sdss.org/dr12/en/home.aspx}}), whereas the
remaining four objects had to be verified spectroscopically (see next section).
This is necessary to verify the computed photometric
redshift and to calculate the extinction of each of the quasars by analyzing the
amount of reddening from the slope of the spectral energy distribution
(SED) and the photometric data points.
In Table~\ref{tab:comptab} all candidate and confirmed quasars are listed by name together with the
respective techniques by which they were selected. The letters
denoting each of the selection techniques, as shown in the table, are used
in the following sections. The names are listed either as Q+coordinates for
known quasars or CQ+coordinates for the candidate quasars.

\begin{table*}
\centering
\caption{Number, object name, selection technique(s), {\large $z_{\mathrm{spec}}$}, {\large $z_{\mathrm{phot}}$}, $A_V$, brightness in $J$-band, and additional notes for each of the confirmed quasars in the parent sample. Following the notation in Sect. \ref{subsec:haqsel} -- \ref{subsec:sbsel} for the selection techniques; K = KX selection, X = X-ray detection, R = radio detection, M = MIR selection, S/B = SDSS/BOSS photometric selection, and H = HAQ selection.}
\begin{tabular*}{1.0\textwidth}{@{\extracolsep{\fill}}l r l c c c c c l}
\hline\hline
\# & Object & Selection &{\large $z_\mathrm{spec}$} & {\large $z_\mathrm{phot}$} & A$_V$         & A$_V$                       &  $J$-band   & Notes\\
   &            &                       &                                                        &                                                         &     (SMC)      & (Zafar+15)           & (mag)    & \\  
   (1) & (2) & (3) & (4) & (5) & (6) & (7) & (8) & (9) \\

\hline
1  & Q\,095820.5+020304.1  & KXMS & 1.356 & 1.462 & 0.00 & 0.00 & 19.70 & \textit{g-r, r-i} \\ 
2  & CQ\,095833.3+021720.4 & K & 1.910 & 1.750 & 0.23 & 0.23 & 19.99 & \textit{g-r} \\
3  & Q\,095836.0+015157.1  & KXRB & 2.935 & 2.919 & 0.00 & 0.00 & 19.32 & \textit{g-r} \\
4  & Q\,095857.4+021314.5  & KXRH & 1.024 & 1.043 & 0.38 & 0.30 & 19.41 & \textit{} \\
5  & Q\,095858.7+020139.1  & KXMB & 2.448 & 2.418 & 0.00 & 0.00 & 18.70 & \textit{g-r, r-i} \\
6  & Q\,095918.7+020951.7  & KXRMSH & 1.156 & 1.181 & 0.30 & 0.15 & 19.72 & \textit{} \\
7  & Q\,095921.3+024412.4  & KXRBH & 1.004 & 0.030 & 0.98 & 0.60 & 19.37 & \textit{} \\
8  & CQ\,095924.4+020842.6 & K &  ---  & 1.150 & ---  & ---  & 19.43  & \textit{g-r}, blended SED, non-quasar \\
9  & Q\,095924.5+015954.3  & KXMS & 1.241 & 1.281 & 0.00 & 0.00 & 18.49 & \textit{g-r, r-i} \\
10 & Q\,095938.3+020450.1  & KX & 2.802 & 2.779 & 0.23 & 0.08 & 19.74 & \textit{H-K$_s$} \\
11 & CQ\,095938.6+023316.7 & X & 0.740 & 0.742 & 1.21 & 0.68 & 19.47 & \textit{g-r} \\
12 & Q\,095940.1+022306.7  & KX & 1.131 & 1.110 & 0.15 & 0.15 & 19.93 & \textit{g-r, r-i} \\
13 & Q\,095940.8+021938.7  & KXRH & 1.454 & 1.465 & 0.30 & 0.30 & 19.74 & \textit{} \\
14 & Q\,095943.4+020707.4  & KXRB & 2.194 & 2.306 & 0.00 & 0.00 & 19.45 & \textit{g-r, r-i} \\
15 & Q\,095949.4+020141.0  & KXS & 1.753 & 1.753 & 0.00 & 0.00 & 18.99 & \textit{g-r} \\
16 & CQ\,100008.9+021440.7 & XRMH & 2.680 & 2.663 & 0.15 & 0.15 & 18.96 & \textit{} \\
17 & Q\,100012.9+023522.8  & KXMS & 0.698 & 0.702 & 0.00 & 0.08 & 18.47 & \textit{g-r} \\
18 & Q\,100014.1+020054.5  & KXS & 2.498 & 2.469 & 0.00 & 0.00 & 19.45 & \textit{g-r, r-i} \\
19 & Q\,100024.4+015054.0  & KXS & 1.661 & 1.669 & 0.30 & 0.30 & 19.61 & \textit{g-r, H-K$_s$} \\
20 & Q\,100024.5+020619.8  & KXB & 2.288 & 2.287 & 0.00 & 0.00 & 19.51 & \textit{g-r, r-i} \\
21 & Q\,100024.6+023149.1  & KXS & 1.319 & 1.362 & 0.00 & 0.00 & 19.36 & \textit{g-r, r-i, H-K$_s$} \\
22 & Q\,100025.3+015852.1  & KXMS & 0.372 & 0.372 & 0.00 & 0.00 & 19.07 & \textit{g-r} \\
23 & Q\,100047.8+020756.8  & KXB & 2.159 & 2.177 & 0.00 & 0.00 & 19.67 & \textit{g-r, r-i} \\
24 & Q\,100049.9+020500.0  & KXRMBH & 1.236 & 1.272 & 0.30 & 0.30 & 19.04 & \textit{} \\
25 & Q\,100050.1+022854.8  & KXBH & 3.365 & 3.378 & 0.00 & 0.00 & 19.64 & \textit{} \\
26 & Q\,100055.4+023441.4  & KXRS & 1.402 & 1.414 & 0.08 & 0.08 & 19.74 & \textit{g-r, r-i} \\
27 & Q\,100058.8+015400.3  & KXS & 1.560 & 1.557 & 0.00 & 0.00 & 19.87 & \textit{g-r, r-i} \\
28 & Q\,100114.3+022356.7  & KXMS & 1.802 & 1.777 & 0.00 & 0.00 & 19.38 & \textit{g-r} \\
29 & Q\,100116.8+014053.6  & KXS & 2.055 & 2.050 & 0.00 & 0.00 & 19.76 & \textit{g-r} \\
30 & Q\,100120.3+020341.2  & KX & 0.903 & 0.906 & 0.00 & 0.00 & 19.77 & \textit{g-r, r-i} \\
31 & Q\,100151.1+020032.5  & KX & 0.967 & 1.000 & 0.23 & 0.11 & 19.79 & \textit{g-r, r-i} \\
32 & Q\,100159.8+022641.7  & KX & 2.030 & 2.011 & 0.00 & 0.00 & 19.27 & \textit{g-r} \\
33 & Q\,100201.5+020329.4  & KXMS & 2.016 & 1.880 & 0.23 & 0.23 & 18.30 & \textit{g-r} \\
34 & Q\,100210.7+023026.2  & KXS & 1.160 & 1.340 & 0.00 & 0.00 & 19.17 & \textit{g-r, r-i}, XMM-COSMOS source \\
\hline
\end{tabular*}
\label{tab:comptab}
\end{table*}

\section{Spectroscopic observations and analysis}    \label{sec:obs}

\subsection{Observations and data reduction}

The observations of the four candidate quasars were carried out during an observing run with the Nordic Optical Telescope (NOT) on La Palma on February 12--15, 2016. The spectra were obtained using the Andalucia Faint Object Spectrograph and Camera (ALFOSC). In Table~\ref{tab:obs_setup} the setup for the spectroscopic follow-up of these four candidate quasars is summarized. Grism~4 covers the wavelengths $3200-9100$~\AA~with a spectral resolution of $\sim$300, whereas grism~18 is limited to the blue part of the spectrum covering the wavelengths of $3530-5200$~\AA. This was only used for CQ095924.4+020842.6 since it appeared to have few to no emission features in the spectrum obtained with grism~4. Blocking filter 94 was used in combination with grism 4 to eliminate second-order contamination from wavelengths shorter than 3560 \AA. The spectra were taken by aligning the slit at parallactic angle.

The spectra were processed using a combination of IRAF\footnote{IRAF is the Image Reduction and Analysis Facility, a general purpose software system for the reduction and analysis of astronomical data. IRAF is written and supported by
the National Optical Astronomy Observatories (NOAO) in Tucson, Arizona. NOAO is operated by the Association of Universities for Research in Astronomy (AURA), Inc. under cooperative agreement with the National Science Foundation} and MIDAS\footnote{ESO-MIDAS is a copyright protected software product of the European Southern Observatory. The software is available under the GNU General Public License.} tasks for low-resolution spectroscopy. To reject cosmic rays, we used the LA-Cosmic software\footnote{Cosmic-Ray Rejection by Laplacian Edge Detection} developed by \cite{vanDokkum01}. The flux calibration was done using a spectrophotometric standard star observed on the same night as the science spectra. We corrected the spectra for Galactic extinction using the extinction maps of \cite{Schlegel98}. To improve the absolute flux calibration, we scaled the spectra to be consistent with the $r$-band photometry from SDSS.

\subsection{Spectroscopic analysis}

\begin{figure*} [!ht] 
\centering
         \begin{minipage}[t]{0.45\textwidth}
            \centering
            \includegraphics[width=\textwidth]{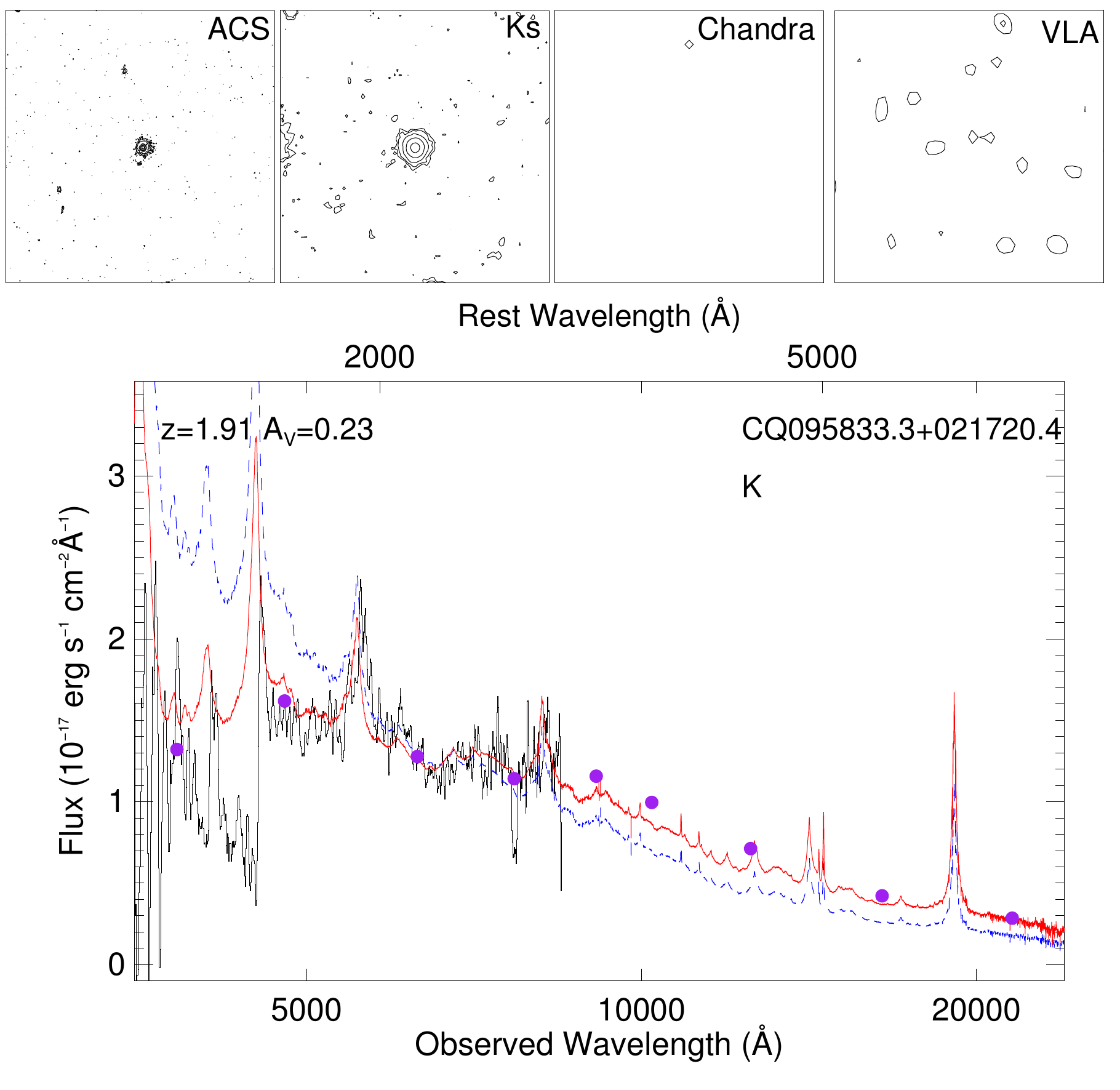}
            {\textbf{CQ: 2}} 
            \label{fig:quasar2}
\end{minipage} %
\hspace{1cm}%
         \begin{minipage}[t]{0.45\textwidth}
            \centering
            \includegraphics[width=\textwidth]{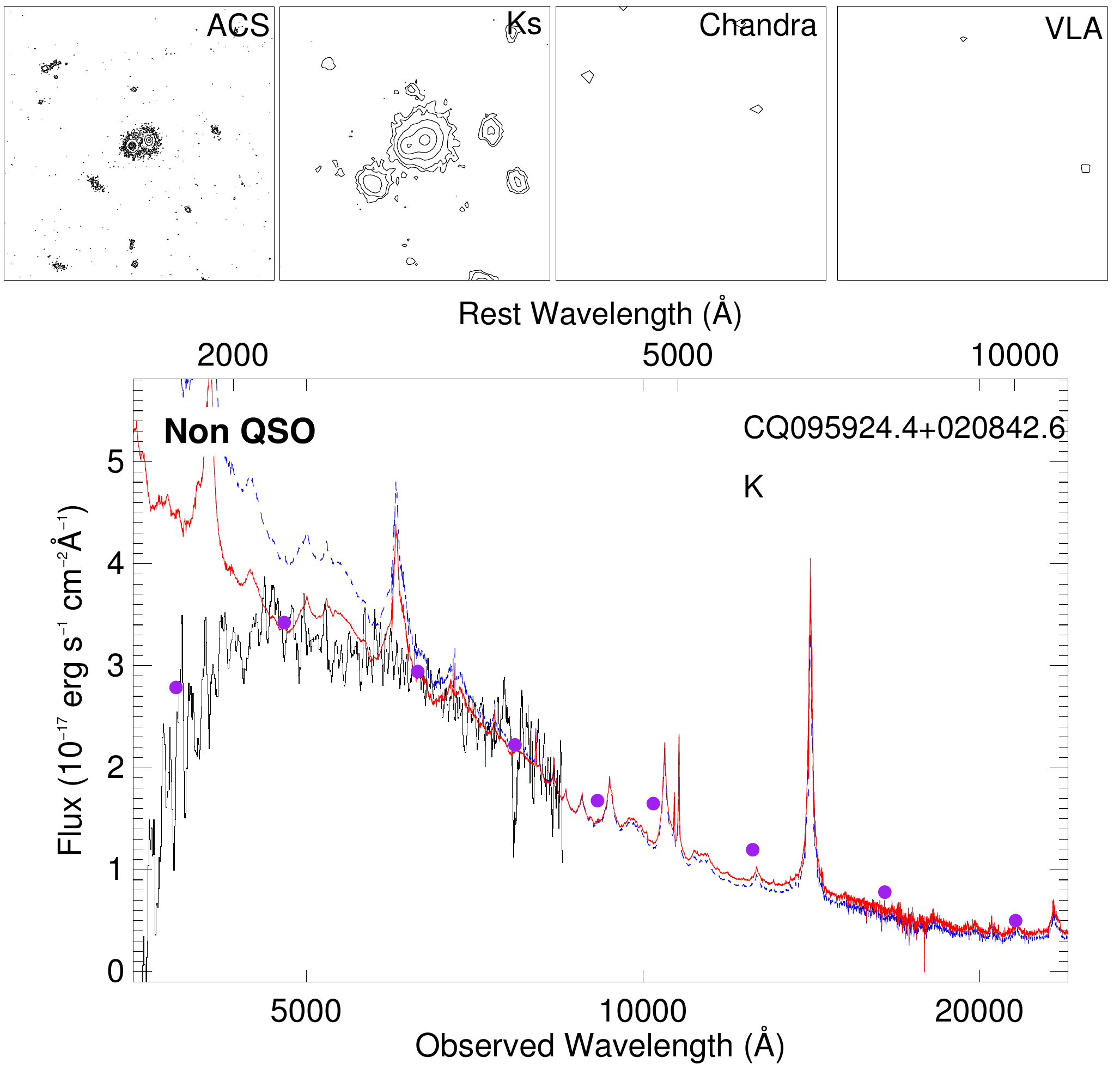}
            {\textbf{CQ: 8}}
            \label{fig:quasar8}
\end{minipage} \\[20pt]
         \begin{minipage}[t]{0.45\textwidth}
            \centering
            \includegraphics[width=\textwidth]{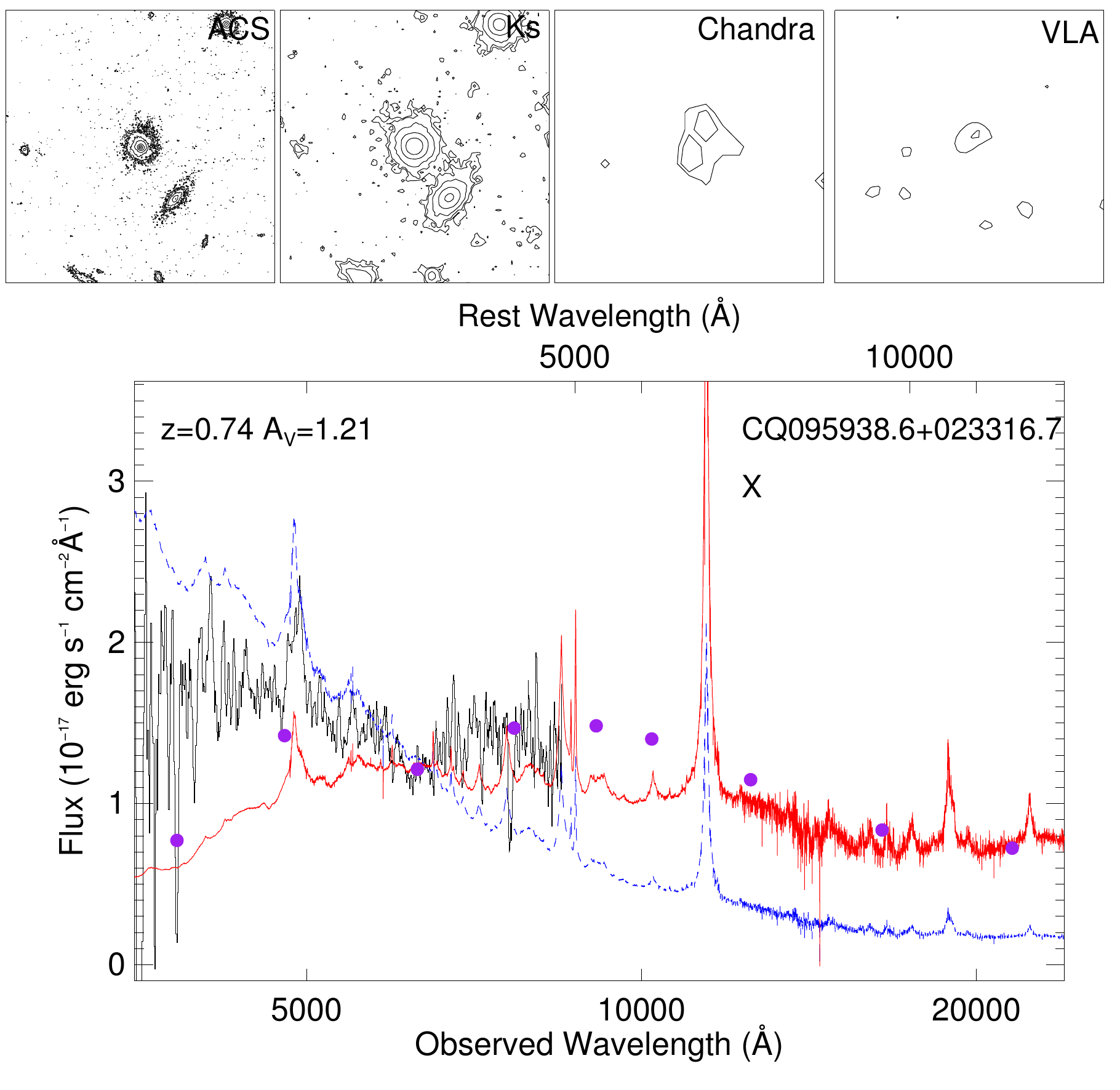}
            {\textbf{CQ: 11}}
            \label{fig:quasar11}
\end{minipage} %
\hspace{1cm}%
         \begin{minipage}[t]{0.45\textwidth}
            \centering
            \includegraphics[width=\textwidth]{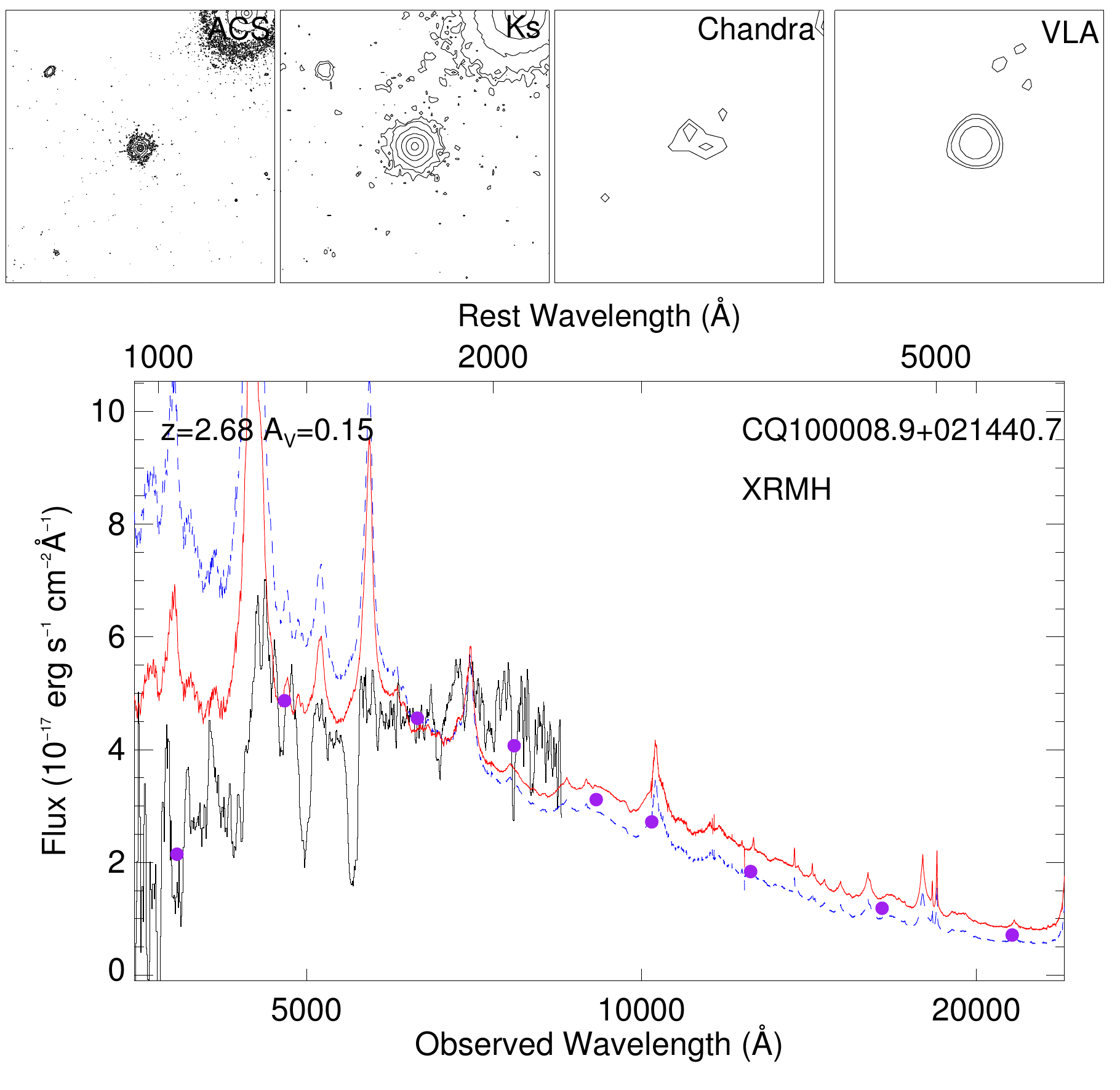}
            {\textbf{CQ: 16}}
            \label{fig:quasar16}
\end{minipage} %
        \caption[] %
                {Extracted 1D spectra and images in four wavelength bands of the four candidate quasars observed for the purpose of this study. The three objects CQ095833.3+021720.4, CQ095938.6+023316.7, and CQ100008.9+021440.7 are confirmed to be quasars, whereas object CQ095924.4+020842.6 is not. The three confirmed quasars all show a pronounced amount of reddening and BAL features. {\it Upper panels:} Images of the objects shown in the optical F814W (I-band) filter from the HST-ACS mosaic, the near-infrared $K_s$ filter from the KPNO mosaic, the X-ray {\it Chandra} merged energy band image, and the radio emission as seen by the VLA. Each of the images are $15\times 15$ arcsec where north is up and east is to the left.
                {\it Bottom panels:} The solid black lines show the observed spectra from the NOT. Overplotted are the photometric data points shown by the purple dots from the merged optical/near-infrared photometric COSMOS catalog together with the composite quasar template spectrum from \cite{Selsing16} with (red solid line) and without (blue dashed line) reddening, respectively. The redshift and amount of reddening, $A_V$, for each of the objects are shown in the upper left corner, while the name and selection method (from Col. 3 in Table~\ref{tab:comptab}) by which they were identified are shown in the upper right corner.} 
        \label{fig:specimgCQ1}
    \end{figure*}

In Fig.~\ref{fig:specimgCQ1} we show the one-dimensional spectra of the four candidate quasars along with the photometry from SDSS and UltraVISTA and the contours of the belonging images in four different wavelength bands: The optical F814W (I-band) filter from the HST-ACS mosaic, the near-infrared $K_s$ filter from the KPNO mosaic, the X-ray \textit{Chandra} merged energy band image, and the radio contours as observed by the VLA. We show the $K_s$ imaging from KPNO since this (and not the UltraVISTA imaging) is available at the NASA/IPAC-IRSA web page. Overplotted is a composite quasar template by \cite{Selsing16} with and without reddening (solid red and dashed blue lines, respectively). The composite was constructed from luminous blue quasars at $1 < z < 2.1$ selected from SDSS and has a slightly steeper spectral slope than other existing templates ($\alpha_{\lambda}=1.70\pm 0.01$ assuming a power-law continuum). This is explained for
instance by the broader spectral wavelength coverage because the slope can more easily be determined without strong contamination quasar emission lines or by intrinsic host galaxy emission \citep[see][for further details]{Selsing16}. The template was matched to the visible emission lines to compute the spectroscopic redshifts. The template was constructed by combining spectra of luminous blue quasars at $1 < z < 2.1$ selected from the SDSS. The spectra combined to make the composite were observed with the X-shooter instrument at ESO/VLT, which allows for a simultaneous observation of the entire spectral range from the $U$ to the $K$ band. Hence, the observed effect of variability in quasars is not a problem in this composite. For the object CQ095924.4+020842.6 the composite is overplotted at the photometrically computed redshift ($z_{\mathrm{phot}}=1.15$) since no emission line features are unambiguously present in the spectrum.

The amount of reddening for each of the objects was calculated assuming both an SMC-like extinction curve and a steeper extinction curve for intrinsically reddened quasars \citep{Zafar15}. For the SMC we used the extinction curve as described by \cite{Gordon03}, but with a modification for wavelengths greater than 4400~\AA~\citep{Fitzpatrick05}, following the same procedure as in \citet{Urrutia09,Fynbo13}. The extinction curve computed by \cite{Zafar15} was determined from 16 reddened quasars from the original HAQ survey, where the SMC could not provide a good solution to the observed spectral SEDs. We found that for most of the dust-reddened quasars the Zafar+15 extinction curve template computes a lower value of $A_V$ than when assuming SMC-like extinction (see Table~\ref{tab:comptab}). However, none of the two extinction curve templates were preferred over the other, meaning that both of the templates were in general good matches to the observed spectra. In the following we therefore consider only the $A_V$ assuming SMC-like extinction.

\begin{table} 
\caption{Setup for the spectroscopic follow-up of the remaining four candidate quasars. Blocking filter 94 was used in combination with grism 4 for all observations.}
{\begin{tabular*}{1.0\columnwidth}{@{\extracolsep{\fill}}l l c r}
\hline\hline
Object          & Grism & Slit width & Exp. time\\
                &               & (arcsec)       & (sec)\\ 
\hline
CQ095833.3+021720.4  & \#4       & 1.3    & $3\times 1200$ \\
CQ095924.4+020842.6  & \#4~+~\#18 & 1.3/1.0   & $6\times 900$  \\
CQ095938.6+023316.7  & \#4        & 1.0           & $3\times 1200$ \\
CQ100008.9+021440.7  & \#4        & 1.3           & $3\times 600$  \\
\hline
\end{tabular*}}
\label{tab:obs_setup}
\end{table}

The redshifts and the extinction was determined by visually matching the composite quasar template to the photometric data points and the observed spectra. For simplicity we assumed that the dust is located at the redshift of the quasar. The objects that do not show signs of reddening agree well with the unreddened composite quasar template. The reddened quasars can be identified by the discrepancy in the unreddened template to the observed spectra where the extinction parameter is simply a measurement of the amount of divergence. We disregarded photometric data blueward of the \lya~emission line, and in case of strong absorption at individual data points, these were excluded as well. The photometric data points are shown as the purple dots at the $u,g,r,i,z,Y,J,H,K_s$ effective wavelengths. The redshifts and amounts of reddening are shown for each of the objects, together with the object name, coordinates, and the respective selection technique by which the object was detected. In the Appendix, Fig.~\ref{fig:specimgquasar}, the spectra of the remaining quasars are shown together with the corresponding images of each of the individual objects. Their extinction was determined following the same procedure as described above. Below each of the objects the SDSS/BOSS selection flags are stated.

We were able to obtain a secure confirmation of three objects: CQ095833.3+021720.4, CQ095938.6+023316.7, and CQ100008.9+021440.7. The visible emission lines in the three spectra allow for a clear redshift determination and agreed well with the computed photometric redshift. Two of the objects, CQ095833.3+021720.4 and CQ100008.9+021440.7, show clear broad absorption line (BAL) features. Quasar templates, both with and without reddening, show a poor fit to the spectrum of object CQ095924.4+020842.6. The photometry at MIR wavelengths disagrees with typical quasar MIR colors \citep{Nikutta14} as
well. Because several stellar absorption lines at $z=0.05$ are detected in the spectrum, we can identify a low-redshift quiescent galaxy as contribution to the SED. However, the full multiwavelength photometry cannot be fit well with BC03 stellar population synthesis models \citep{Bruzual03}. The observed NIR and especially the MIR photometry shows excess compared to that expected from good model fits to the optical photometry. This most likely indicates that an additional source at higher redshift contributes to the observed SED, whose light would dominate at longer wavelengths. This scenario is supported by the HST imaging, which shows two close line-of-sight compact sources. For this object we do not list redshift or $A_V$ in Fig.~\ref{fig:specimgCQ1}, but mark it as unclassified.

\section{Results}    \label{sec:res}

\begin{figure} 
    \centering
                \includegraphics[width=\columnwidth]{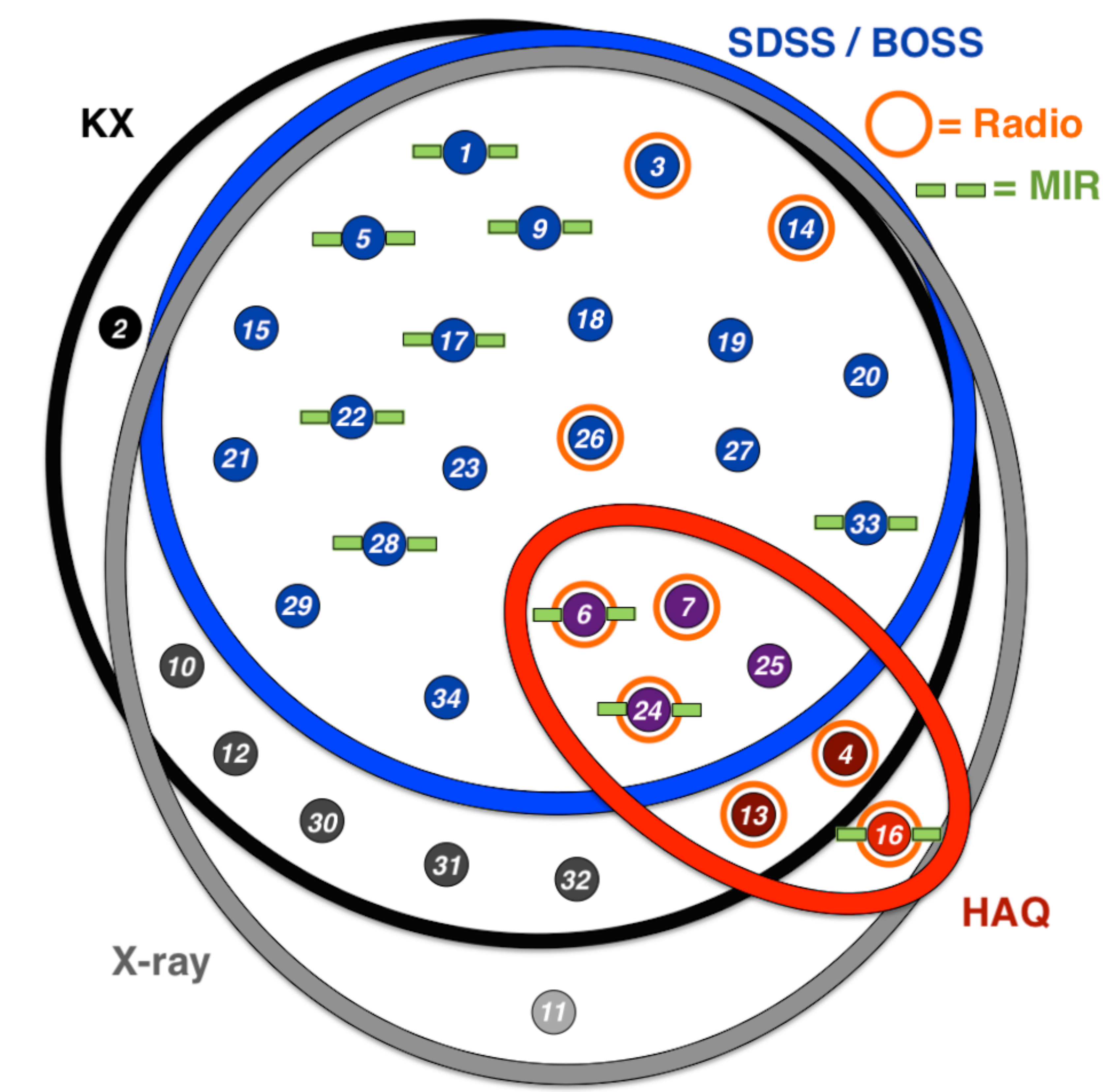}
                \caption[Network2]%
                {Quantity diagram (a so-called Venn diagram) of the respective sub-populations of the 33 confirmed quasars. The numbers in each of the symbols represents the objects with the same numbering as in Table~\ref{tab:comptab}. The parent sample of quasars consists of $7/33\sim 21\%$ HAQs. Only quasars 2 and 11 have been selected by one technique alone from KX selection or X-ray detection, respectively. Of the 13 quasars with $A_V>0.1$, 6 were selected by the HAQ criteria, 11 were identified by the KX selection, 12 with X-ray detection, 6 had radio detections, 5 were from the photometric selection of SDSS/BOSS, and 4 of the reddened quasars were identified by the MIR selection.}
                \label{fig:vdia}
        \end{figure}

We found 33 bright spatially unresolved quasars within the C-COSMOS sub-field brighter than $J=20$ magnitude (corresponding to 37 quasars per square degree, see Sect.~\ref{subsec:ps}). We assume that our sample is a good representation of the full underlying population of quasars, although small (see below), since the bias should be negligible when using the multiple selection approaches, each of which target different aspects of quasar properties. In Fig.~\ref{fig:vdia} a quantity diagram of the sub-populations of all the confirmed quasars is shown sorted according to the techniques through which they were selected and how they coincide. The numbering in the figure follows the same notation as in Table~\ref{tab:comptab}. 

\begin{figure} 
        \centering
            \includegraphics[width=\columnwidth]{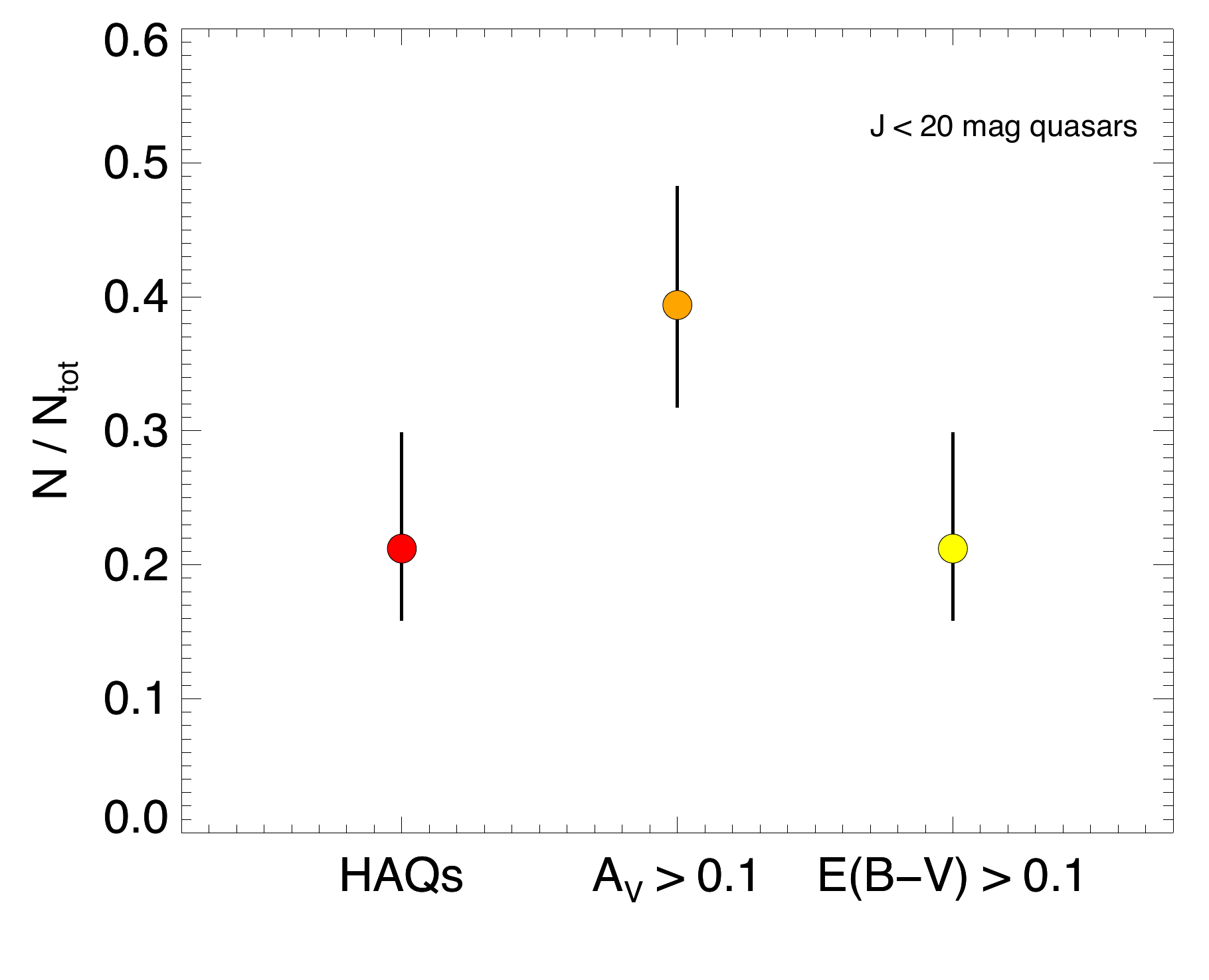}
            \caption[Network2]%
            {Fractions of the HAQ-selected quasars, the quasars with $A_V>0.1,$ and the quasars with $E(B-V)>0.1$ compared to the total. The error bars correspond to the $68\%$ confidence interval.}
            \label{fig:plotfrac}
    \end{figure}

Out of the total number of quasars in the parent sample we found that seven optical point sources from the photometric catalog were selected by the specific optical/NIR HAQ criteria, see Eq.~\ref{eq:haq}. This yields a fraction of $f_{\mathrm{HAQ}}=N_{\mathrm{HAQ}}/N_{\mathrm{total}}=0.21^{+0.09}_{-0.05}$ ($21\%^{+9}_{-5}$) using the small number statistics formulated in \cite{Cameron11} with error bars corresponding to the $68\%$ confidence interval, see Fig~\ref{fig:plotfrac}. The fraction of all reddened quasars in our sample, defined as having $A_V>0.1$, was found to be, again with error bars corresponding to the $68\%$ confidence interval, $f_{\mathrm{A_V>0.1}}=N_{\mathrm{A_V>0.1}}/N_{\mathrm{total}}=0.39^{+0.09}_{-0.08}$ ($39\%^{+9}_{-8}$). Previously, \citet{Glikman07,Glikman12,Urrutia09} have defined red quasars as having $E(B-V)>0.1$. We found that seven of the quasars in our parent sample satisfy this criterion, so that the fraction is again $f_{E(B-V)>0.1}=21\%^{+9}_{-5}$, which is consistent with the findings of \citet{Glikman07,Glikman12,Urrutia09}, see Sect.~\ref{sec:conc}. The SDSS/BOSS photometric selection found 23 quasars, 5 of which have $A_V>0.1$. This means that roughly 22\% of the SDSS/BOSS photometrically selected quasars are classified as reddened, while 40\% are expected. Hence an incompleteness of 45\% in this particular optical quasar sample is observed.

In Table~\ref{tab:comptab} we show the redshifts, both spectroscopic and photometric, along with the calculated reddening, $A_V$, both assuming SMC-like extinction and the extinction curve found by \cite{Zafar15}, and the corresponding reddening in terms of $E(B-V)$. The notes list the specific colors in which the individual objects do not meet the HAQ criteria.

The computed photometric redshifts generally agree well with the spectroscopically determined redshifts, both for the objects with public spectra and for the quasars we observed. We find a maximum difference of $\vert z_{\mathrm{spec}}-z_{\mathrm{phot}}\vert=0.18$ (except for the extreme case of Q\,095921.3+024412.4, see below), whereas for most objects the spectroscopic and photometric redshifts are determined well within $\vert z_{\mathrm{spec}}-z_{\mathrm{phot}}\vert < 0.1$. All but one of the seven HAQs show a pronounced amount of reddening ranging from $0.15 < A_V < 0.98$. Following the parameterization from \cite{Gordon03}, where (for SMC-like extinction)
\begin{equation}
R_V\equiv A_V/E(B-V),~R_V=2.74, 
\end{equation}
this corresponds to $0.05 < E(B-V) < 0.36$. The HAQs reddened by dust are at redshifts in the range $1.00 < z < 2.66$. The only HAQ with no sign of dust extinction is the object Q100050.1+022356.7 at $z=3.37$. The distribution of redshifts and $A_V$ of the HAQs from our parent sample agree well with the findings of \cite{Krogager15}, see Fig.~3 of their paper, for example. As mentioned, only four of the seven HAQs have been identified by the SDSS-III/BOSS program, which again is similar to the fraction found in the original HAQ survey, although from a much larger sample, where $409/901\sim 45\%$ of the HAQ candidates had been observed as part of the SDSS (DR7) program before their study. In the spectrum of object Q095938.3+020450.1 (quasar 10) we found an indication of a damped \lya~absorption feature blueward of the strong \lya~emission line in the \lya~forest. The search for quasars reddened by intervening absorbers was the primary goal of the HAQ survey, but this object evaded selection by being slightly too blue ($H-K_s=-0.02$) for the HAQ criterion of $H-K_s > 0$. The optical colors are otherwise within the HAQ selection criteria, indicating the red nature ($A_V=0.23$) of this object.

\begin{figure*} 
\centering
        \begin{minipage}[b]{0.33\textwidth}
            \centering
            \includegraphics[width=\textwidth]{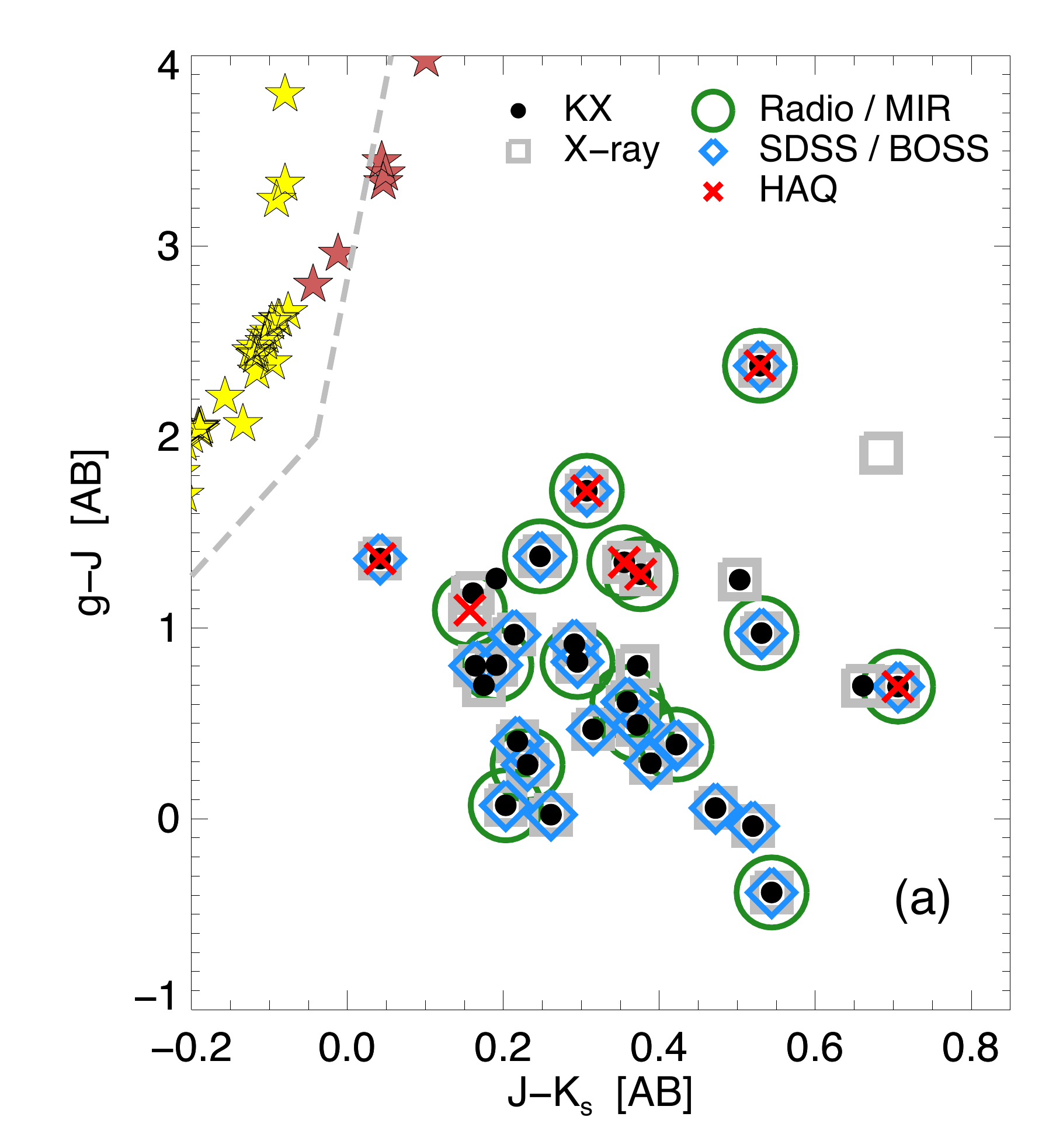}
            \label{fig:compquasara}
        \end{minipage}
        \hfill
        \begin{minipage}[b]{0.33\textwidth}  
            \centering 
            \includegraphics[width=\textwidth]{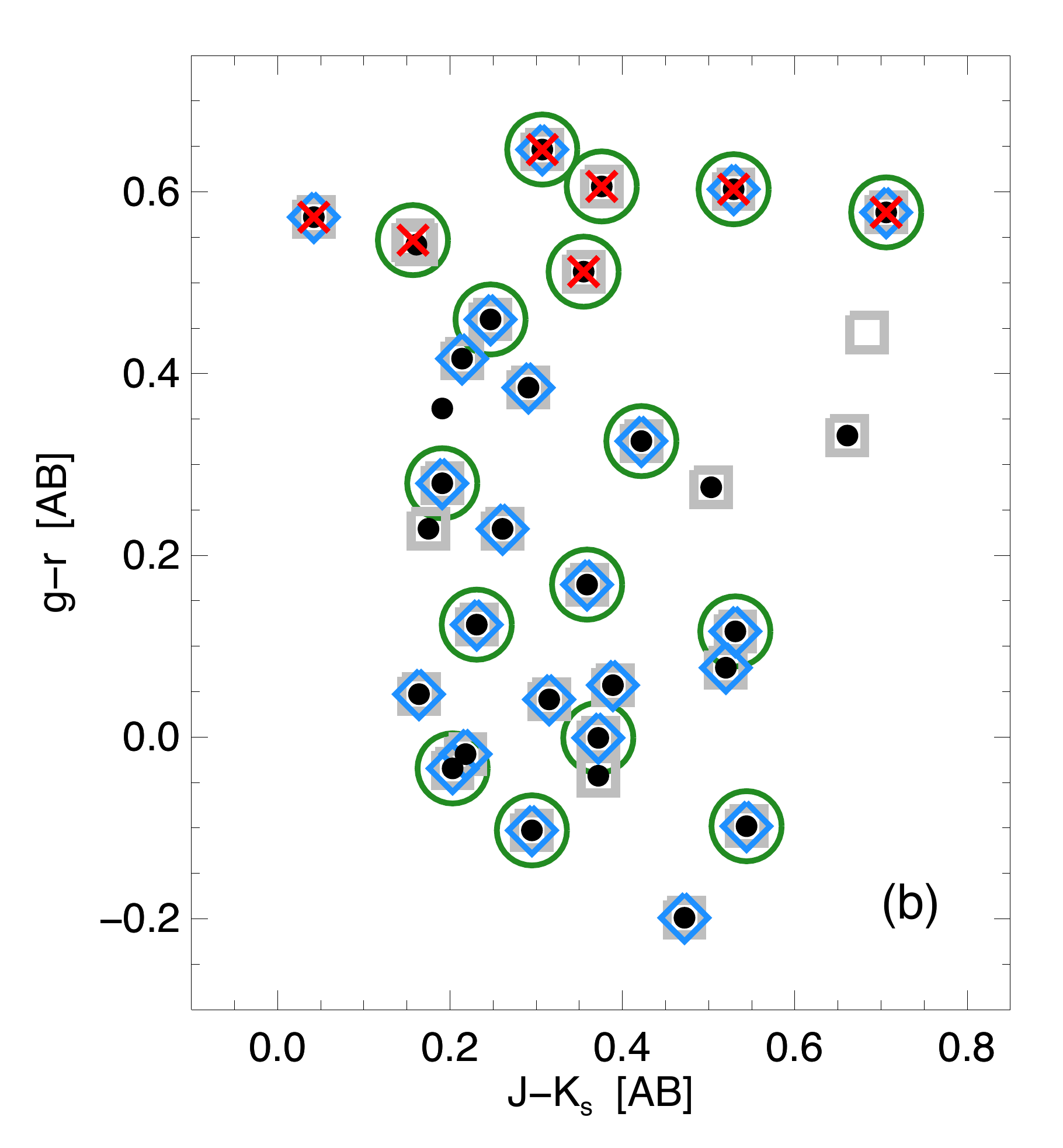}
            \label{fig:compquasarb}
        \end{minipage}
        \hfill
        \begin{minipage}[b]{0.33\textwidth}   
            \centering 
            \includegraphics[width=\textwidth]{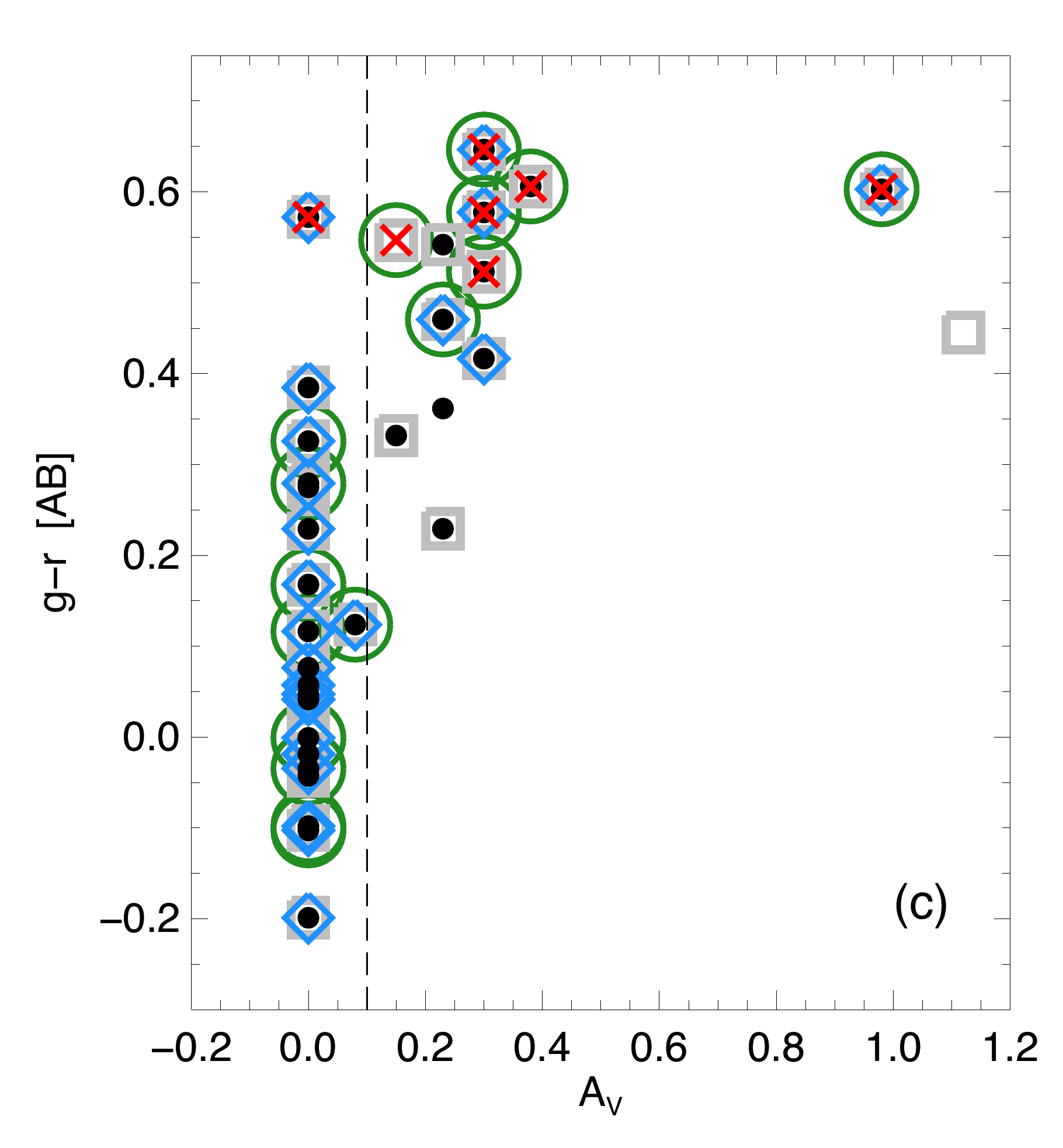}
            \label{fig:compquasarc}
        \end{minipage}
        \caption[] %
        {{\it Panel (a) and (b):} Optical/NIR color-color diagrams of the 33 confirmed quasars. The respective symbols represent the individual selection method used to identify each quasar as defined by the legend. The yellow and red star symbols show the photometry of main-sequence and M-dwarf stars from the \cite{Hewett06} catalog, respectively. {\it Panel (c):} The reddening, $A_V$, of the confirmed quasars as a function of their color in $g-r$. The HAQs contribute to about half of the population of reddened quasars defined as having $A_V>0.1$ (black dashed line).} 
        \label{fig:compquasar}
    \end{figure*}

In Fig.~\ref{fig:compquasar}, panels (a) and (b), we show the photometric colors of all the quasars from the parent sample in two optical/NIR color-color diagrams and in panel (c) the $g-r$ colors as a function of the reddening, $A_V$, assuming SMC-like extinction. Again, the respective symbols represent the individual selection methods used to identify each quasar as shown in panel (a), upper right corner. For comparison we show the photometry of main-sequence stars and M-dwarf stars in panel (a), illustrated by the yellow and red stars, respectively. In panel (b) the stellar track is outside the plotting region and is therefore absent. The stellar sources were obtained from the \cite{Hewett06} catalog, where the yellow stellar track in general represents the densest region of the stellar locus. When
we plot all the defined point sources from the full photometric COSMOS catalog used in our study, the main location in color-color space follows an equal track. In panel (a) the gray dashed line represents the selection line of the KX method formulated in Eq.~\ref{eq:kx}. Panel (c) illustrates that the HAQs indeed are a part of the most reddened quasars in our sample. The quasars that were not selected by the HAQ criteria but still show a pronounced amount of reddening are either red in the $gJK_s$ or the $J-K_s$ colors, suggesting that by refining the HAQ criteria, a larger and more complete sample of the reddest quasars can be obtained. 

For one of the quasars, Q095921.3+024412.4 (quasar~7), we had to discard the spectroscopic redshift of $z=3.623$ as determined by the BOSS program. This quasar is obscured and shows high reddening ($A_V=0.98$) with only a few weak emission line signatures apparent. The available notes at the SDSS DR12 website report a poor $\chi^2$ fit for the redshift determination of this quasar. For comparison, the photometric redshift from \cite{Salvato11} was computed as $z_{\mathrm{phot}}=0.030$. The redshift we suggest of $z=1.004$, found when visually matching the spectrum to the composite quasar template, was also listed by \cite{Trump09}, who determined a value of $z=1.0037\pm 0.0054$.

\section{Survey reliability} \label{sec:disc}

The success of a survey hinges critically upon obtaining the optimal balance between efficiency and completeness. In other words, being able to select a sample of candidates, which includes close to all of the objects of the targeted type, and close to nothing else. Below we review the individual surveys used to build our parent sample and discuss the completeness and reliability of each of the approaches.

\subsection{Efficiency}

Our parent sample was produced by only considering the brightest and spatially unresolved population of quasars obtained from six distinctive selection techniques. The conservative criteria we used of star-like morphology and $J\leq 20$ magnitude have introduced some significant bias (by design) when considering the general and complete AGN population. We required that the quasars were optically unresolved to remove objects with contributing host galaxy light and to be consistent with previous photometric selection approaches. The $J=20$ magnitude cut ensured that our selection was complete down to the limiting depths in magnitude for each of the individual surveys. When comparing the different selection techniques, this cut allows the selection of quasars based on relatively shallow surveys to still be relevant.

\begin{figure} 
        \centering
            \includegraphics[width=\columnwidth]{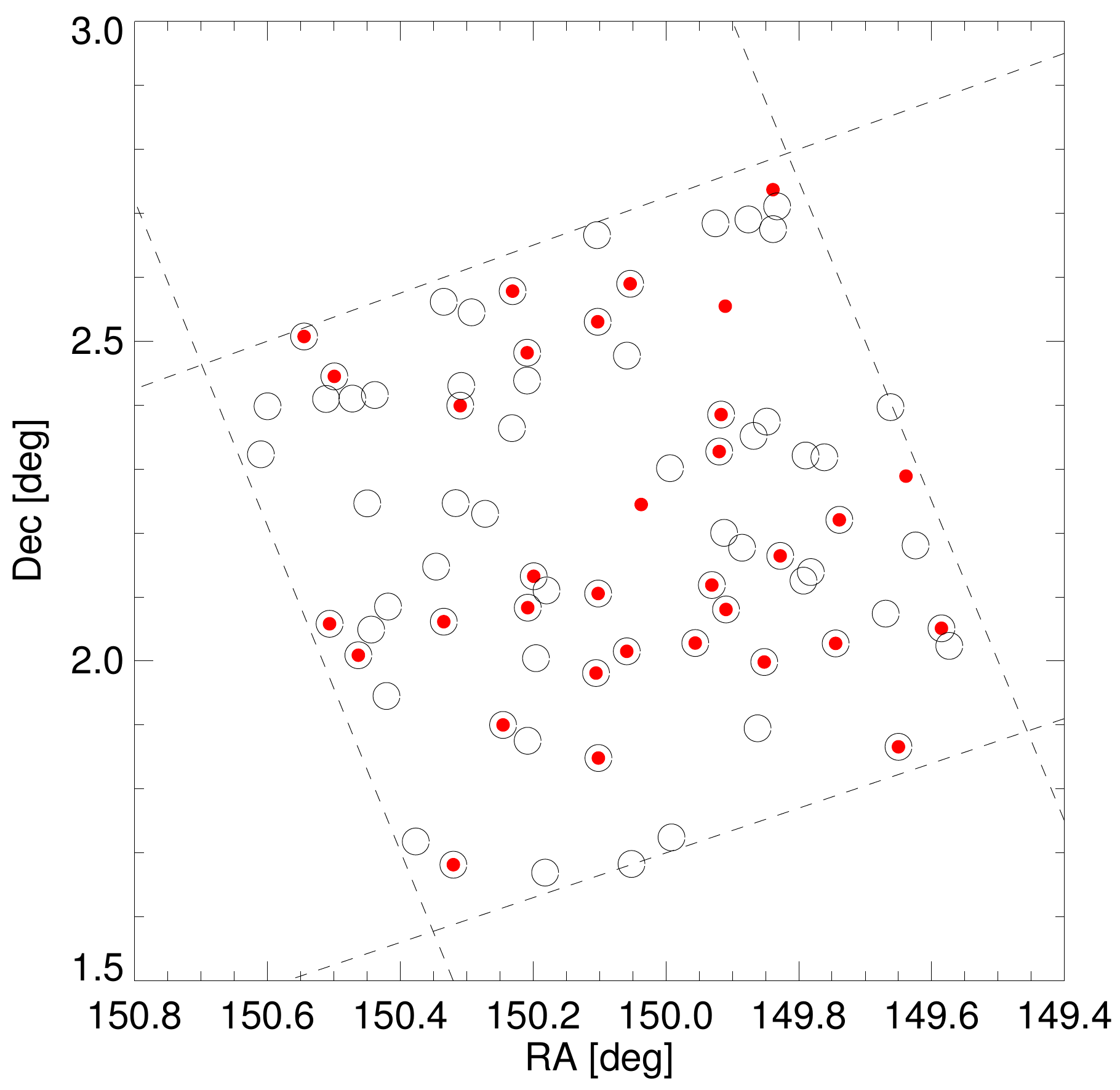}
            \caption[Network2]%
            {Full SDSS-III/BOSS DR12Q sample (black circles) within the C-COSMOS field together with the confirmed quasars from our parent sample (red dots). About $60\%$ of the DR12Q sample have not been selected by our criteria in morphology and/or brightness, where approximately 60\%~vs.~40\% was due to $J>20$ mag and optical resolved morphology, respectively.}
            \label{fig:fulldr12Qs}
    \end{figure}

To compare the SDSS/BOSS selected quasars from our parent sample to the full SDSS-III/BOSS sample, we extracted all the quasars from the DR12Q sample (P\^{a}ris et al. 2016, in prep.) within the C-COSMOS field, see Fig.~\ref{fig:fulldr12Qs}. Here we show the full SDSS/BOSS DR12Q sample together with the confirmed quasars from our study to directly compare the two samples. In total, 42 quasars from the DR12Q sample have not been selected from our criteria but all had a counterpart in the photometric catalog. Of these, 25 quasars avoided selection by being too faint in the $J$ band. However, most of the objects in the DR12Q sample have $J\leq 20.5$ mag, indicating that the cut in brightness is close to the limiting depth in magnitude up to which the SDSS/BOSS survey is complete. Furthermore, we found that 18 of 42 quasars in the DR12Q sample miss the selection because of extended spatial morphology in the photometric catalog, where most of these appear to be Seyfert galaxies in the SDSS database. 

The survey that was suppressed the most by our requirement of optically unresolved morphology was the approach of X-ray selection.
Without the point source criteria of the photometric catalog applied to the X-ray selected sample, we obtained 224 X-ray detected AGNs all brighter than $J=20$ magnitude within the C-COSMOS field. That is, only $\sim 14\%$ of the X-ray selected sample is recovered using our morphology criteria, meaning that they have unresolved optical/NIR counterparts. Moreover, the MIR color-selected sample relying on \textit{WISE} colors only detects 18 quasars in total, where only 10 (55\%) of these have been considered in our study because of their unresolved morphology. This agrees with the findings of \citet{Gabor09,Griffith10,Stern12}, who have examined the optical morphologies of AGNs in the COSMOS field where $\sim 50\%$ of the brightest MIR color-selected population is extended or unresolved, respectively. The high density of the population of optically resolved X-ray selected AGNs is higher than has previously been reported. \cite{Griffith10} found that $72\%$ of the X-ray selected sample is optically resolved, but only for the faintest population based on \textit{XMM-Newton} observations (well beyond $J=20$ mag). Some of the discrepancy for the brightest population could be explained by the more sensitive {\it Chandra} satellite, but the fraction of resolved X-ray AGNs is still incomparably higher than expected. Since we only considered the brightest and spatially unresolved population of quasars, the sub-population of reddened and extremely obscured quasars can be even higher when examining the entire population of optically unresolved and extended X-ray and/or MIR selected quasars, see, for example, \cite{Gavignaud06,Fiore08}. 

\begin{figure} 
        \centering
            \includegraphics[width=\columnwidth]{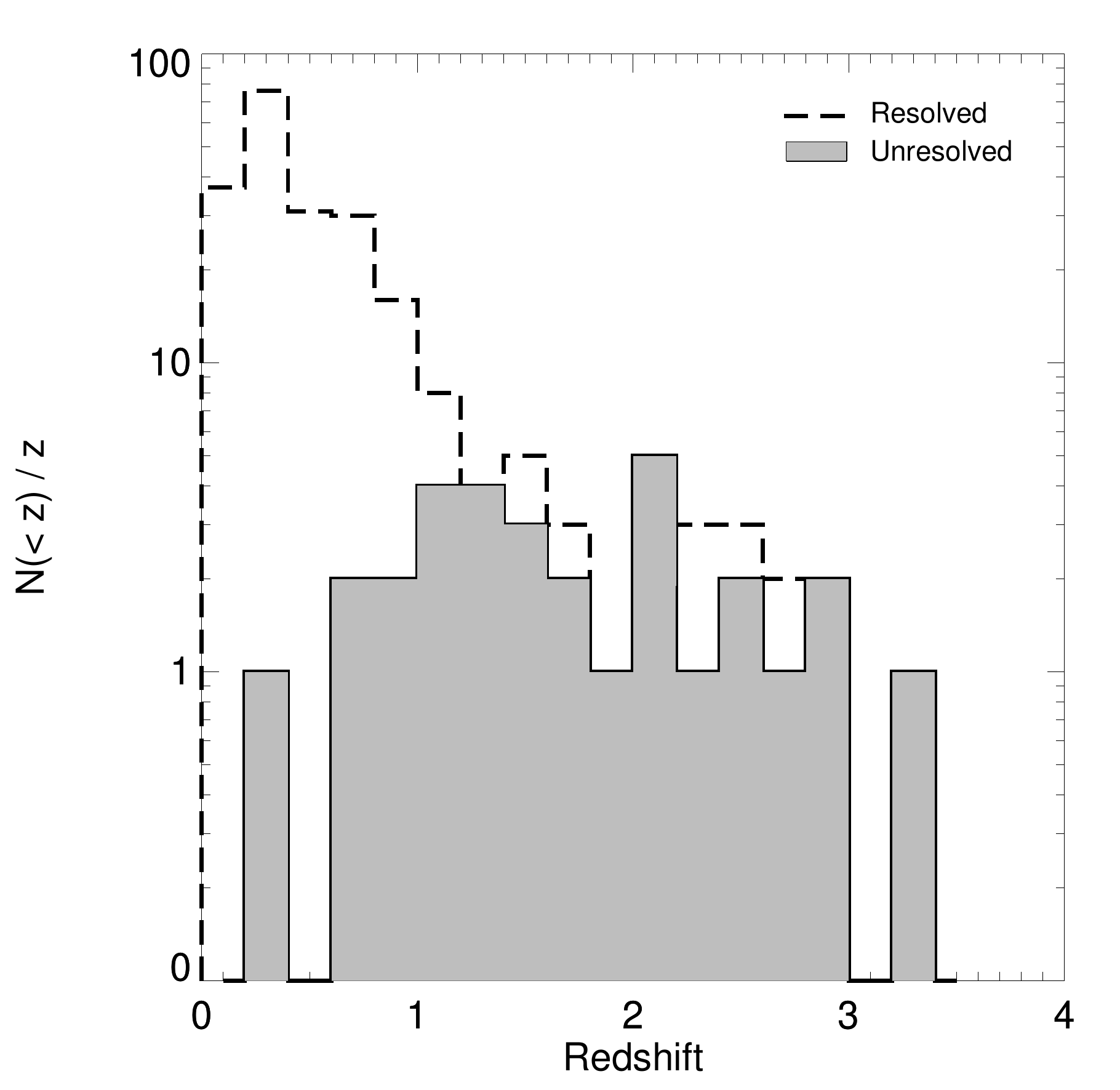}
            \caption[Network2]%
            {Histogram of the redshift distribution of the brightest ($J<20$ mag) optically resolved and unresolved {\it Chandra} X-ray detected quasars in the COSMOS field, respectively. At redshifts below $z \lesssim 1$ the incompleteness of our parent sample due to morphology is most severe, while at $z > 1$ the unresolved population of quasars is representative of the full population.}
            \label{fig:xray_comp_unres}
    \end{figure}

To show them in context to each other, we plotted the optically extended and point source X-ray detected quasars, respectively, as a function of redshift in Fig.~\ref{fig:xray_comp_unres}. To be consistent, only the brightest population ($J\leq 20$ mag) of quasars, again within the C-COSMOS field, is shown. Below redshifts $z \lesssim 1$, the fraction of AGNs with dominating light from the host galaxy causing observed extended optical morphology (at the resolution of the CFHT+Subaru $i$ band) is more dominant than the spatially unresolved population. At higher redshifts we found that the optically unresolved population of bright quasars is representative of the full underlying population with only a few optically extended quasars rejected for our parent sample. Most of these appear compact in the HST images, but still show signs of elliptical or spiral-like morphology.

For the SDSS/BOSS photometrically selected quasars the invoked limit in brightness for our parent sample was close to the completeness limit for this particular survey. For some of the deepest COSMOS surveys the limit down to which these are complete is several magnitudes deeper. For example, 172 of the X-ray selected quasars (when only considering the optically unresolved population) have magnitudes fainter than $J\ge 20$ mag, whereas only 33 of the radio selected and 11 of the mid-infrared selected quasars are fainter than this magnitude limit. By applying the cut at NIR wavelengths, the selection becomes less biased against quasars reddened by dust than in optical magnitude limited studies. We highlight that none of the defined selection techniques detect all the quasars from our parent sample. The X-ray selection technique found the most sources, where this particular approach of identifying quasars has been credited as being the most reliable \citep{Brandt05}, with the only exception being the quasar CQ095833.3+021720.4. A high absorption is observed in this object, which could explain the non-detection. The reason might also be that
like Q100210.7+023026.2, this quasar is located in a region of short exposure.

\subsection{Degree of contamination}

Even though the X-ray selection technique detects most of the bright AGNs (regardless of optical morphology), great caution has to be applied when building large X-ray selected quasar samples. When considering all the optical/NIR point source counterparts (brighter than $J=20$ mag) of the \textit{Chandra} X-ray sample, we found an additional 31 sources that were either photometrically or spectroscopically classified as stars. That is, when the brightest population of optically unresolved X-ray selected quasars is
observed blindly, $\sim 50\%$ will show stellar contamination. According to the photometric study of the X-ray selected COSMOS sample by \cite{Salvato09,Salvato11}, these could be discarded before selection either by having existing spectra classifying them as stars or by having seemingly stellar-like colors in optical/NIR color-color space. In addition, two radio-loud sources with an
optically unresolved optical/NIR counterpart brighter than $J=20$ mag were also initially present in our sample, but were removed in the same way. We chose to eliminate all objects located in the stellar locus to obtain a conservative estimate of the population of reddened quasars. However, since selection of quasars with apparent optical/NIR colors similar to that of stars is only possible at short (e.g., X-ray) or long (e.g., radio) wavelengths, these would not be detected by any standard optical/NIR color selection techniques, and rejection of them has to be considered
carefully. 

When we applied the KX selection line in $gJK_s$ color-color space, that is, when we separated objects with and without $K_s$-band excess, ten additional contaminating sources were obtained. The contaminants with existing spectra consist of one L dwarf, one white dwarf, one $A0$ star, and one carbon star. The remaining KX candidates were eliminated after being processed by the photo-$z$ algorithm. The majority of these were classified as stars by the code, while only two were classified as blue galaxies. The reduced $\chi_q^2$ from the best fit obtained with the AGN templates in the photometric COSMOS catalog all show relative poor fits as well. This was not the primary motivation for eliminating these sources, however, since the AGN templates are in no way fully inclusive and therefore not reliable \citep{Ilbert13}. The two objects CQ095938.6+203316.7 and CQ100008.9+021440.7 were discarded by this process. For CQ095938.6+203316.7, we simply disregarded it before running the candidates through the algorithm because of its extended morphology in the SDSS imaging. The other quasar, CQ100008.9+021440.7, although selected by the general KX color criteria, was classified as being a star by the photo-$z$ code. This object was spectroscopically confirmed as being a quasar, but indeed with high absorption and very red (and star-like) optical colors, especially in $u-g$. We chose to follow it up spectroscopically since it was identified by many other selection techniques, see, for example, Table~\ref{tab:comptab}. Although higher contamination is observed without applying this algorithm to the KX-selected objects, the code makes the specific output sample less complete.

Even though the HAQ survey is in no way complete, the purity of the sample without further considerations of  template fitting,
for instance, is unparalleled by any of the other surveys. All of the seven HAQ candidates selected from optical/NIR photometry alone were confirmed as being quasars. In the other samples, for example, the X-ray detected, the KX-selected, and the radio-detected quasars, additional photometric $\chi^2$ fitting was executed to effectively remove stellar contamination. Selecting quasars in the COSMOS field also benefits from the previous extensive observations from numerous surveys, which had spectroscopically classified the majority of the stellar contamination.

\section{Discussion and conclusions} \label{sec:conc}

We used the available multiwavelength data of various COSMOS surveys to
define a complete parent sample of bright and spatially unresolved quasars
that are representative of the full underlying population. Specifically, the sample was
produced by using six distinctive techniques to identify quasars ranging
from X-ray to radio wavelengths. From this parent sample of quasars, we
determined the fraction of high $A_V$ quasars (HAQs), a sub-population of
quasars shown to be underrepresented in optical surveys
\citep{Fynbo13,Krogager15}. We found that the total population of quasars in
our sample consists of $f_{\mathrm{HAQ}}=21\%^{+9}_{-5}$ HAQs. The general
population of reddened quasars with $A_V>0.1$ constitutes
$f_{\mathrm{A_V>0.1}}=39\%^{+9}_{-8}$ of the parent sample. The calculated
fractions agree well with other similar studies reported in the
literature. We chose the HAQ criteria to define the optical/NIR photometrically
selected sample as one representation and quasars with $A_V>0.1$ as another.
We found a value slightly above estimates from optically selected samples, see, for instance, the study by \citet{Richards03}. Here they determined that only $\sim$6\%
of a subsample of quasars selected from the SDSS is reddened following a
SMC-like extinction curve, but also that $\sim 15\%$ of reddened quasars
will be missing in the SDSS. 
In our sample, the SDSS/BOSS photometric selection functions detect 23 quasars,
out of which 5 have $A_V>0.1$, meaning that about 22\% of the quasars
in SDSS are reddened, while 40\% are expected, which translates
into an incompleteness of 46\%.
For the photometric sample of SDSS/BOSS, the completeness fraction is then
$5/(23\times 0.4)=54\%$.

\begin{figure} 
        \centering
            \includegraphics[width=\columnwidth]{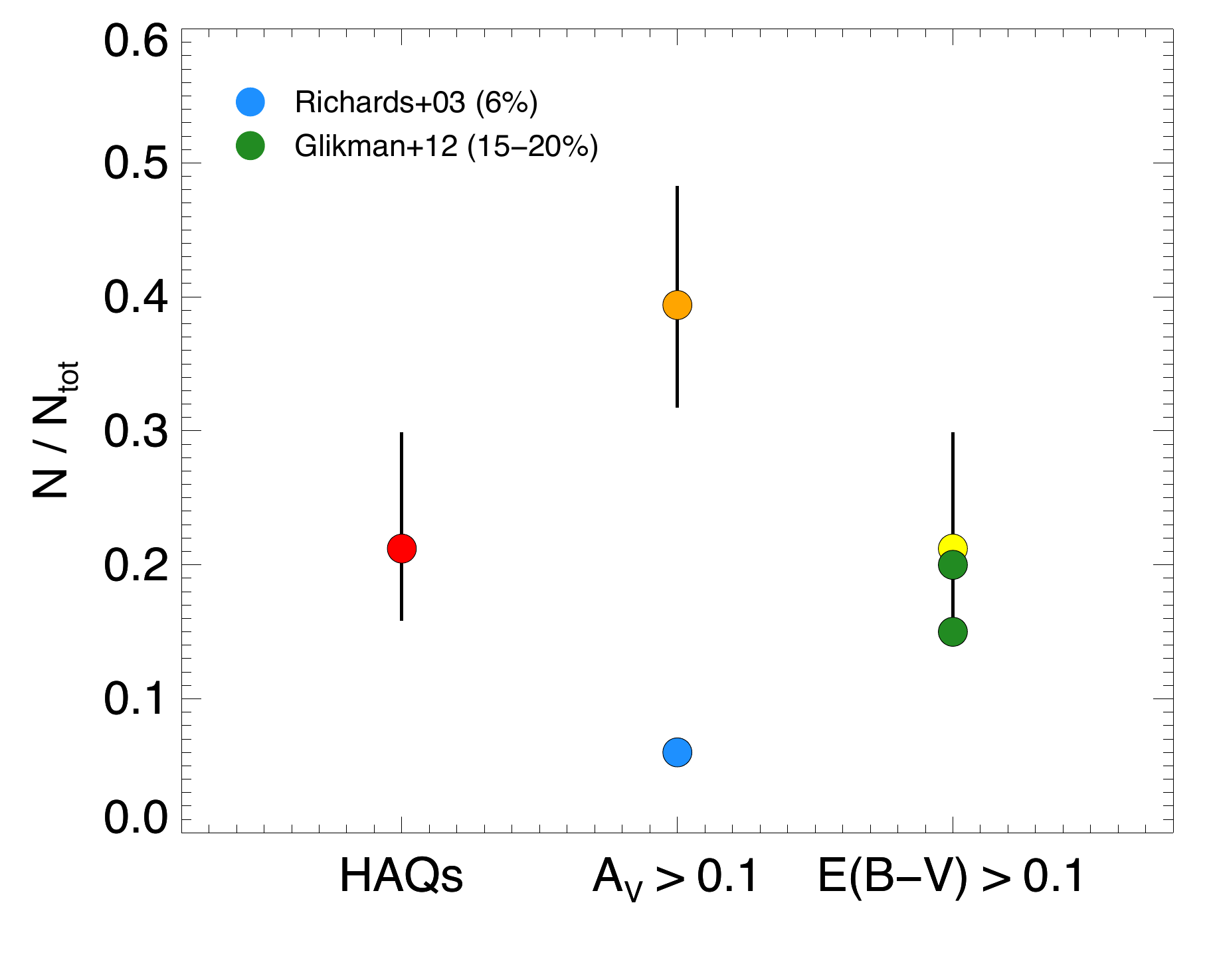}
            \caption[Network2]%
            {Same as Fig.~\ref{fig:plotfrac}, but with comparisons to the fractions of reddened quasars found by \cite{Richards03} and \cite{Glikman12}. \cite{Richards03} considered sources with $E(B-V)>0.04\sim A_V>0.1$ as reddened, while \cite{Glikman12} used the definition of $E(B-V)>0.1$ to identify reddened quasars.}
            \label{fig:plotfraccomp}
    \end{figure}

Selections of reddened quasars detected in radio with FIRST and matched to the
NIR survey 2MASS have been executed as well, developed as another approach of
alleviating the known bias against dust-reddened quasars in optically selected
samples.
\cite{Glikman04} estimated the missing population of reddened quasars in optical
surveys to be in the range $\sim 3-20\%$. Subsequently,
\citet{Glikman07,Urrutia09} estimated a higher fraction of $>\sim 20\%$ based
on the spectroscopic follow-up of their FIRST-2MASS-selected reddened quasars. \cite{Glikman12} found that reddened quasars make up $\lesssim
15\%-20\%$ of the radio-emitting bright quasar population from their sample,
defined as quasars with $E(B-V)>0.1$. For comparison, we found that
$21\%^{+9}_{-5}$ of the quasars from our parent sample had $E(B-V)>0.1$,
assuming the extinction curve of the SMC (see, e.g., Fig~\ref{fig:plotfraccomp} for
a graphical comparison). This is of course still dependent on how the parent
population is defined and how deep the limiting brightness is (since fainter
samples will allow for a selection of even more reddened, i.e., optically faint
quasars). \cite{Richards06} estimated an upper limit of $<30\%$ for the
fraction of reddened quasars that are missed in optical samples
such as the SDSS, based on
the $u-g$ and $g-r$ colors of a MIR color-selected sample of quasars. They
based this on the conclusion that $\sim 70\%$ of the MIR color-selected sample
had blue optical colors consistent with what has previously been reported in
optical UVX surveys (having $u-g<0.6$ and $g-r<0.6$, respectively). This contradicts our findings in Sect.~\ref{sec:res}, however. In
Fig.~\ref{fig:compquasar} we showed that only about half of the reddest
population of quasars, with $A_V>0.1$, have red optical colors of $g-r>0.5$
(and only three of these quasars have $g-r>0.6$), indicating that the $30\%$
upper limit in the density of reddened quasars is not reliable.

\begin{figure} 
        \centering
            \includegraphics[width=\columnwidth]{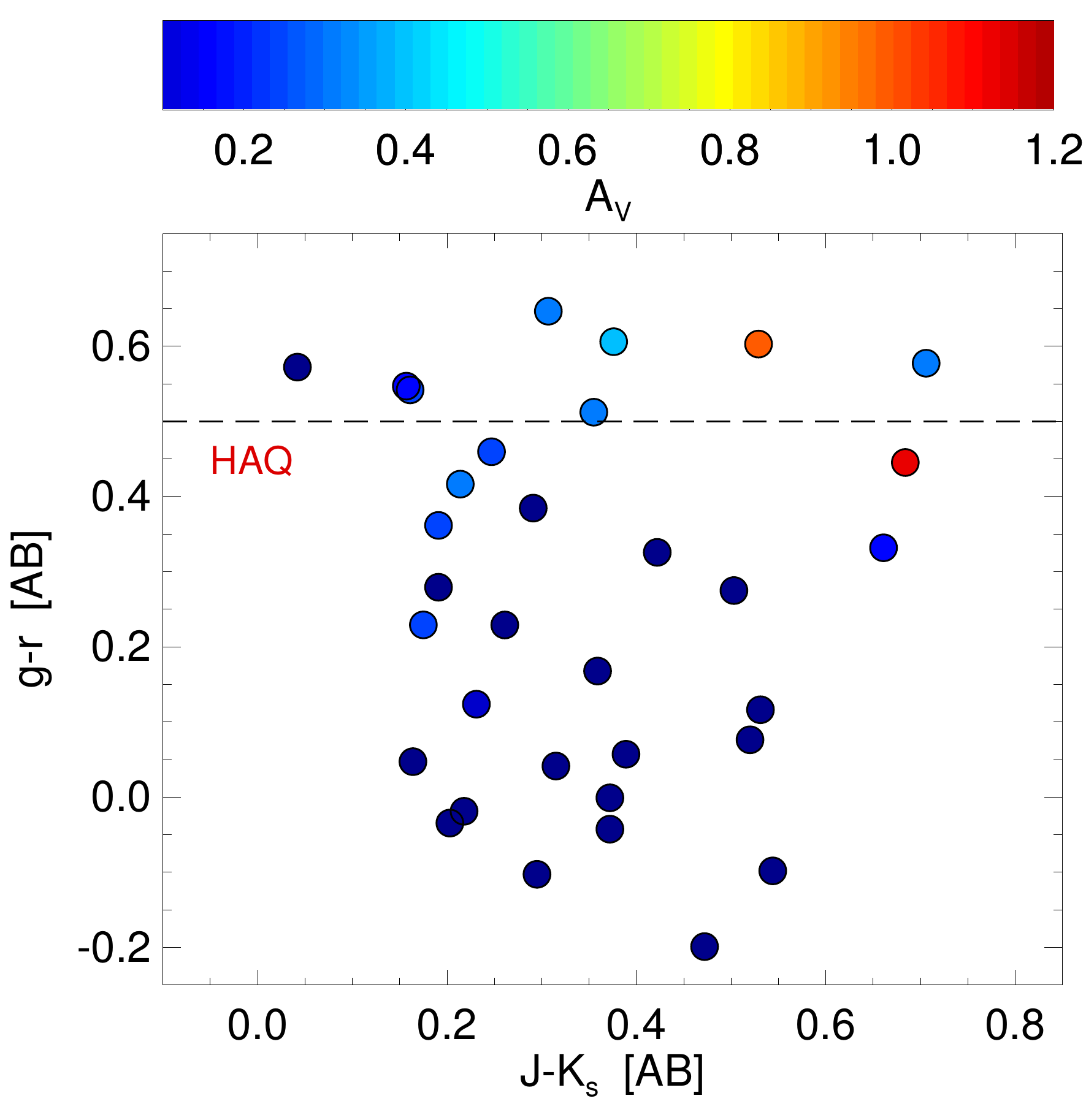}
            \caption[Network2]%
            {Optical/NIR color-color diagram of the 33 quasars from the parent sample,
color-coded as a function of their reddening, $A_V$ (see the top legend). The
HAQ $g-r$ color cut is shown by the dashed line. This conservative cut yields a
reliable sample of reddened quasars only, while more relaxed color criteria are
needed to obtain a more complete sample of reddened quasars, see also 
Fig.~\ref{fig:compquasar} (middle and right panel).}
            \label{fig:Avtrack}
    \end{figure}

Based on our results, we conclude that the HAQ selection is reliable in
detecting reddened quasars. Seven out of the 13 reddened quasars are not 
selected, but these are all except for one quasars with too little reddening to pass
the $g-r$ color cut, see Fig.~\ref{fig:Avtrack}. One apparently reddened 
quasar, CQ095938.6+023316.7, looks blue in the optical spectrum, but the NIR 
photometry has substantial excess over that expected from an unreddened quasar 
template. This could be explained by a significant contribution of light from 
the host galaxy to the NIR photometry. This is consistent with the morphology 
of the object in the HST image, which has significant extended fuzz around the 
point spread function of the quasar. The KX, X-ray, and MIR selections are more complete, but 
not as tailored to select exclusively reddened quasars.

At redshifts below $z \lesssim 1$, spatially resolved quasars are dominant, meaning that our parent sample of quasars in this redshift range is not representative of the total population of AGNs. At higher redshifts, however, we found that the population of spatially unresolved sources is representative of the full underlying population of quasars. The HAQ and in general the reddened sub-population of quasars thus make up a considerable fraction of the brightest optically unresolved sources. 
Studies of the nature and evolution of quasars (and in general SMBHs) through cosmic time must include this in their considerations to remove the otherwise distorted perspective originating from optical quasar surveys.

\begin{acknowledgements}
We would like to thank the anonymous referee for a constructive and insightful report that gave valuable suggestions to present the results of this paper in the best possible way.
Furthermore, we would like to thank D. Malesani and C. Grillo for carrying out the observations of the candidate quasars at the NOT. Otherwise we would not have succeeded in obtaining a 100\% redshift completeness of our parent sample. We also wish to thank I. P\^aris for her help with the SDSS-III/BOSS DR12Q sample and the general SDSS/BOSS selection functions.
The research leading to these results has received funding from the European
Research Council under the European Union's Seventh Framework Program
(FP7/2007-2013)/ERC Grant agreement no. EGGS-278202.
JKK acknowledges support from the European Union’s Seventh Framework Programme for research and innovation 
under the Marie-Curie grant agreement no. 600207 with reference DFF-MOBILEX--5051-00115. MV gratefully acknowledges support from the Danish Council for Independent Research via grant no. DFF 4002-00275.
The data presented here were obtained with ALFOSC, which is provided by the Instituto de Astrofisica de Andalucia (IAA) under a joint agreement with the University of Copenhagen and NOTSA.
Based on data products from observations made with ESO Telescopes at the La Silla Paranal Observatory under ESO programme ID 179.A-2005 and on data products produced by TERAPIX and the Cambridge Astronomy Survey Unit on behalf of the UltraVISTA consortium.
Funding for SDSS-III has been provided by the Alfred P. Sloan Foundation, the Participating Institutions, the National Science Foundation, and the U.S. Department of Energy Office of Science. The SDSS-III web site is http://www.sdss3.org/.

SDSS-III is managed by the Astrophysical Research Consortium for the Participating Institutions of the SDSS-III Collaboration including the University of Arizona, the Brazilian Participation Group, Brookhaven National Laboratory, Carnegie Mellon University, University of Florida, the French Participation Group, the German Participation Group, Harvard University, the Instituto de Astrofisica de Canarias, the Michigan State/Notre Dame/JINA Participation Group, Johns Hopkins University, Lawrence Berkeley National Laboratory, Max Planck Institute for Astrophysics, Max Planck Institute for Extraterrestrial Physics, New Mexico State University, New York University, Ohio State University, Pennsylvania State University, University of Portsmouth, Princeton University, the Spanish Participation Group, University of Tokyo, University of Utah, Vanderbilt University, University of Virginia, University of Washington, and Yale University.
\end{acknowledgements}

\bibliographystyle{aa}
\bibliography{ref}

\section{Appendix} \label{sec:app}

\textit{See next page.}


\begin{figure*} [!ht]
        \centering
\caption[] %
        {Here we present all of the images (at the top), spectra (black solid line), and the photometric data points (purple dots in the $u,g,r,i,z,Y,J,H,K_s$ filters) for each of the confirmed quasars. Overplotted is the composite quasar template with and without reddening (red and blue lines, respectively). The redshift and amount of reddening are shown in the upper left corner, while the object name and technique used to select this particular object are listed in the upper right corner. Below each object
we show the SDSS/BOSS selection flags.} 
        \begin{minipage}[c]{0.45\textwidth}
            \centering
\includegraphics[width=\textwidth]{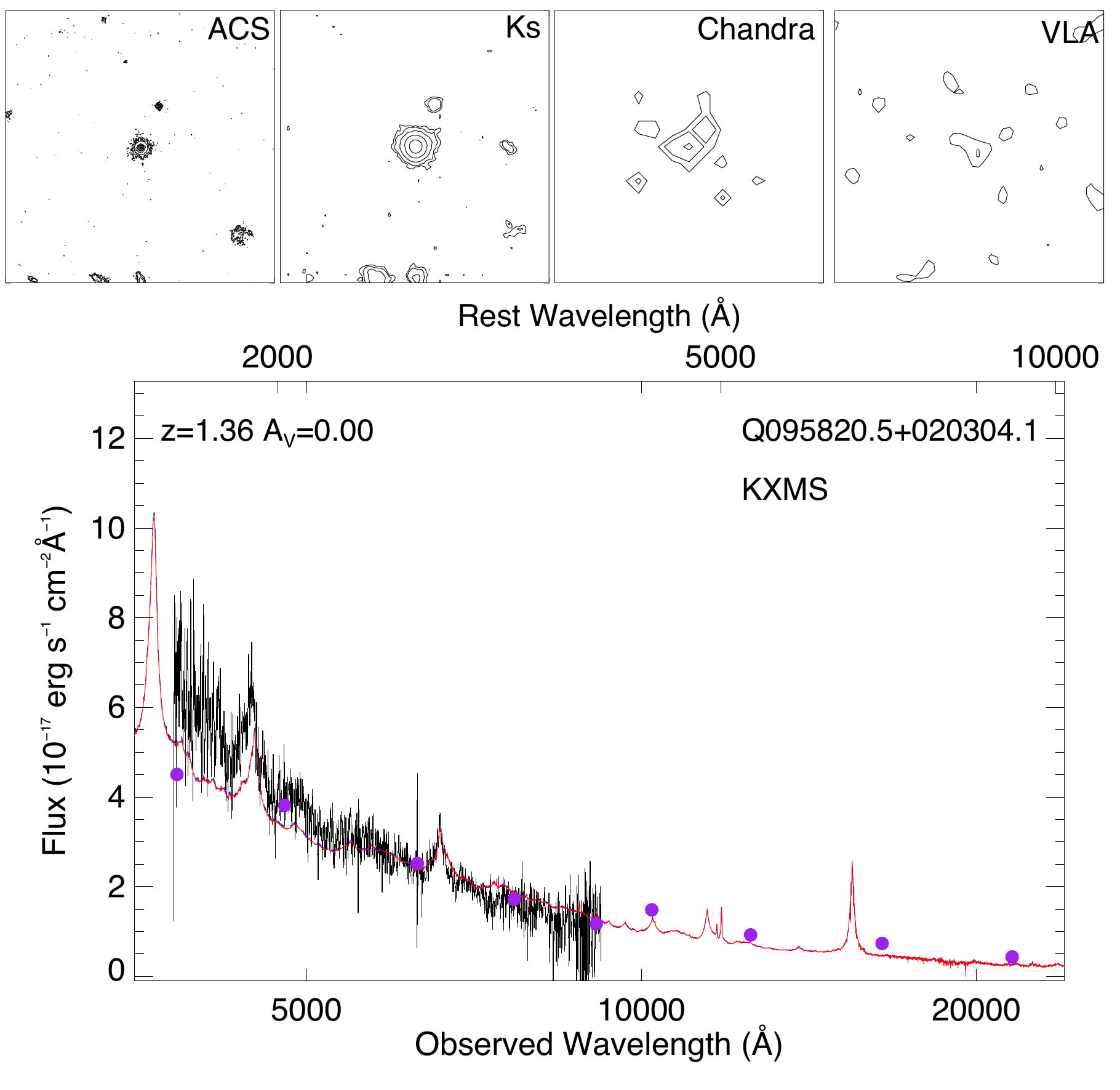} 
            {1: QSO\_FAINT, SERENDIP\_BLUE
            } 
        \end{minipage}
\hspace{1cm}%
        \begin{minipage}[c]{0.45\textwidth}  
            \centering 
                        \includegraphics[width=\textwidth]{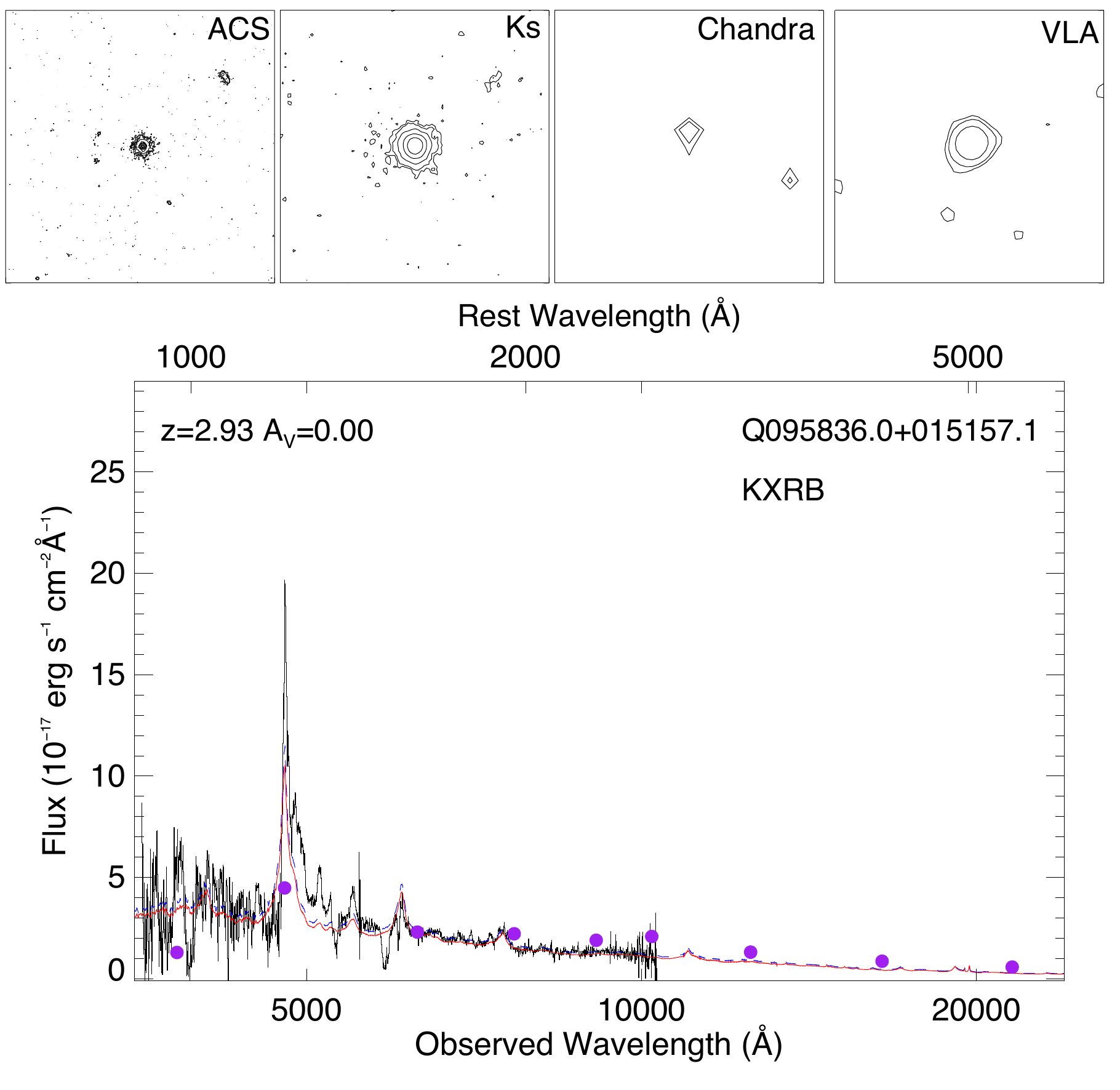} 
                        {3: QSO\_KNOWN\_MIDZ, QSO\_NN, QSO\_KDE, QSO\_CORE\_MAIN}
        \end{minipage} \\[20pt]
        \begin{minipage}[c]{0.45\textwidth}   
            \centering 
            \includegraphics[width=\textwidth]{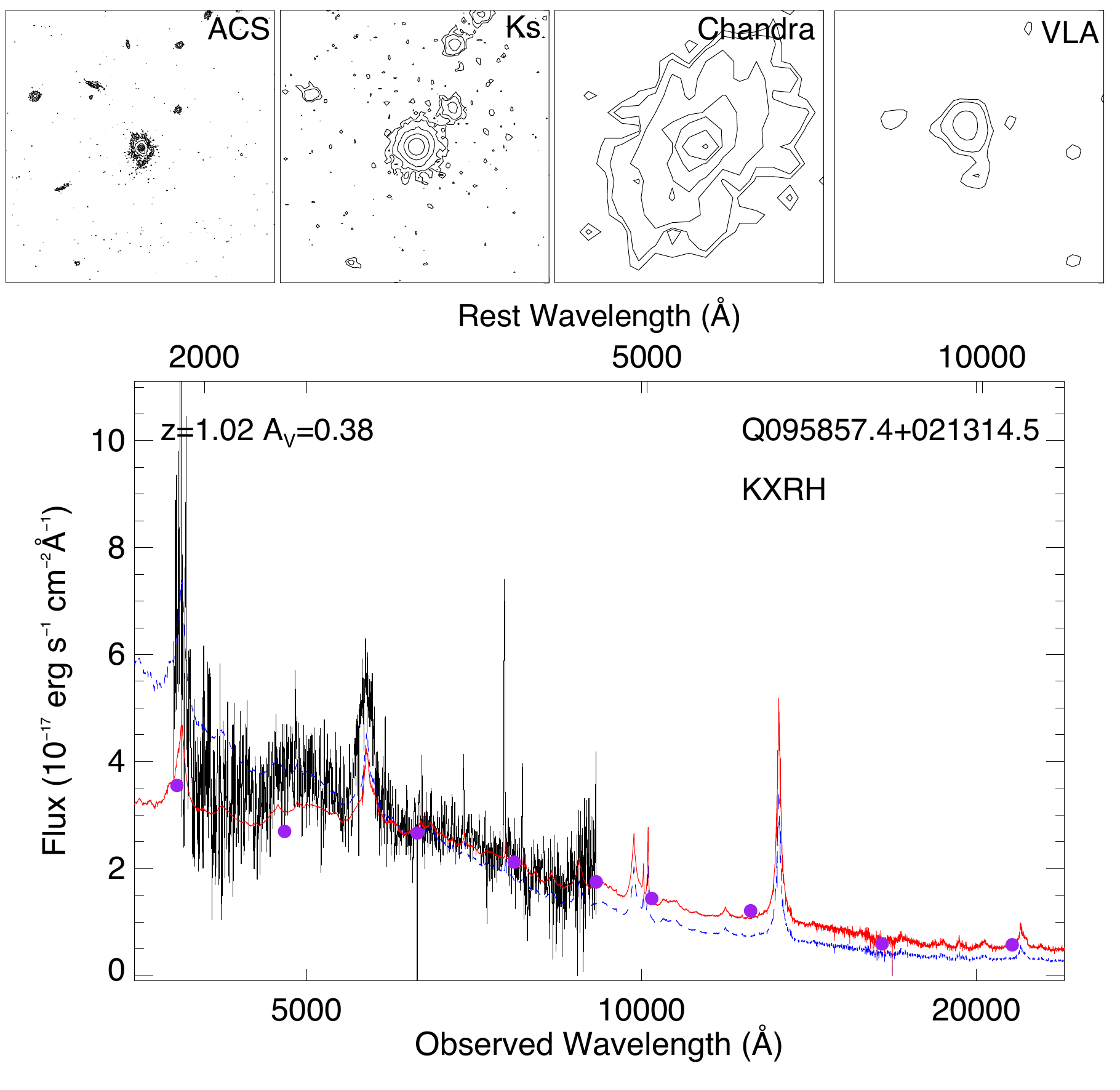} 
                        {4: QSO\_FAINT, ROSAT\_D, ROSAT\_C, ROSAT\_B}
        \end{minipage}
\hspace{1cm}%
        \begin{minipage}[c]{0.45\textwidth}   
            \centering 
            \includegraphics[width=\textwidth]{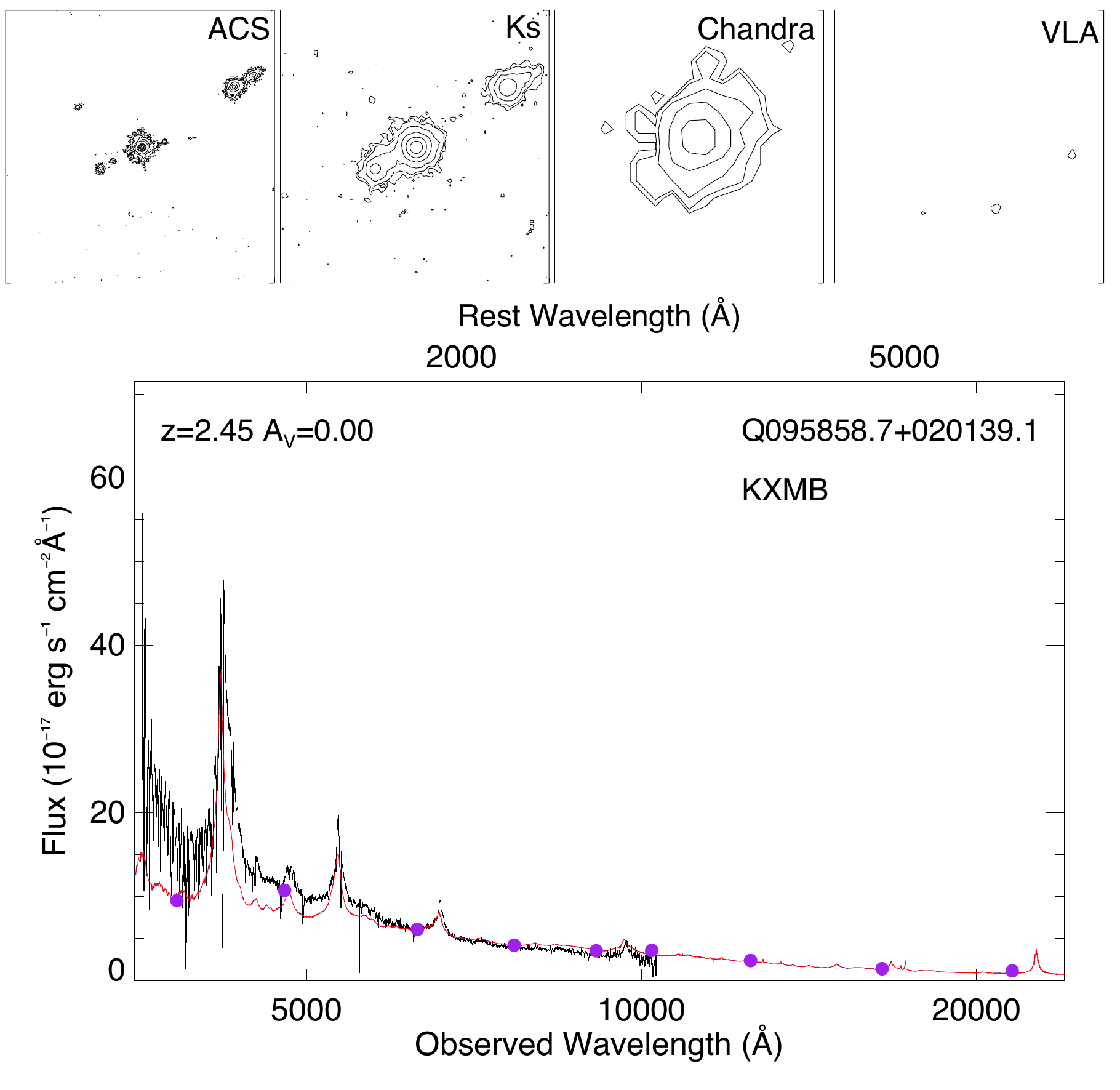} 
            {5: QSO\_KNOWN\_MIDZ, QSO\_NN, QSO\_KDE, QSO\_CORE\_MAIN}    
        \end{minipage}
        \label{fig:specimgquasar}
    \end{figure*}

\begin{figure*}[!ht]
\ContinuedFloat
        \centering
        \caption[] %
        {\textit{Continued}} 
                \begin{minipage}[c]{0.45\textwidth}
                    \centering
                    \includegraphics[width=\textwidth]{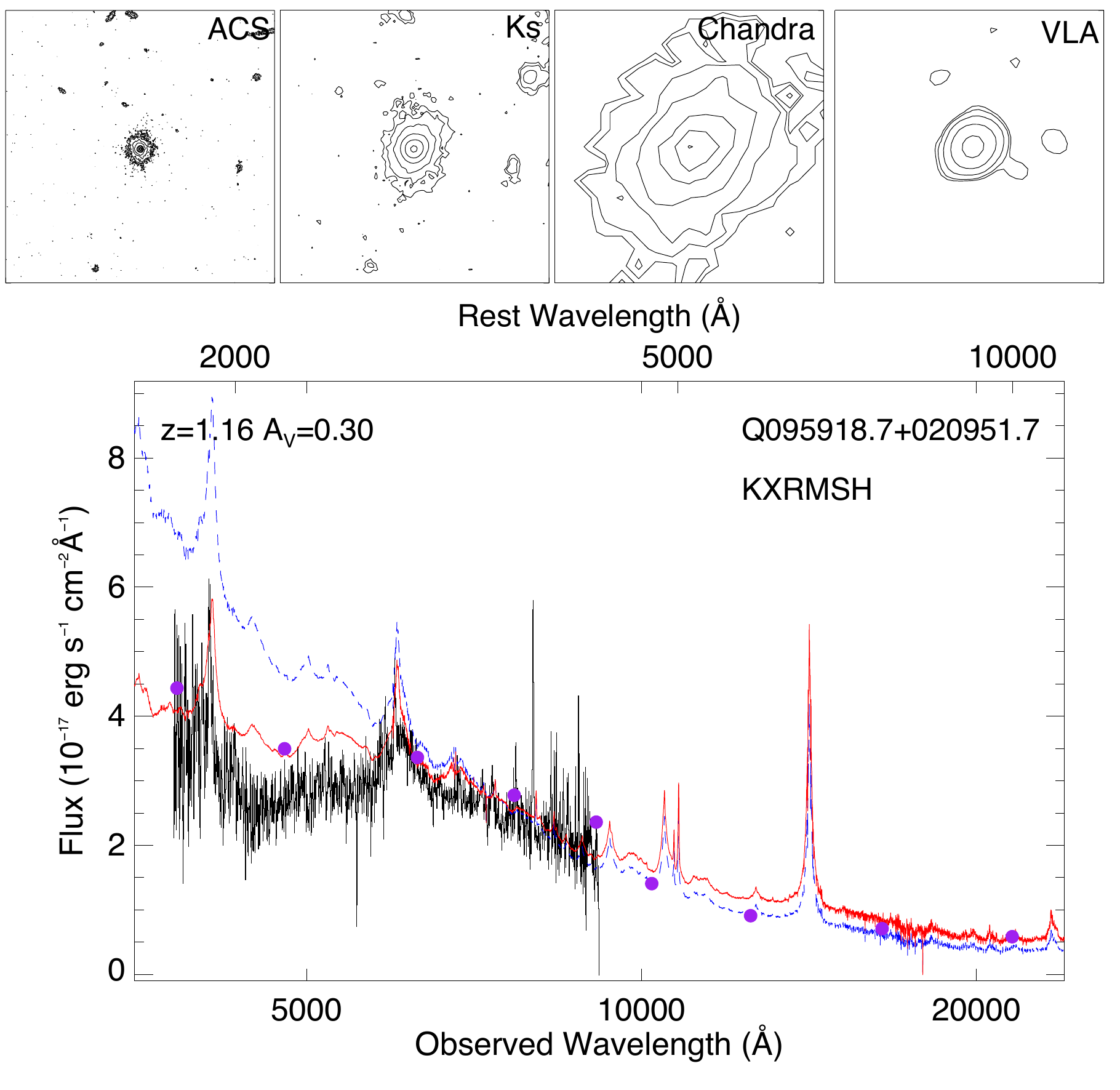}
                    
                    {6: QSO\_FAINT, SERENDIP\_FIRST, SERENDIP\_BLUE}    
                \end{minipage}
                \hspace{1cm}%
        \begin{minipage}[c]{0.45\textwidth}
            \centering
            \includegraphics[width=\textwidth]{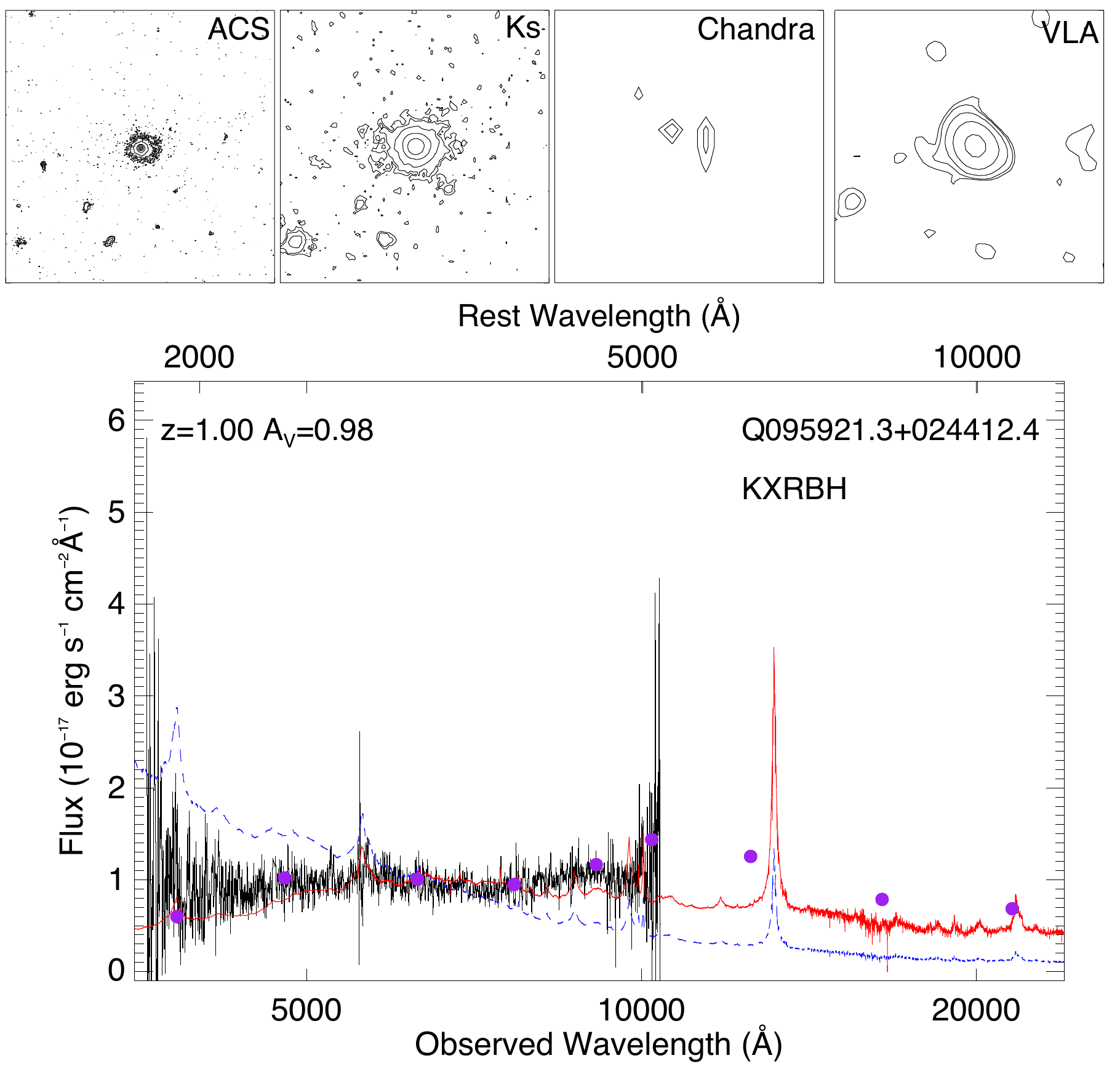}
           
            {7: QSO\_FIRST\_BOSS,~QSO\_KDE,~QSO\_BONUS\_MAIN}    
        \end{minipage} \\[20pt]
        \begin{minipage}[c]{0.45\textwidth}  
            \centering 
            \includegraphics[width=\textwidth]{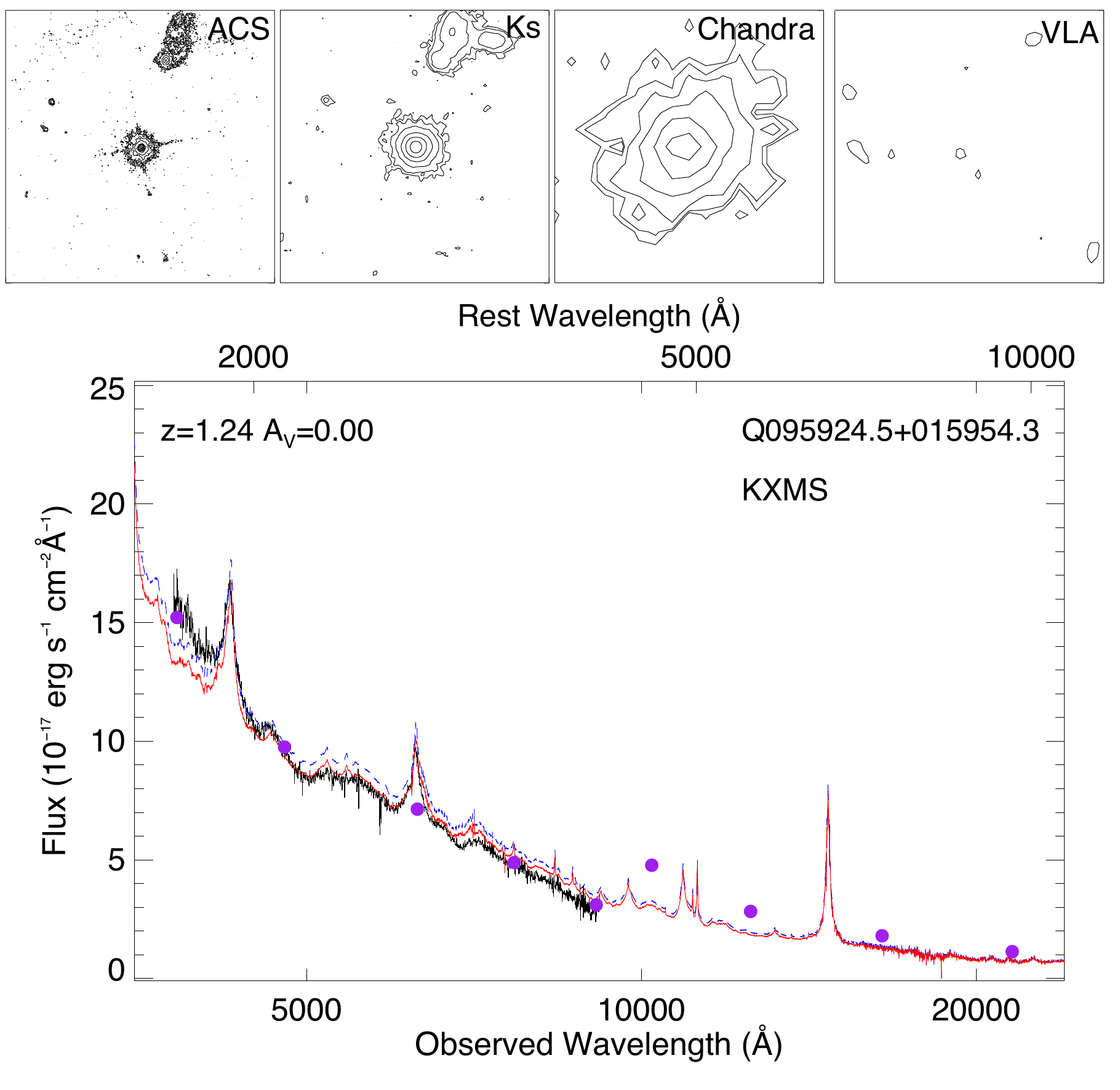}
            
            {9: SEGUE2\_CHECKED} 
        \end{minipage}
                \hspace{1cm}%
        \begin{minipage}[c]{0.45\textwidth}   
            \centering 
            \includegraphics[width=\textwidth]{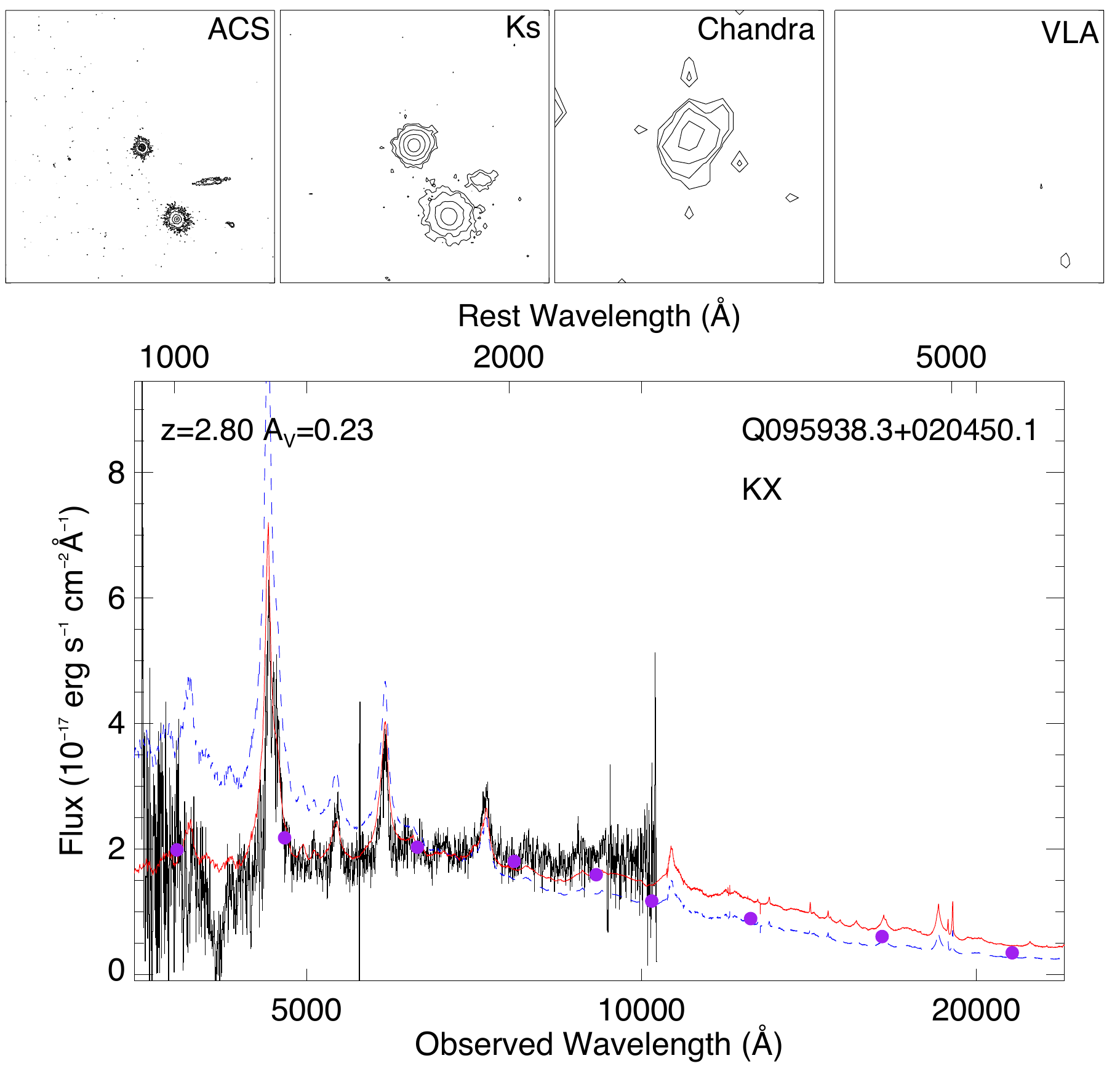}

            {10: CHANDRAv1}
        \end{minipage}
    \end{figure*}

\begin{figure*}[!ht]
\ContinuedFloat
        \centering
        \caption[] %
        {\textit{Continued}} 
                \begin{minipage}[c]{0.45\textwidth}   
                    \centering 
                    \includegraphics[width=\textwidth]{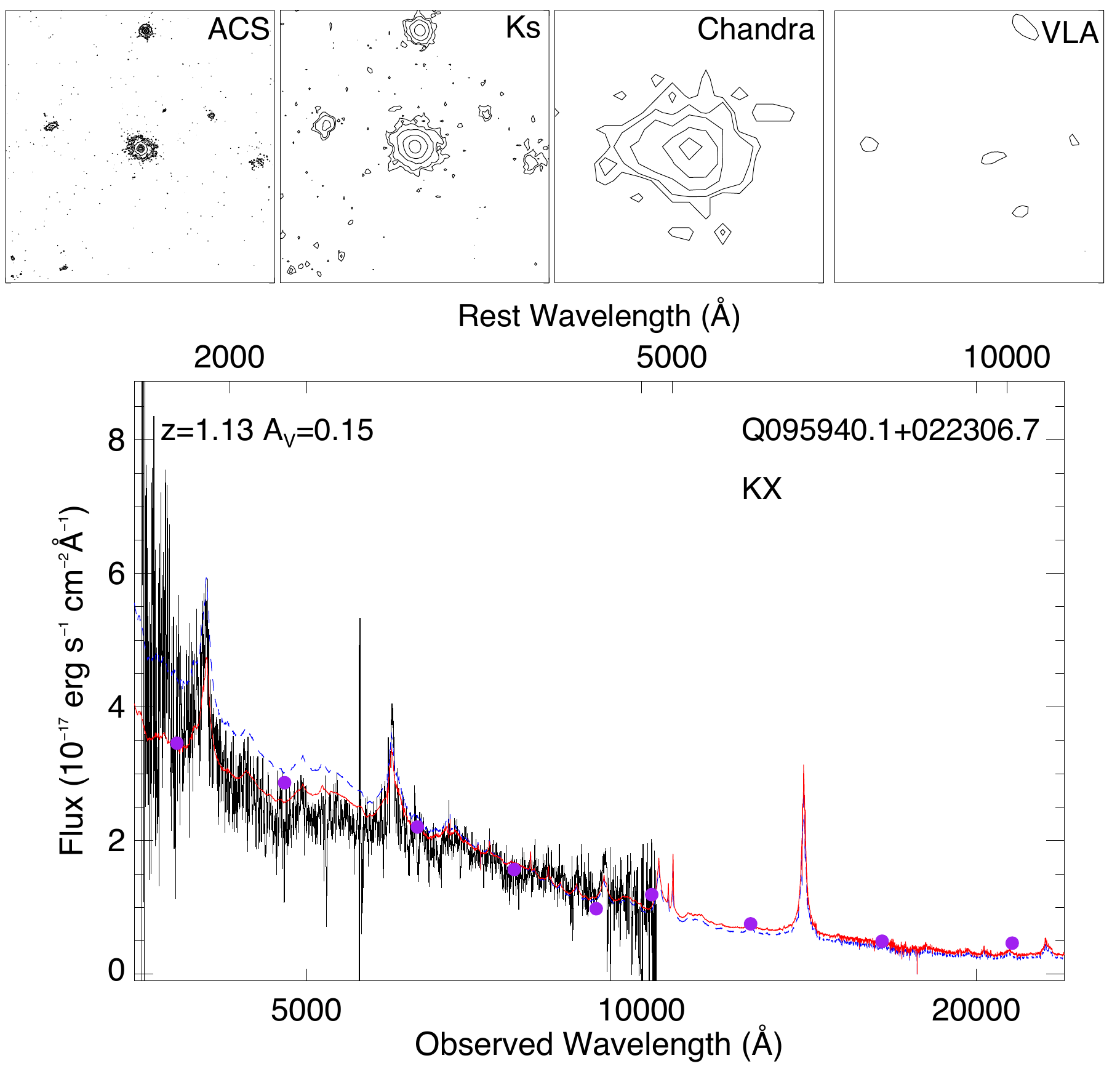}
                    \caption*
                    {12: CHANDRAv1}
                \end{minipage}
                \hspace{1cm}%
        \begin{minipage}[c]{0.45\textwidth}
            \centering
            \includegraphics[width=\textwidth]{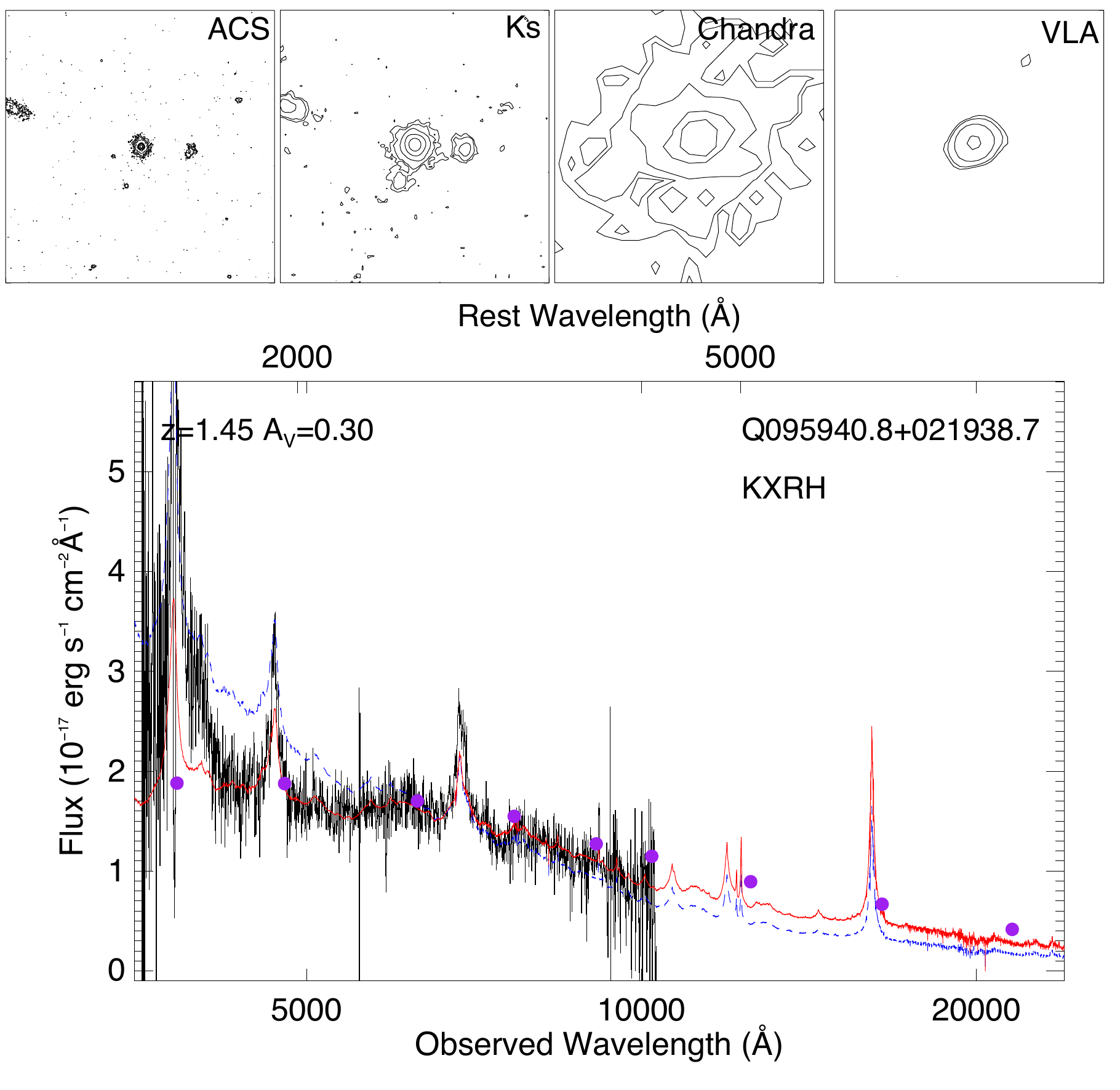}

            {13: CHANDRAv1}
        \end{minipage} \\[20pt]
        \begin{minipage}[c]{0.45\textwidth}  
            \centering 
            \includegraphics[width=\textwidth]{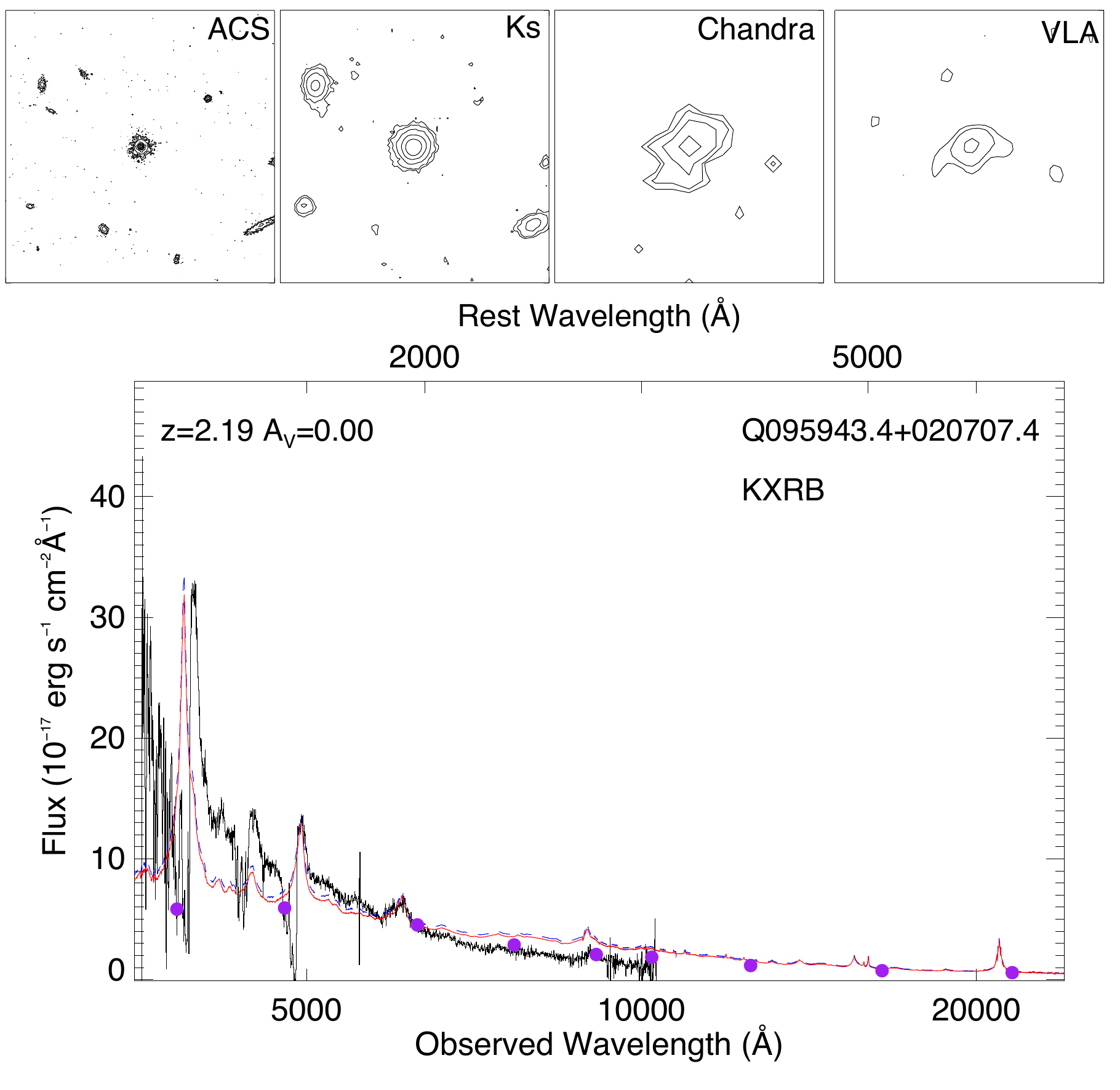}

            { 14: QSO\_KDE, QSO\_BONUS\_MAIN, CHANDRAv1}
        \end{minipage}
        \hspace{1cm}%
        \begin{minipage}[c]{0.45\textwidth}   
            \centering 
            \includegraphics[width=\textwidth]{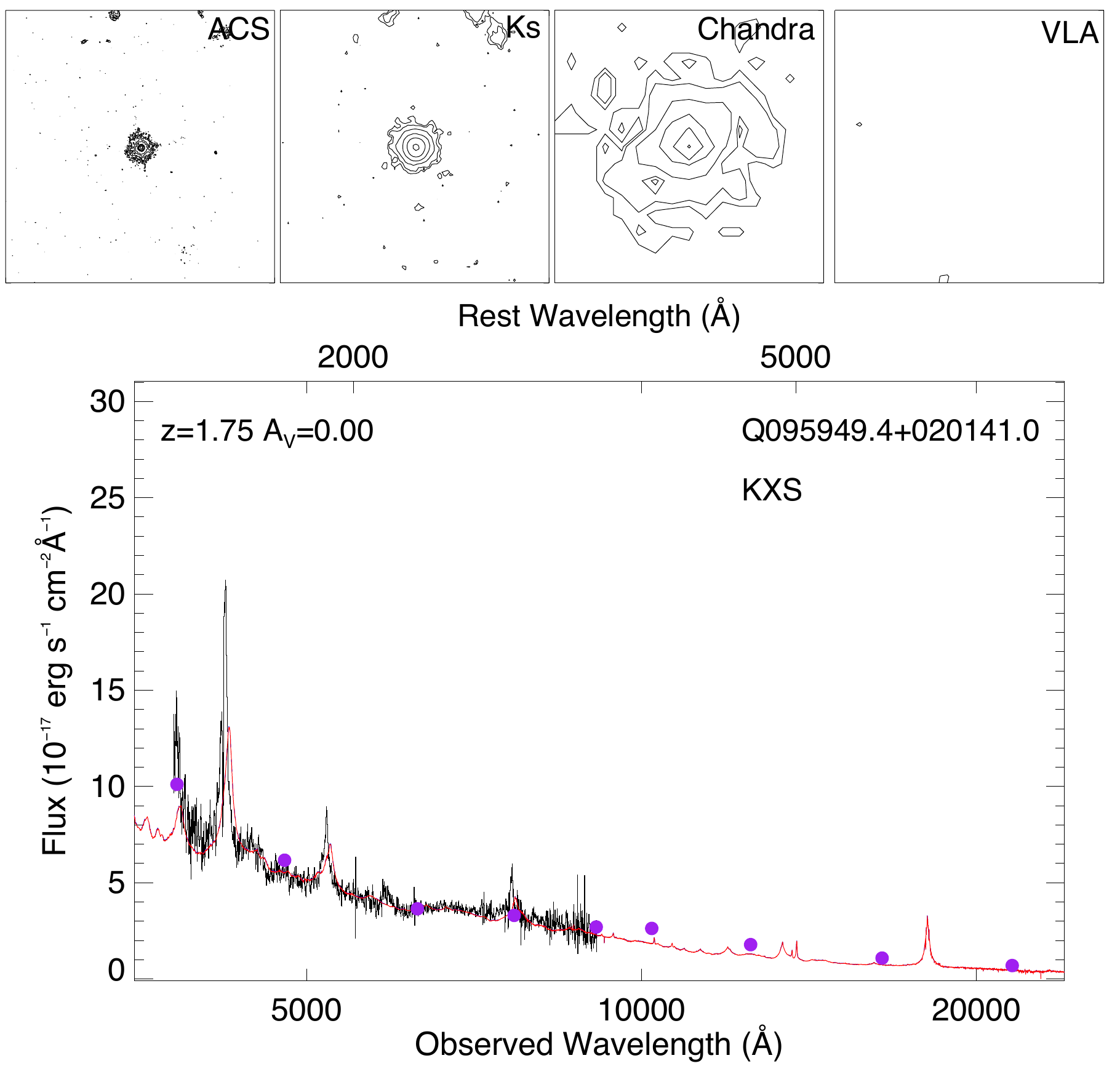}

            {15: SERENDIP\_BLUE, QSO\_SKIRT}
        \end{minipage}
    \end{figure*}
    
\begin{figure*}[!ht]
\ContinuedFloat
        \centering
        \caption[] %
        {\textit{Continued}} 
                \begin{minipage}[c]{0.45\textwidth}   
                    \centering 
                    \includegraphics[width=\textwidth]{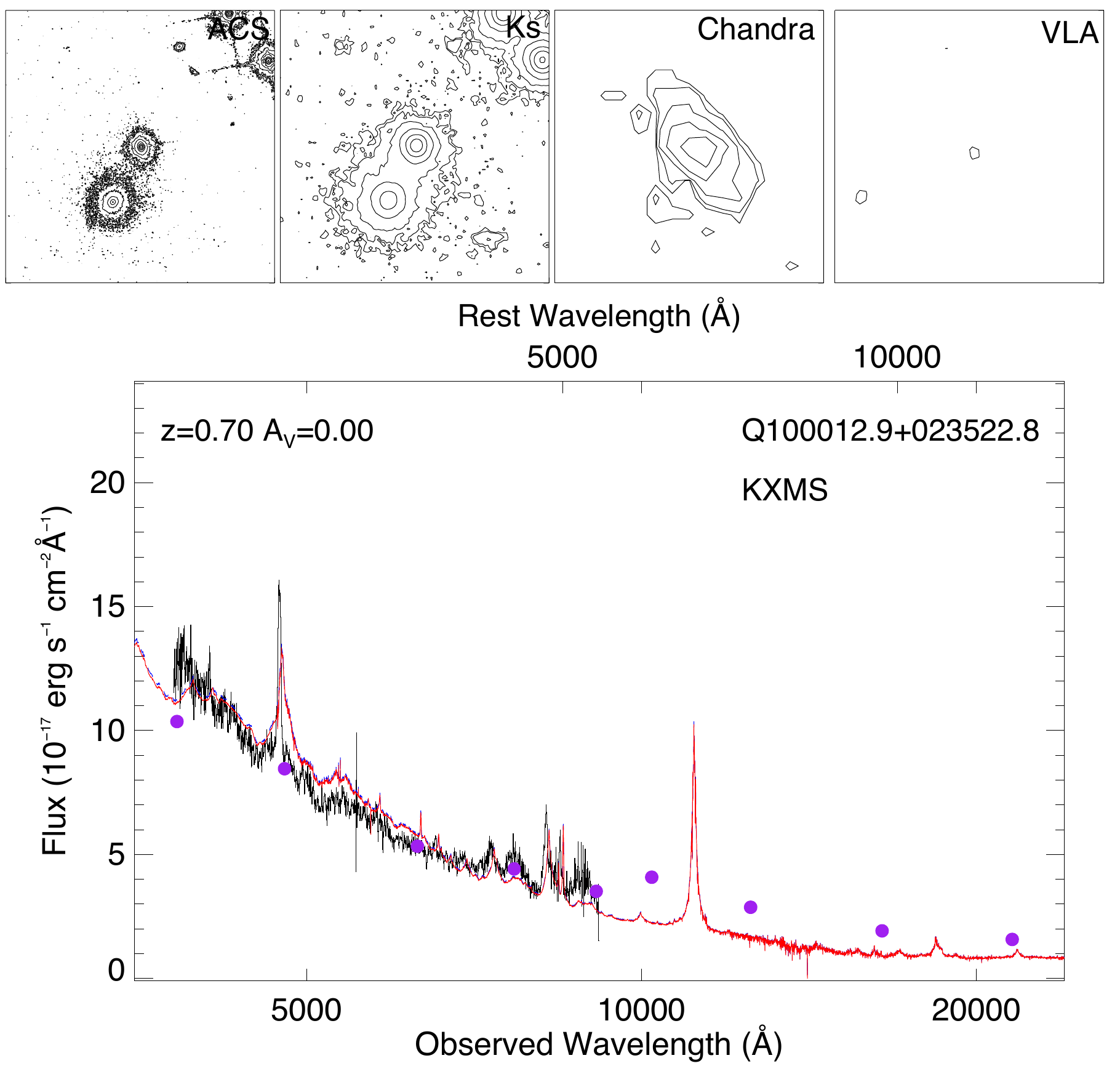}
                    \caption*
                    {17: QSO\_SKIRT}
                \end{minipage}
                \hspace{1cm}%
        \begin{minipage}[c]{0.45\textwidth}
            \centering
            \includegraphics[width=\textwidth]{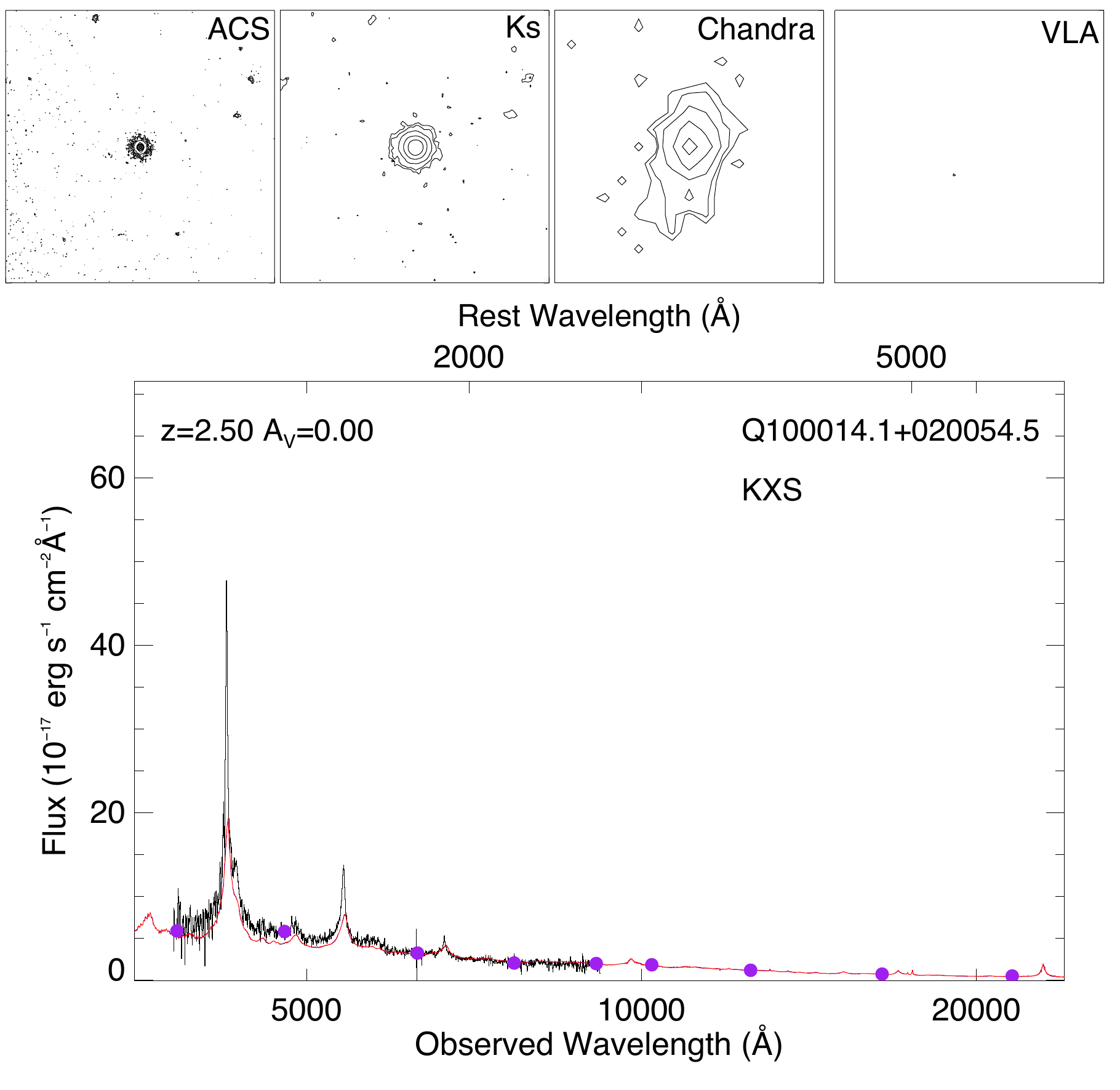}

            {18: QSO\_FAINT, SERENDIP\_BLUE
            }
        \end{minipage} \\[20pt]
        \begin{minipage}[c]{0.45\textwidth}  
            \centering 
            \includegraphics[width=\textwidth]{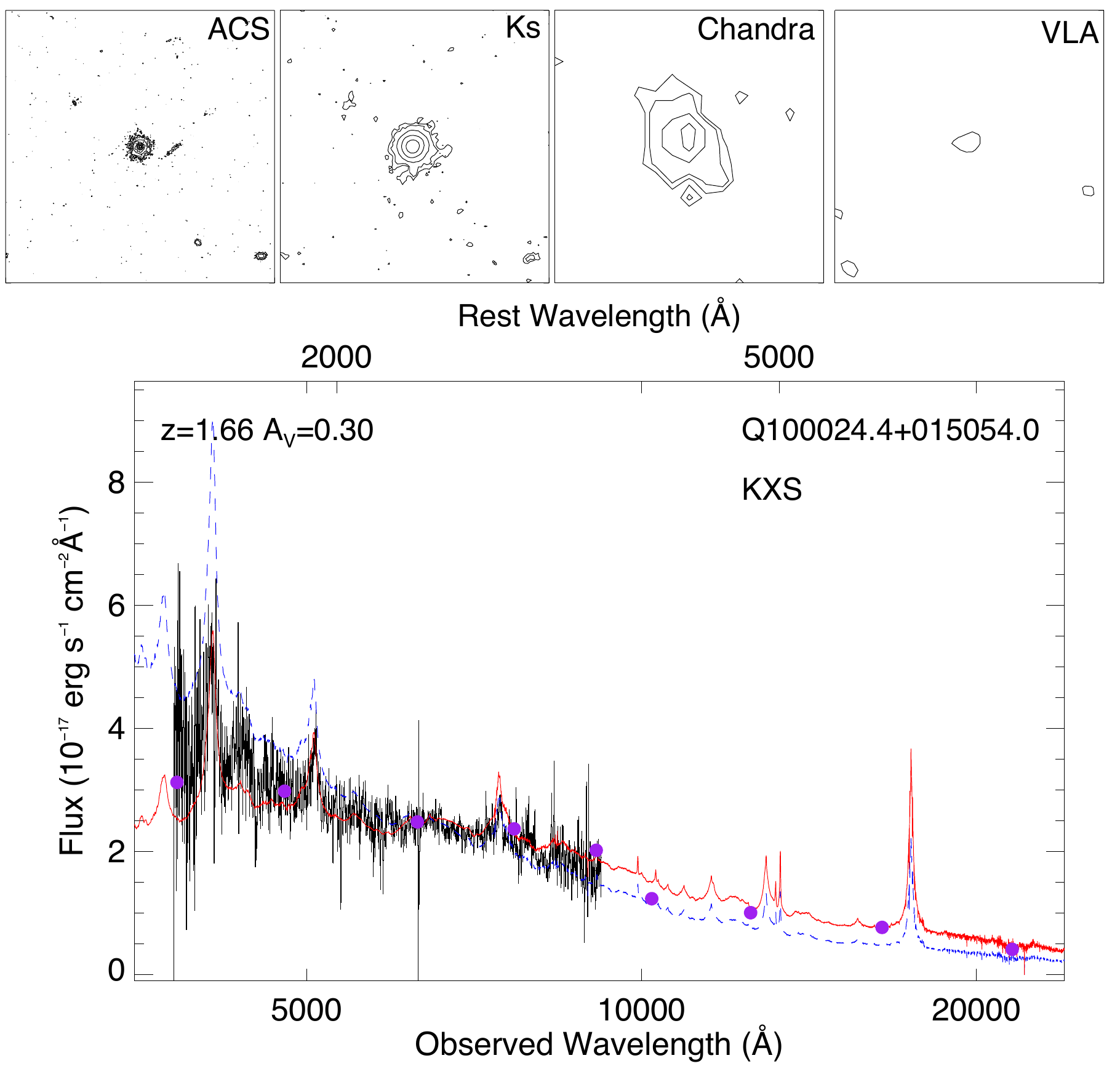}

            {19: QSO\_FAINT, SERENDIP\_BLUE}
        \end{minipage}
\hspace{1cm}%
        \begin{minipage}[c]{0.45\textwidth}   
            \centering 
            \includegraphics[width=\textwidth]{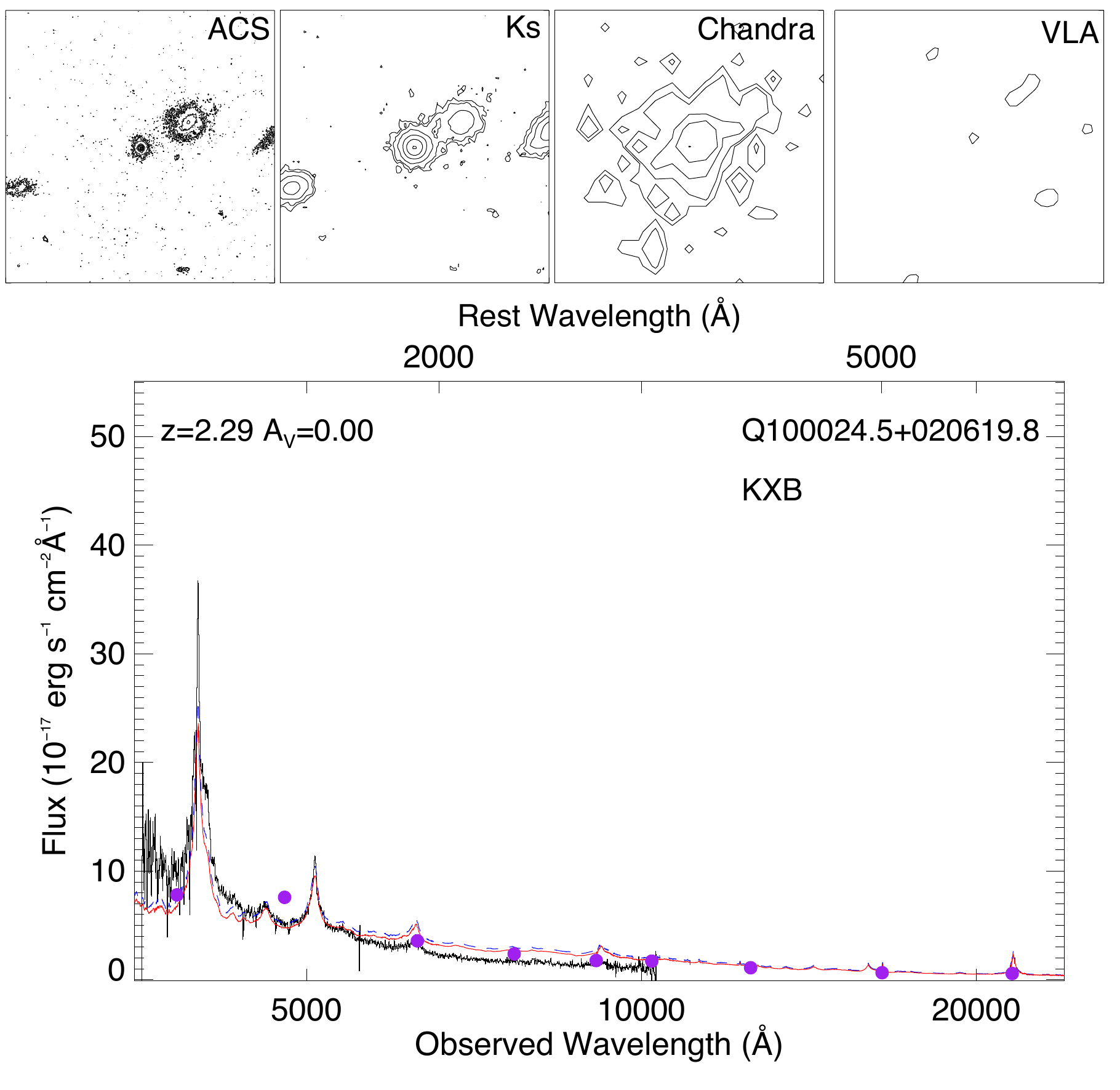}

            {20: QSO\_NN, QSO\_KDE, QSO\_CORE\_MAIN, QSO\_BONUS\_MAIN, CHANDRAv1}
        \end{minipage}
    \end{figure*}
    
\begin{figure*}[!ht]
\ContinuedFloat
        \caption[] %
        {\textit{Continued}} 
        \centering
                \begin{minipage}[c]{0.45\textwidth}   
                    \centering 
                     \includegraphics[width=\textwidth]{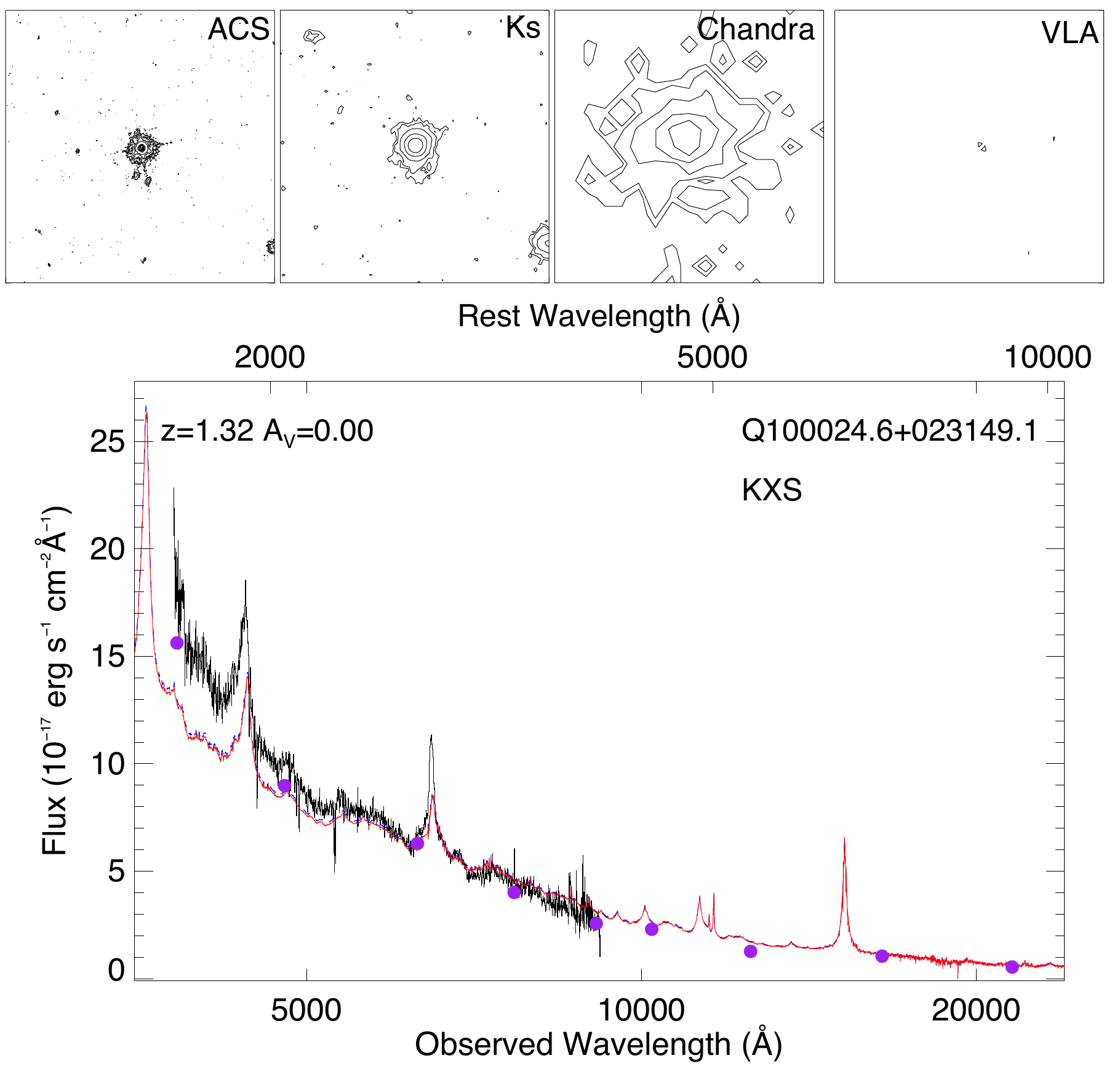}
                     \caption*
                     {21: SERENDIP\_BLUE, QSO\_SKIRT}
                \end{minipage}
                \hspace{1cm}%
        \begin{minipage}[c]{0.45\textwidth}
            \centering
            \includegraphics[width=\textwidth]{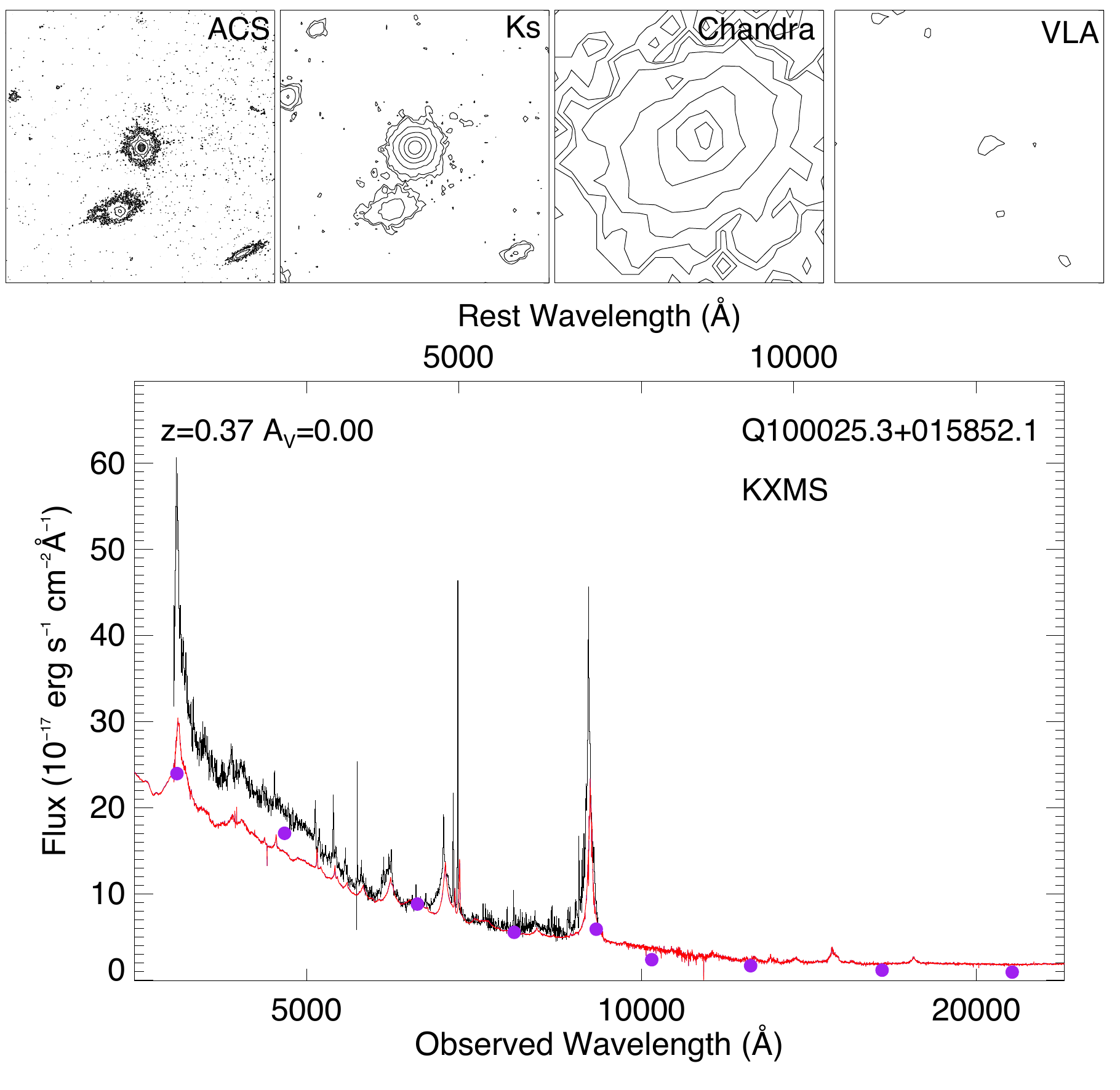}

            {22: QSO\_SKIRT, QSO\_HIZ}
        \end{minipage} \\[20pt]
        \begin{minipage}[c]{0.45\textwidth}  
            \centering 
            \includegraphics[width=\textwidth]{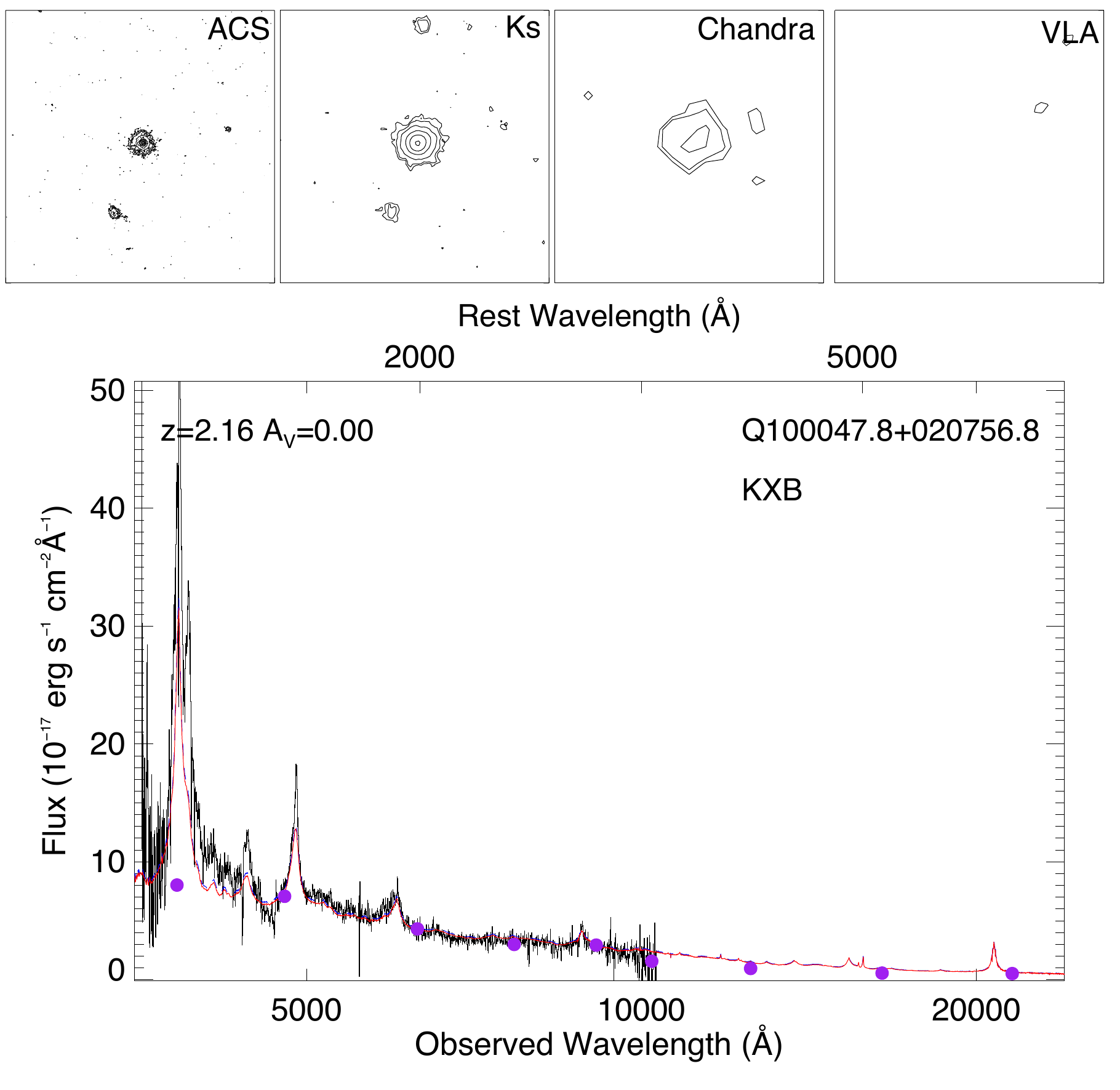}

            {23: QSO\_NN, QSO\_KDE, QSO\_BONUS\_MAIN, CHANDRAv1}
        \end{minipage}
\hspace{1cm}%
        \begin{minipage}[c]{0.45\textwidth}   
            \centering 
            \includegraphics[width=\textwidth]{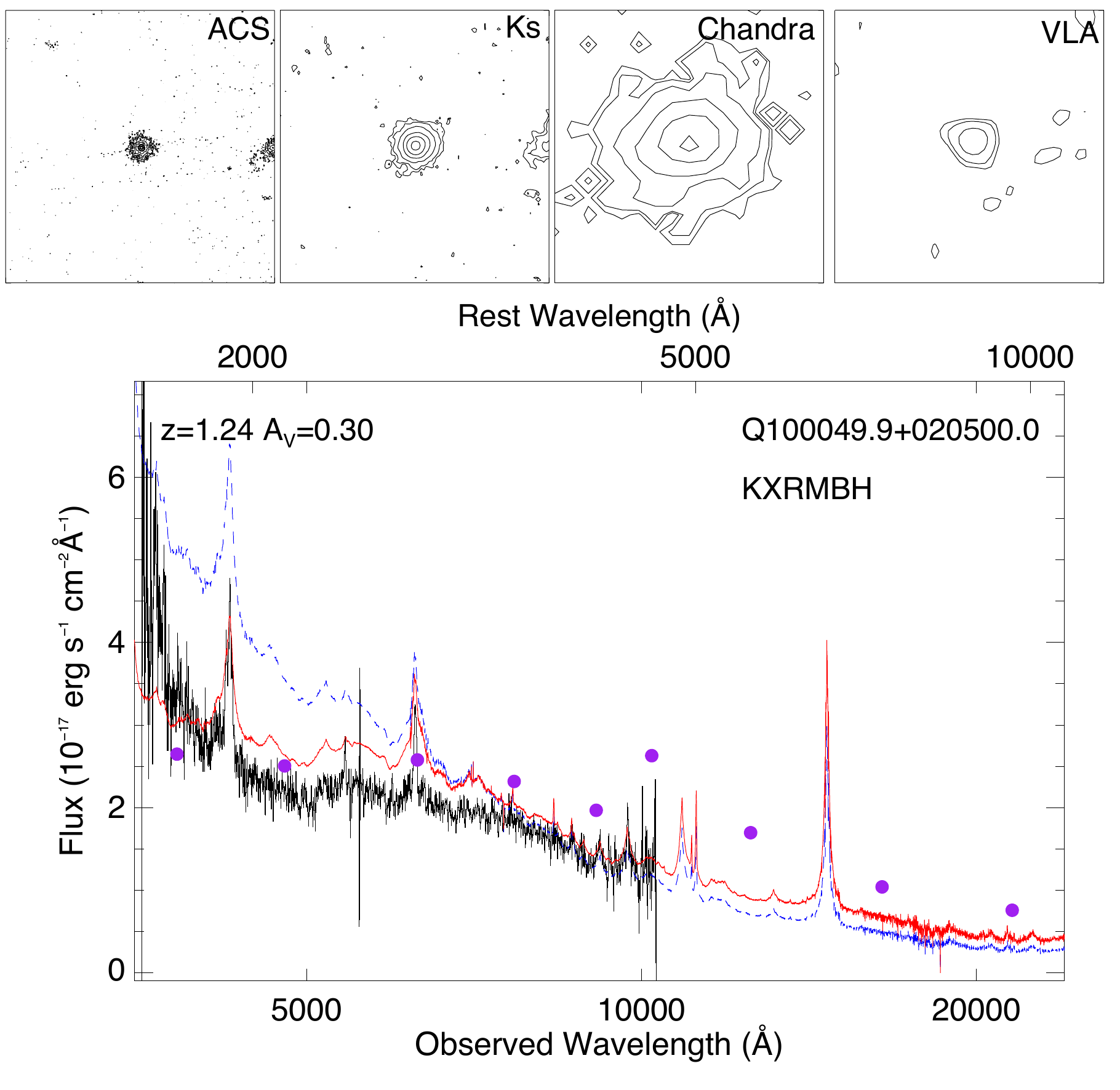}

            {24: QSO\_BONUS\_MAIN, CHANDRAv1}
        \end{minipage}
    \end{figure*}

\begin{figure*}[!ht]
\ContinuedFloat
        \caption[] %
        {\textit{Continued}} 
        \centering
                \begin{minipage}[c]{0.45\textwidth}   
                    \centering 
                    \includegraphics[width=\textwidth]{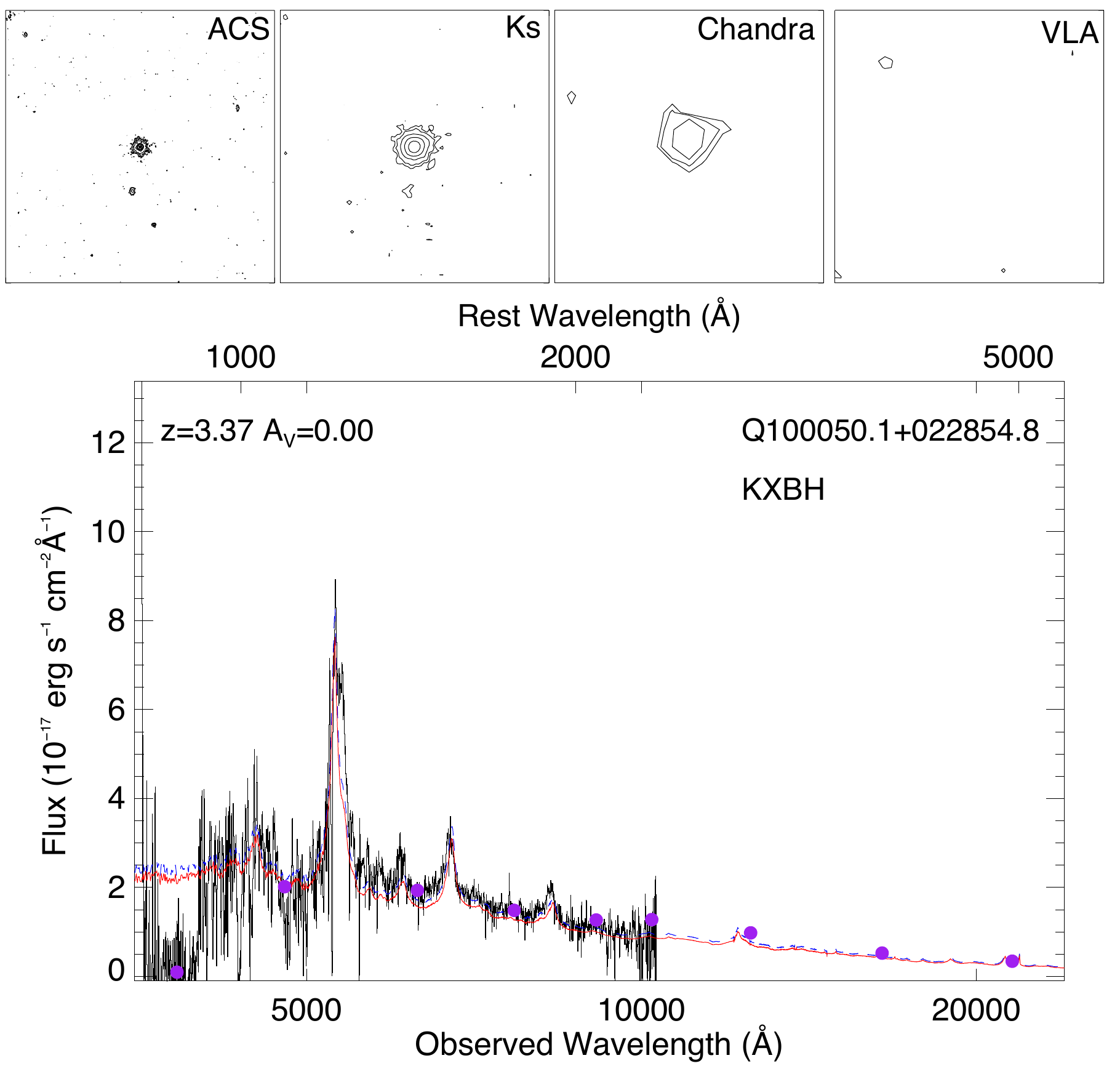}
                    \caption*
                    {25: QSO\_NN, QSO\_KDE, QSO\_CORE\_MAIN, QSO\_BONUS\_MAIN, CHANDRAv1}
                \end{minipage}
                \hspace{1cm}%
        \begin{minipage}[c]{0.45\textwidth}
            \centering
            \includegraphics[width=\textwidth]{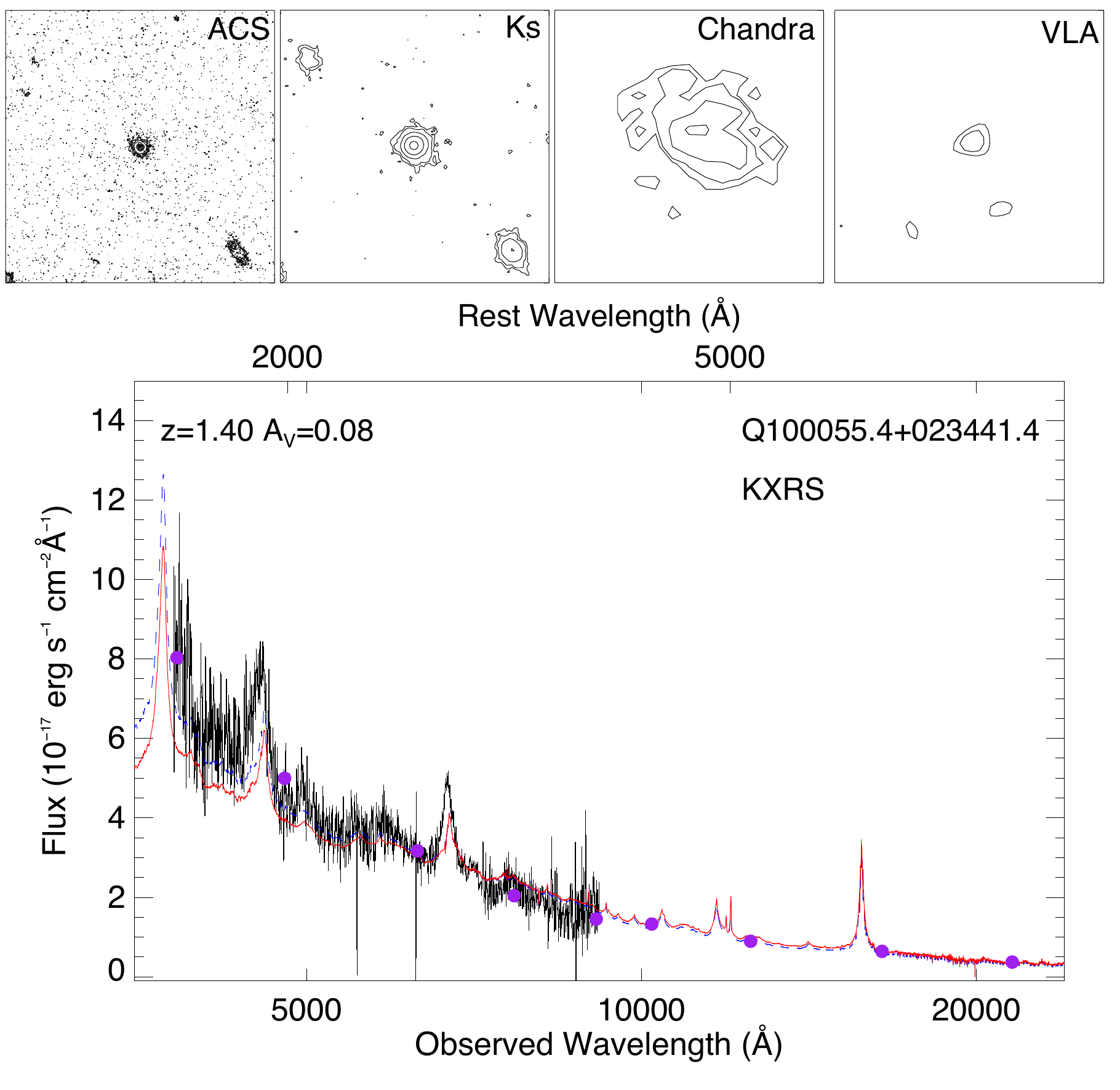}

            {26: QSO\_FAINT, SERENDIP\_BLUE}
        \end{minipage} \\[20pt]
        \begin{minipage}[c]{0.45\textwidth}  
            \centering 
            \includegraphics[width=\textwidth]{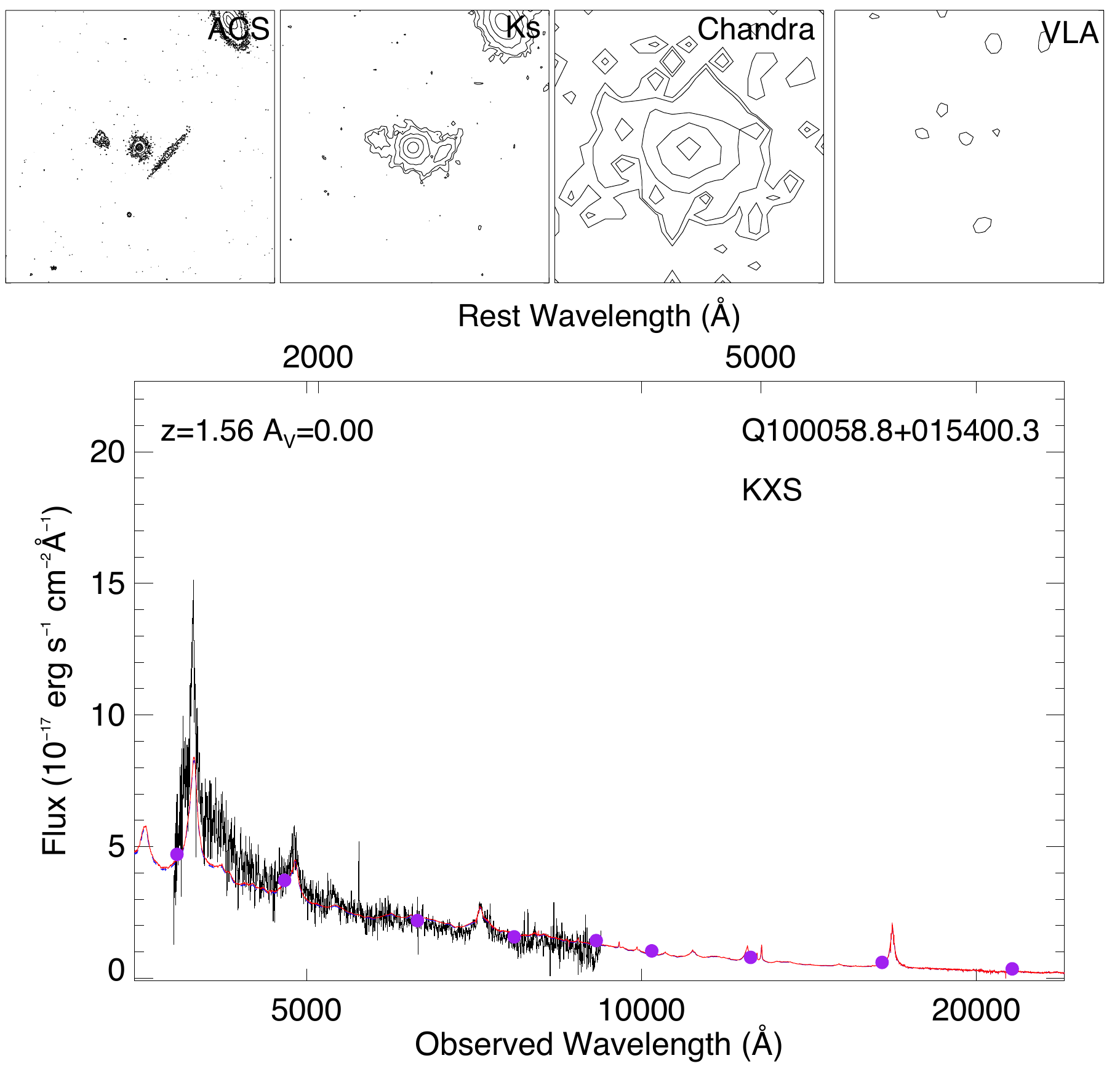}

            {27: QSO\_FAINT, SERENDIP\_BLUE}
        \end{minipage}
\hspace{1cm}%
        \begin{minipage}[c]{0.45\textwidth}   
            \centering 
            \includegraphics[width=\textwidth]{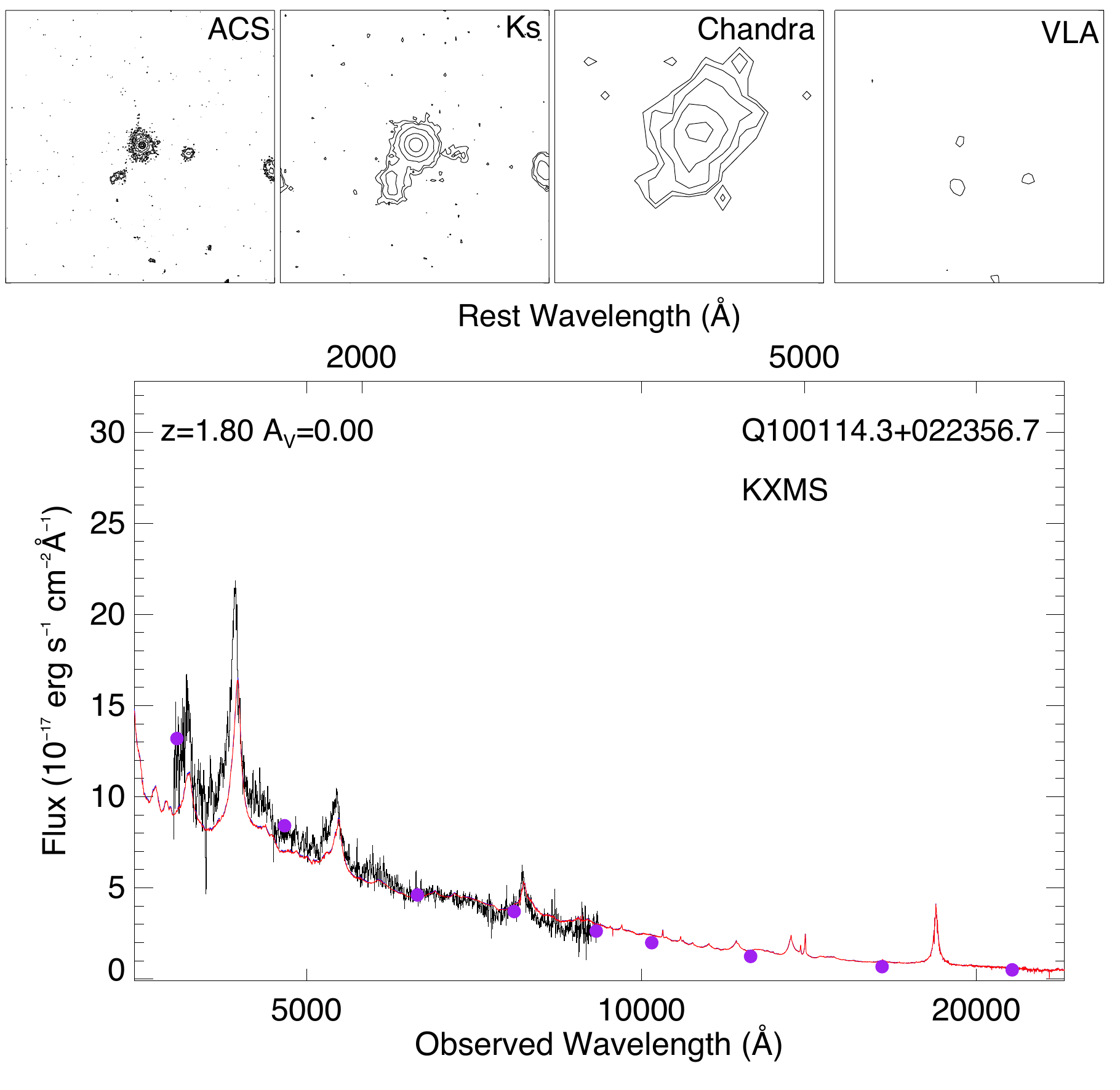}

            {28: SERENDIP\_BLUE, QSO\_SKIRT}
        \end{minipage}
    \end{figure*}
    
\begin{figure*}[!ht]
\ContinuedFloat
        \caption[] %
        {\textit{Continued}} 
        \centering
                \begin{minipage}[c]{0.45\textwidth}   
                    \centering 
                    \includegraphics[width=\textwidth]{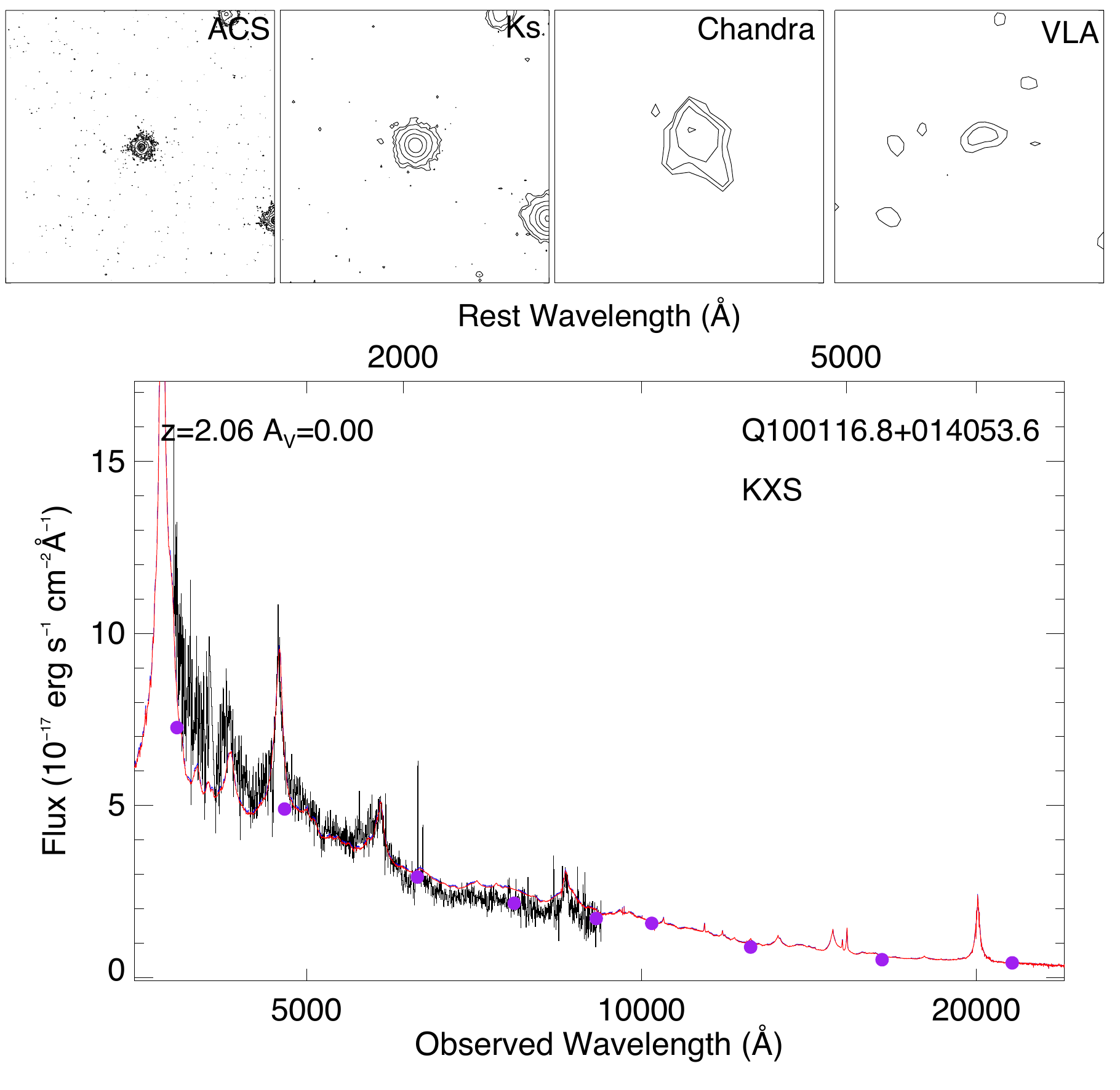}
                    \caption*
                    {29: QSO\_FAINT, SERENDIP\_BLUE}
                \end{minipage}
                \hspace{1cm}%
        \begin{minipage}[c]{0.45\textwidth}
            \centering
            \includegraphics[width=\textwidth]{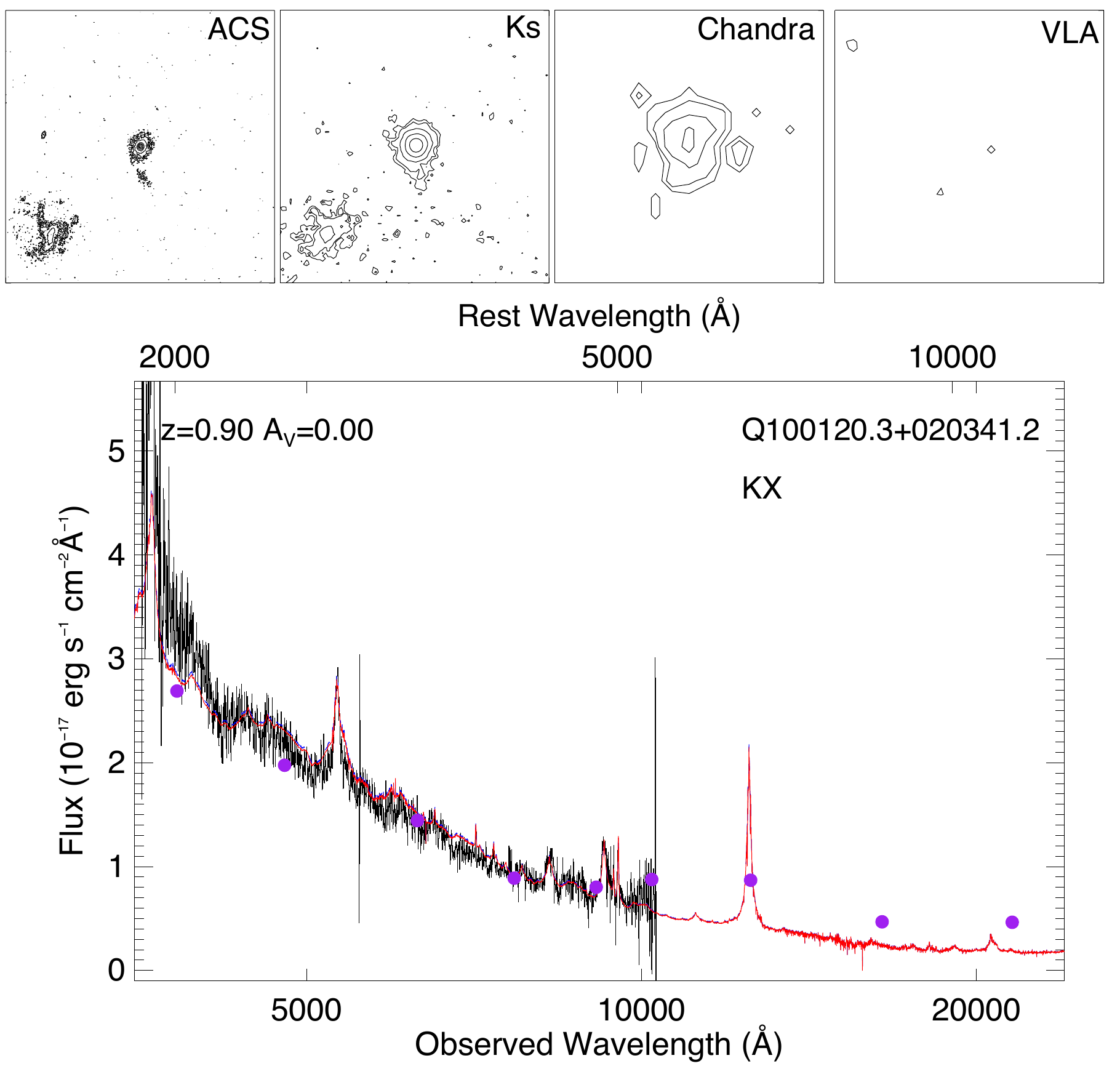}

            {30: CHANDRAv1}
        \end{minipage} \\[20pt]
        \begin{minipage}[c]{0.45\textwidth}  
            \centering 
             \includegraphics[width=\textwidth]{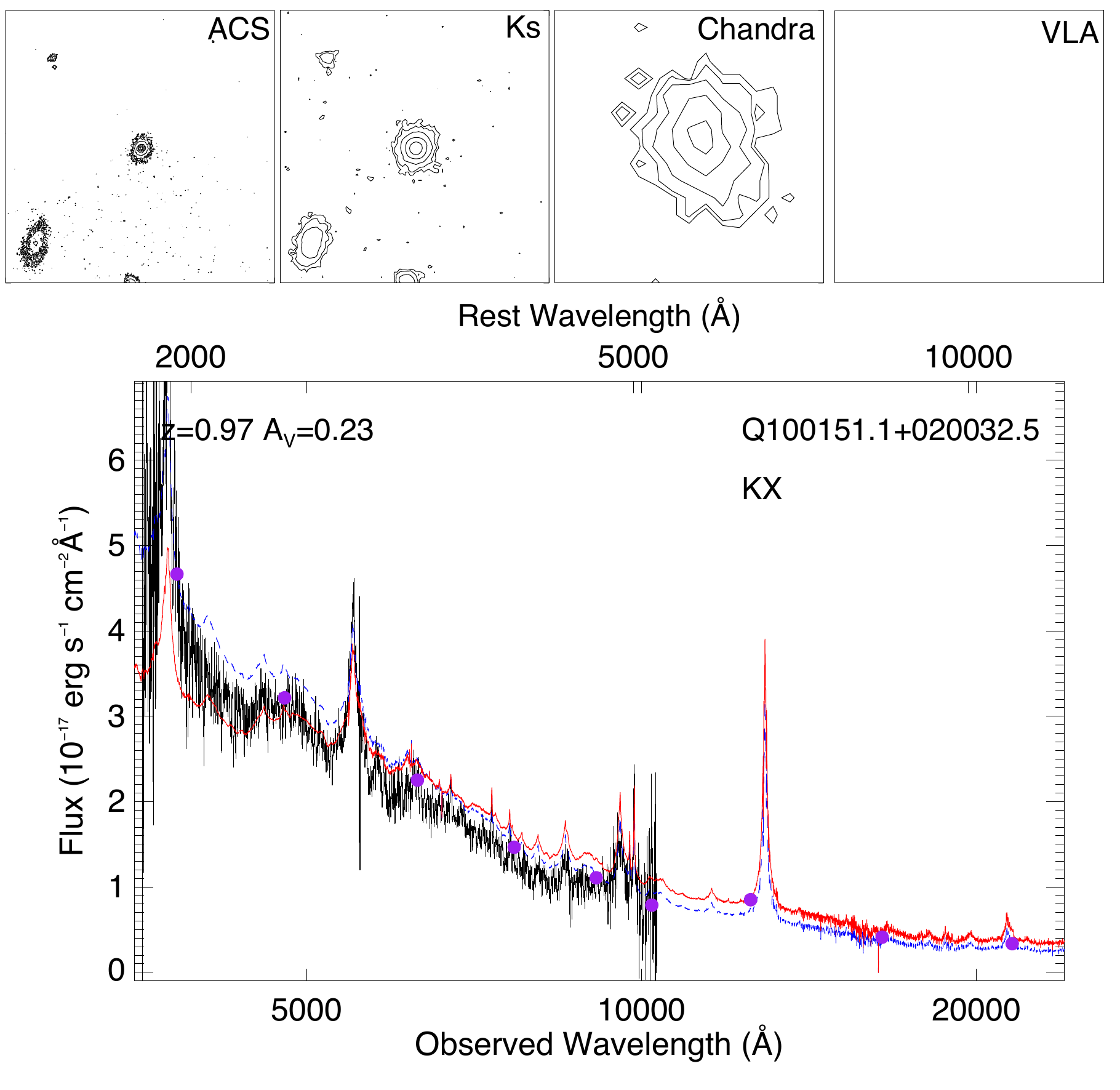}

             {\#31: CHANDRAv1}
        \end{minipage}
\hspace{1cm}%
        \begin{minipage}[c]{0.45\textwidth}   
            \centering 
            \includegraphics[width=\textwidth]{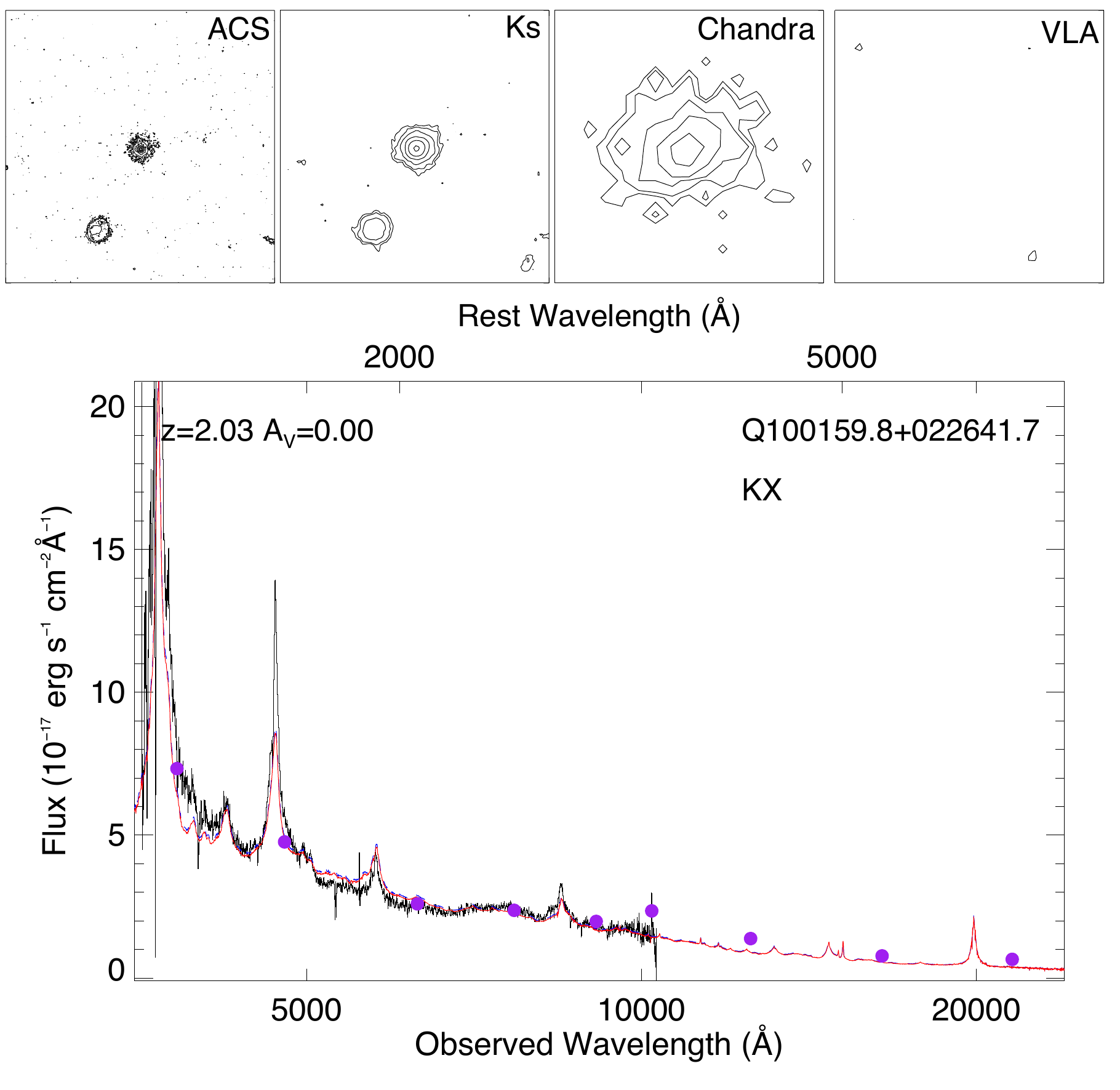}

            {32: XMMBRIGHT}
            \end{minipage}
\end{figure*}

\begin{figure*}[!ht]
\ContinuedFloat
        \caption[] %
        {\textit{Continued}} 
        \centering
                \begin{minipage}[c]{0.45\textwidth}   
                    \centering 
                    \includegraphics[width=\textwidth]{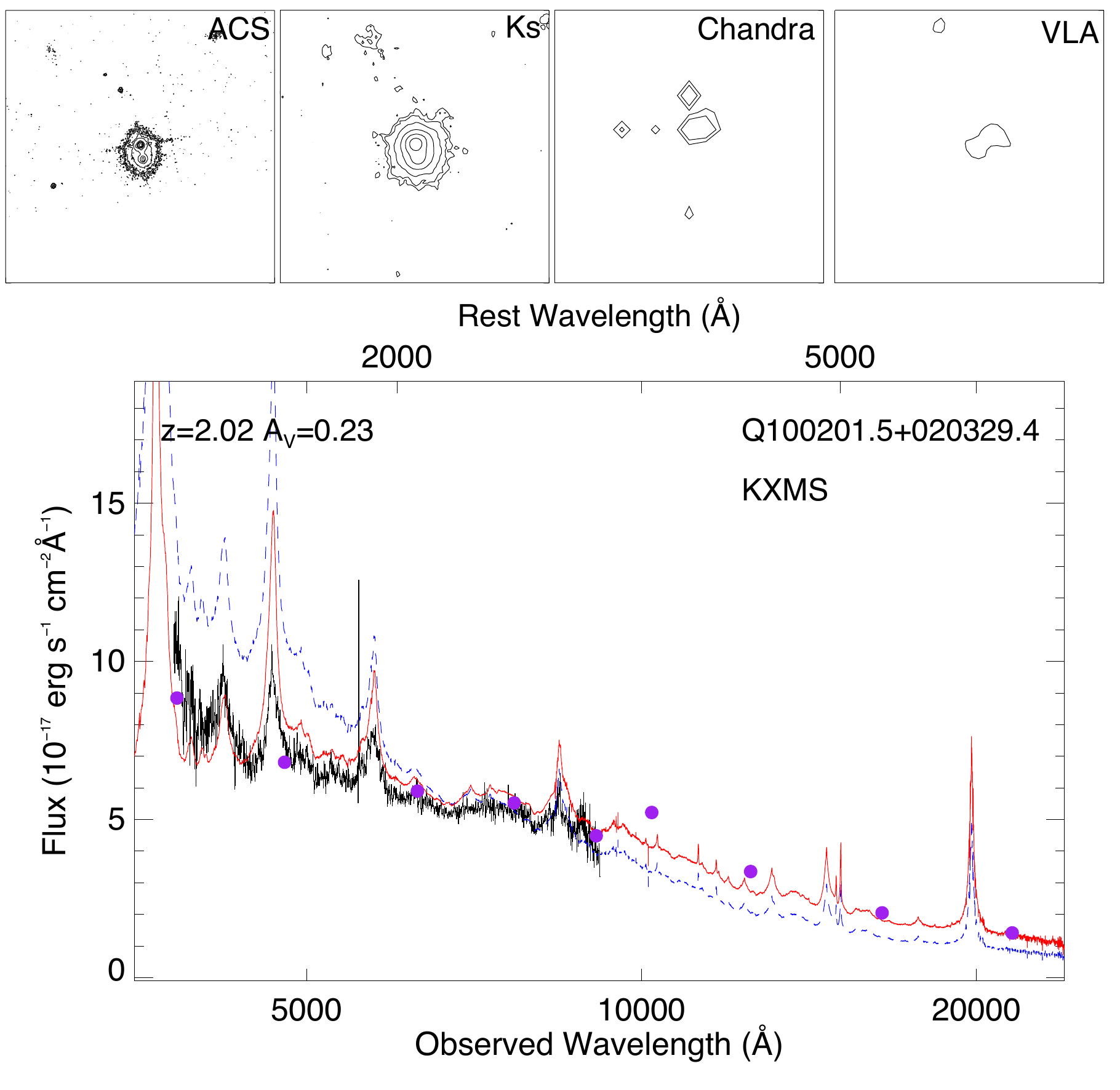}

                    {33: SERENDIP\_BLUE, QSO\_SKIRT}
                \end{minipage}
                \hspace{1cm}%
        \begin{minipage}[c]{0.45\textwidth}
            \centering
            \includegraphics[width=\textwidth]{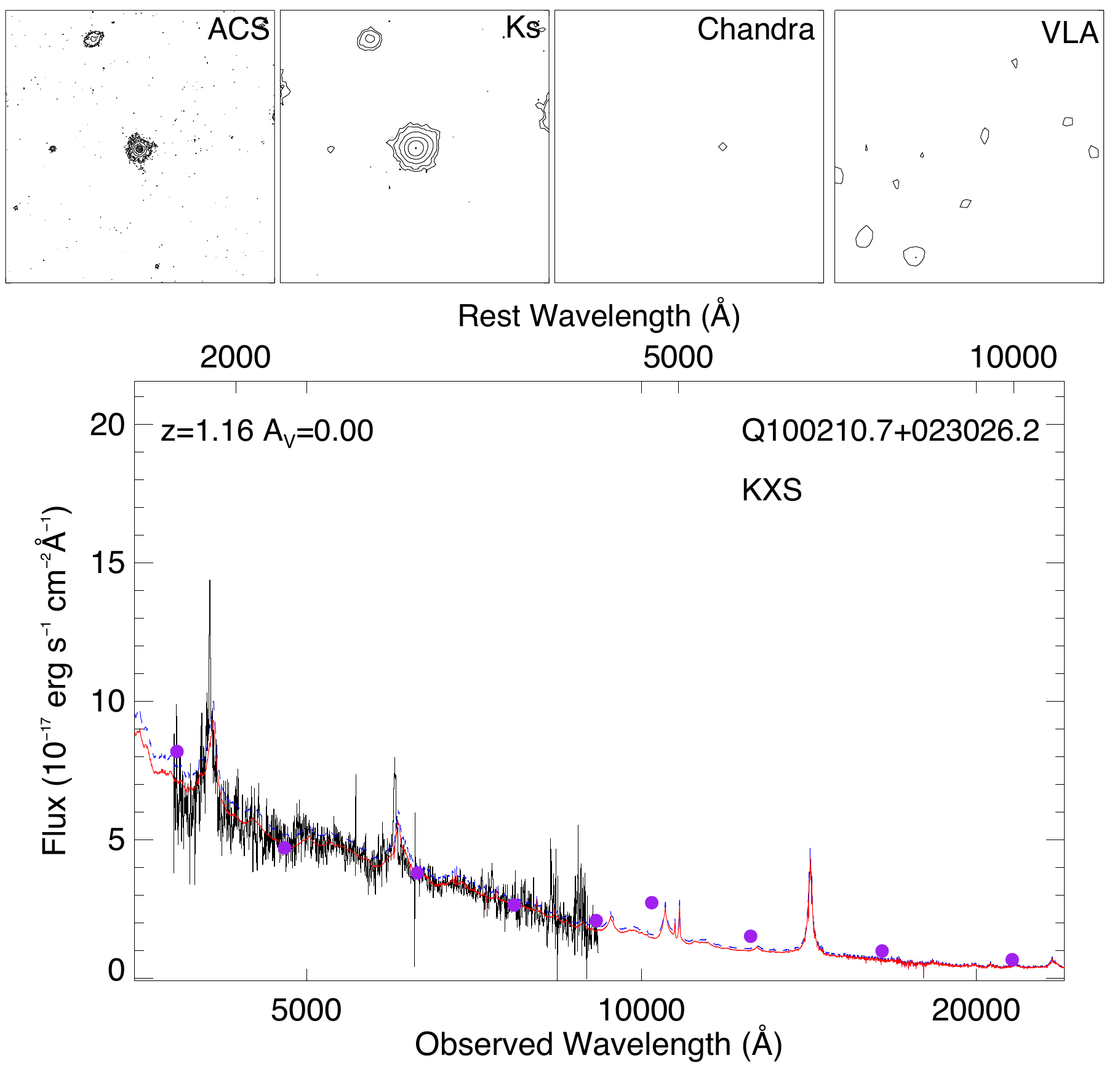}

            {34: QSO\_FAINT, SERENDIP\_BLUE.}
        \end{minipage}
    \end{figure*}

\end{document}